\documentclass[acmsmall,nonacm]{acmart}
\usepackage{cleveref}
\usepackage[inline]{enumitem}
\usepackage{extarrows}
\usepackage{apxproof}
\usepackage{longtable}
\usepackage{adjustbox}
\usepackage{url}
\usepackage{multicol}
\usepackage{stmaryrd}
\usepackage{proof}
\usepackage{bbold}
\usepackage{subcaption}
\usepackage[utf8]{inputenc}
\usepackage{newunicodechar}
\usepackage{underoverlap}

\usepackage{amssymb}
\usepackage{newtxmath}

\usepackage{agda}
\AgdaNoSpaceAroundCode
\renewcommand{\AgdaCodeStyle}{\footnotesize}

\usepackage{tikz}
\usetikzlibrary{decorations.markings}
\usetikzlibrary{quotes,fit,positioning}
\usetikzlibrary{knots}
\tikzstyle{func}=[rectangle,draw,fill=black!20,minimum size=1.9em,text width=2.4em, text centered]
\usetikzlibrary{braids}
\tikzset{>=latex}
\usepackage{tikzit}

\tikzstyle{control}=[fill=black, draw=black, shape=circle, scale=0.5]
\tikzstyle{not}=[fill=white, draw=black, shape=circle]

\tikzstyle{new edge style 0}=[{-|}, {{-{|[scale=2.0]}}}]
\tikzstyle{new edge style 1}=[{|-}, {{{|[scale=2.0]}-}}]
\tikzstyle{new edge style 2}=[<-, {{{Stealth[round,scale=1.3]}-}}]
\tikzstyle{new edge style 3}=[->, {{-{Stealth[round,scale=1.3]}}}]
\tikzstyle{new edge style 4}=[-, draw={rgb,255: red,128; green,128; blue,128}, densely dashdotted]
\tikzstyle{new edge style 5}=[-, draw=red]
\tikzstyle{new edge style 6}=[-, dotted, draw=red]

\usepackage{quiver}

\usepackage{macros}
\usepackage{hott}
\usepackage{syntax}

\usepackage[nocenter]{qtree}

\newtheorem{theorem}{Theorem}
\newtheorem{corollary}[theorem]{Corollary}
\newtheorem{lemma}[theorem]{Lemma}
\newtheorem{definition}[theorem]{Definition}

\newtheoremrep{proprep}[theorem]{Proposition}

\newtheoremrep{theorem}{Theorem}[section]
\newtheoremrep{corollary}[theorem]{Corollary}
\newtheoremrep{proposition}[theorem]{Proposition}
\newtheoremrep{lemma}[theorem]{Lemma}
\newtheoremrep{definition}[theorem]{Definition}

\DeclarePairedDelimiter{\denot}{\llbracket}{\rrbracket}

\newcommand{\scalef}{0.85}
\newcommand{\alt}{~|~}
\newcommand{\zerot}{\mathbb{0}}
\newcommand{\onet}{\mathbb{1}}

\newcommand{\Acon}[1]{\AgdaInductiveConstructor{#1}}
\newcommand{\Afun}[1]{\AgdaFunction{#1}}
\newcommand{\inlv}[1]{\ensuremath{\Acon{inj₁} \; #1}}
\newcommand{\inrv}[1]{\ensuremath{\Acon{inj₂} \; #1}}
\newcommand{\idc}{\AgdaInductiveConstructor{id⟷₁}}

\newcommand{\identlp}{\AgdaInductiveConstructor{unite₊l}}
\newcommand{\identrp}{\AgdaInductiveConstructor{uniti₊l}}

\newcommand{\swapp}{\AgdaInductiveConstructor{swap₊}}
\newcommand{\assoclp}{\AgdaInductiveConstructor{assocl₊}}
\newcommand{\assocrp}{\AgdaInductiveConstructor{assocr₊}}
\newcommand{\identlt}{\AgdaInductiveConstructor{unite⋆l}}
\newcommand{\identrt}{\AgdaInductiveConstructor{uniti⋆l}}

\newcommand{\swapt}{\AgdaInductiveConstructor{swap⋆}}
\newcommand{\assoclt}{\AgdaInductiveConstructor{assocl⋆}}
\newcommand{\assocrt}{\AgdaInductiveConstructor{assocr⋆}}
\newcommand{\absorbr}{\AgdaInductiveConstructor{absorbr}}

\newcommand{\factorzl}{\AgdaInductiveConstructor{factorzl}}
\newcommand{\dist}{\AgdaInductiveConstructor{dist}}
\newcommand{\factor}{\AgdaInductiveConstructor{factor}}

\newcommand{\isoone}{\ensuremath{\leftrightarrow}_1}
\newcommand{\isotwo}{\ensuremath{\leftrightarrow}_2}
\newcommand{\Rule}[4]{
\makebox{{\rm #1}
$\displaystyle
\frac{\begin{array}{l}#2 \\\end{array}}
{\begin{array}{l}#3      \\\end{array}}$
 #4}}
\newcommand{\jdg}[3]{#2 \proves_{#1} #3}
\newcommand{\proves}{\vdash}

\newunicodechar{⊸}{$\multimap$}
\newunicodechar{𝕍}{$\mathbb{V}$}
\newunicodechar{𝕃}{$\mathbb{L}$}
\newunicodechar{𝕄}{$\mathbb{M}$}
\newunicodechar{ℝ}{$\mathbb{R}$}
\newunicodechar{𝕌}{$\mathbb{U}$}
\newunicodechar{𝔹}{$\mathbb{B}$}
\newunicodechar{𝕐}{$\mathbb{Y}$}
\newunicodechar{𝔼}{$\mathbb{E}$}
\newunicodechar{𝔽}{$\mathbb{F}$}
\newunicodechar{𝕋}{$\mathbb{T}$}
\newunicodechar{𝕚}{$\mathbb{i}$}
\newunicodechar{𝟘}{$\mathbb{0}$}
\newunicodechar{𝟙}{$\mathbb{1}$}
\newunicodechar{𝟚}{$\mathbb{2}$}
\newunicodechar{𝟛}{$\mathbb{3}$}
\newunicodechar{𝟠}{$\mathbb{8}$}
\newunicodechar{⟷}{$\leftrightarrow$}
\newunicodechar{₀}{$_{0}$}
\newunicodechar{₁}{$_{1}$}
\newunicodechar{₂}{$_{2}$}
\newunicodechar{₃}{$_{3}$}
\newunicodechar{₄}{$_{4}$}
\newunicodechar{■}{$\blacksquare$}
\newunicodechar{⊡}{$\boxdot$}
\newunicodechar{⋆}{$\star$}
\newunicodechar{◎}{$\circledcirc$}
\newunicodechar{⊗}{$\otimes$}
\newunicodechar{⊕}{$\oplus$}
\newunicodechar{₊}{$_{+}$}
\newunicodechar{↔}{$\leftrightarrow$}
\newunicodechar{⇔}{$\Leftrightarrow$}
\newunicodechar{∀}{$\forall$}
\newunicodechar{ϕ}{$\phi$}
\begin{code}[hide]%
\>[0]\AgdaSymbol{\{-\#}\AgdaSpace{}%
\AgdaKeyword{OPTIONS}\AgdaSpace{}%
\AgdaPragma{--without-K}\AgdaSpace{}%
\AgdaPragma{--exact-split}\AgdaSpace{}%
\AgdaPragma{--rewriting}\AgdaSpace{}%
\AgdaPragma{--overlapping-instances}\AgdaSpace{}%
\AgdaSymbol{\#-\}}\<%
\\
\>[0]\AgdaKeyword{module}\AgdaSpace{}%
\AgdaModule{Pi.Examples.ExamplesL}\AgdaSpace{}%
\AgdaKeyword{where}\<%
\\
\\[\AgdaEmptyExtraSkip]%
\>[0]\AgdaKeyword{open}\AgdaSpace{}%
\AgdaKeyword{import}\AgdaSpace{}%
\AgdaModule{HoTT}\AgdaSpace{}%
\AgdaKeyword{hiding}\AgdaSpace{}%
\AgdaSymbol{(}\AgdaOperator{\AgdaDatatype{\AgdaUnderscore{}<\AgdaUnderscore{}}}\AgdaSpace{}%
\AgdaSymbol{;}\AgdaSpace{}%
\AgdaInductiveConstructor{ltS}\AgdaSpace{}%
\AgdaSymbol{;}\AgdaSpace{}%
\AgdaInductiveConstructor{ltSR}\AgdaSpace{}%
\AgdaSymbol{;}\AgdaSpace{}%
\AgdaOperator{\AgdaPrimitive{\AgdaUnderscore{}+\AgdaUnderscore{}}}\AgdaSpace{}%
\AgdaSymbol{;}\AgdaSpace{}%
\AgdaOperator{\AgdaFunction{\AgdaUnderscore{}×\AgdaUnderscore{}}}\AgdaSymbol{)}\<%
\\
\>[0]\AgdaKeyword{import}\AgdaSpace{}%
\AgdaModule{lib.types.Nat}\AgdaSpace{}%
\AgdaSymbol{as}\AgdaSpace{}%
\AgdaModule{N}\<%
\\
\>[0]\AgdaKeyword{import}\AgdaSpace{}%
\AgdaModule{lib.types.Sigma}\AgdaSpace{}%
\AgdaSymbol{as}\AgdaSpace{}%
\AgdaModule{S}\<%
\\
\\[\AgdaEmptyExtraSkip]%
\>[0]\AgdaKeyword{open}\AgdaSpace{}%
\AgdaKeyword{import}\AgdaSpace{}%
\AgdaModule{Pi.UFin.BAut}\<%
\\
\>[0]\AgdaKeyword{open}\AgdaSpace{}%
\AgdaKeyword{import}\AgdaSpace{}%
\AgdaModule{Pi.Common.Misc}\<%
\\
\>[0]\AgdaKeyword{open}\AgdaSpace{}%
\AgdaKeyword{import}\AgdaSpace{}%
\AgdaModule{Pi.Common.Extra}\<%
\\
\\[\AgdaEmptyExtraSkip]%
\>[0]\AgdaKeyword{open}\AgdaSpace{}%
\AgdaKeyword{import}\AgdaSpace{}%
\AgdaModule{Pi.Syntax.Pi+.Indexed}\AgdaSpace{}%
\AgdaSymbol{as}\AgdaSpace{}%
\AgdaModule{Pi+}\<%
\\
\>[0][@{}l@{\AgdaIndent{0}}]%
\>[2]\AgdaKeyword{renaming}\AgdaSpace{}%
\AgdaSymbol{(}\AgdaOperator{\AgdaDatatype{\AgdaUnderscore{}⟷₁\AgdaUnderscore{}}}\AgdaSpace{}%
\AgdaSymbol{to}\AgdaSpace{}%
\AgdaOperator{\AgdaDatatype{\AgdaUnderscore{}⟷₁₊\AgdaUnderscore{}}}\AgdaSymbol{;}\AgdaSpace{}%
\AgdaOperator{\AgdaDatatype{\AgdaUnderscore{}⟷₂\AgdaUnderscore{}}}\AgdaSpace{}%
\AgdaSymbol{to}\AgdaSpace{}%
\AgdaOperator{\AgdaDatatype{\AgdaUnderscore{}⟷₂₊\AgdaUnderscore{}}}\AgdaSymbol{;}\AgdaSpace{}%
\AgdaFunction{!⟷₁}\AgdaSpace{}%
\AgdaSymbol{to}\AgdaSpace{}%
\AgdaFunction{!⟷₁₊}\AgdaSymbol{;}\AgdaSpace{}%
\AgdaDatatype{U}\AgdaSpace{}%
\AgdaSymbol{to}\AgdaSpace{}%
\AgdaDatatype{U+}\AgdaSymbol{;}\<%
\\
\>[2]\AgdaInductiveConstructor{idr◎l}\AgdaSpace{}%
\AgdaSymbol{to}\AgdaSpace{}%
\AgdaInductiveConstructor{idr◎l+}\AgdaSymbol{;}\AgdaSpace{}%
\AgdaInductiveConstructor{swapl₊⟷₂}\AgdaSpace{}%
\AgdaSymbol{to}\AgdaSpace{}%
\AgdaInductiveConstructor{swapl₊⟷₂+}\AgdaSymbol{;}\AgdaSpace{}%
\AgdaFunction{unite₊r}\AgdaSpace{}%
\AgdaSymbol{to}\AgdaSpace{}%
\AgdaFunction{unite₊r+}\AgdaSymbol{)}\<%
\\
\>[0]\AgdaKeyword{open}\AgdaSpace{}%
\AgdaKeyword{import}\AgdaSpace{}%
\AgdaModule{Pi.Syntax.Pi\textasciicircum{}}\AgdaSpace{}%
\AgdaSymbol{as}\AgdaSpace{}%
\AgdaModule{Pi\textasciicircum{}}\<%
\\
\>[0]\AgdaKeyword{open}\AgdaSpace{}%
\AgdaKeyword{import}\AgdaSpace{}%
\AgdaModule{Pi.Syntax.Pi\textasciicircum{}Helpers}\AgdaSpace{}%
\AgdaSymbol{as}\AgdaSpace{}%
\AgdaModule{Pi\textasciicircum{}}\<%
\\
\>[0]\AgdaKeyword{open}\AgdaSpace{}%
\AgdaKeyword{import}\AgdaSpace{}%
\AgdaModule{Pi.Syntax.Pi}\AgdaSpace{}%
\AgdaSymbol{as}\AgdaSpace{}%
\AgdaModule{Pi}\AgdaSpace{}%
\AgdaKeyword{hiding}\AgdaSpace{}%
\AgdaSymbol{(}\AgdaOperator{\AgdaDatatype{\AgdaUnderscore{}⟷₂\AgdaUnderscore{}}}\AgdaSymbol{)}\<%
\\
\>[0]\AgdaKeyword{open}\AgdaSpace{}%
\AgdaKeyword{import}\AgdaSpace{}%
\AgdaModule{Pi.Equiv.Translation2}\<%
\\
\>[0]\AgdaKeyword{import}\AgdaSpace{}%
\AgdaModule{Pi.Equiv.Equiv1}\AgdaSpace{}%
\AgdaSymbol{as}\AgdaSpace{}%
\AgdaModule{Pi+}\<%
\\
\>[0]\AgdaKeyword{import}\AgdaSpace{}%
\AgdaModule{Pi.Equiv.Equiv0Hat}\AgdaSpace{}%
\AgdaSymbol{as}\AgdaSpace{}%
\AgdaModule{Pi\textasciicircum{}}\<%
\\
\>[0]\AgdaKeyword{import}\AgdaSpace{}%
\AgdaModule{Pi.Equiv.Equiv1Hat}\AgdaSpace{}%
\AgdaSymbol{as}\AgdaSpace{}%
\AgdaModule{Pi\textasciicircum{}}\<%
\\
\>[0]\AgdaKeyword{import}\AgdaSpace{}%
\AgdaModule{Pi.Equiv.Equiv0Norm}\AgdaSpace{}%
\AgdaSymbol{as}\AgdaSpace{}%
\AgdaModule{Pi\textasciicircum{}}\<%
\\
\>[0]\AgdaKeyword{import}\AgdaSpace{}%
\AgdaModule{Pi.Equiv.Equiv1Norm}\AgdaSpace{}%
\AgdaSymbol{as}\AgdaSpace{}%
\AgdaModule{Pi\textasciicircum{}}\<%
\\
\>[0]\AgdaKeyword{open}\AgdaSpace{}%
\AgdaKeyword{import}\AgdaSpace{}%
\AgdaModule{Pi.UFin.UFin}\<%
\\
\>[0]\AgdaKeyword{open}\AgdaSpace{}%
\AgdaKeyword{import}\AgdaSpace{}%
\AgdaModule{Pi.UFin.Monoidal}\<%
\\
\>[0]\AgdaKeyword{open}\AgdaSpace{}%
\AgdaKeyword{import}\AgdaSpace{}%
\AgdaModule{Pi.Common.FinHelpers}\<%
\\
\>[0]\AgdaKeyword{open}\AgdaSpace{}%
\AgdaKeyword{import}\AgdaSpace{}%
\AgdaModule{Pi.Lehmer.FinExcept}\<%
\\
\\[\AgdaEmptyExtraSkip]%
\>[0]\AgdaKeyword{open}\AgdaSpace{}%
\AgdaKeyword{import}\AgdaSpace{}%
\AgdaModule{Pi.Examples.Base}\<%
\\
\>[0]\AgdaKeyword{open}\AgdaSpace{}%
\AgdaKeyword{import}\AgdaSpace{}%
\AgdaModule{Pi.Examples.Toffoli}\AgdaSpace{}%
\AgdaKeyword{hiding}\AgdaSpace{}%
\AgdaSymbol{(}\AgdaFunction{cif}\AgdaSymbol{)}\<%
\\
\>[0]\AgdaKeyword{open}\AgdaSpace{}%
\AgdaKeyword{import}\AgdaSpace{}%
\AgdaModule{Pi.Examples.Reset}\AgdaSpace{}%
\AgdaKeyword{hiding}\AgdaSpace{}%
\AgdaSymbol{(}\AgdaFunction{reset}\AgdaSymbol{;}\AgdaSpace{}%
\AgdaFunction{reset2-perm}\AgdaSymbol{)}\<%
\\
\\[\AgdaEmptyExtraSkip]%
\>[0]\AgdaKeyword{private}\<%
\\
\>[0][@{}l@{\AgdaIndent{0}}]%
\>[2]\AgdaKeyword{variable}\<%
\\
\>[2][@{}l@{\AgdaIndent{0}}]%
\>[4]\AgdaGeneralizable{A}\AgdaSpace{}%
\AgdaGeneralizable{B}\AgdaSpace{}%
\AgdaGeneralizable{C}\AgdaSpace{}%
\AgdaGeneralizable{D}\AgdaSpace{}%
\AgdaGeneralizable{E}\AgdaSpace{}%
\AgdaGeneralizable{F}\AgdaSpace{}%
\AgdaSymbol{:}\AgdaSpace{}%
\AgdaDatatype{U}\<%
\end{code}

\newcommand{\cif}{%
\begin{code}%
\>[0]\AgdaFunction{cif}\AgdaSpace{}%
\AgdaSymbol{:}\AgdaSpace{}%
\AgdaSymbol{(}\AgdaBound{c₁}\AgdaSpace{}%
\AgdaBound{c₂}\AgdaSpace{}%
\AgdaSymbol{:}\AgdaSpace{}%
\AgdaGeneralizable{A}\AgdaSpace{}%
\AgdaOperator{\AgdaDatatype{⟷₁}}\AgdaSpace{}%
\AgdaGeneralizable{A}\AgdaSymbol{)}\AgdaSpace{}%
\AgdaSymbol{→}\AgdaSpace{}%
\AgdaSymbol{(}\AgdaFunction{𝟚}\AgdaSpace{}%
\AgdaOperator{\AgdaInductiveConstructor{×}}\AgdaSpace{}%
\AgdaGeneralizable{A}\AgdaSpace{}%
\AgdaOperator{\AgdaDatatype{⟷₁}}\AgdaSpace{}%
\AgdaFunction{𝟚}\AgdaSpace{}%
\AgdaOperator{\AgdaInductiveConstructor{×}}\AgdaSpace{}%
\AgdaGeneralizable{A}\AgdaSymbol{)}\<%
\\
\>[0]\AgdaFunction{cif}\AgdaSpace{}%
\AgdaBound{c₁}\AgdaSpace{}%
\AgdaBound{c₂}\AgdaSpace{}%
\AgdaSymbol{=}\AgdaSpace{}%
\AgdaInductiveConstructor{dist}\AgdaSpace{}%
\AgdaOperator{\AgdaInductiveConstructor{◎}}\AgdaSpace{}%
\AgdaSymbol{((}\AgdaInductiveConstructor{id⟷₁}\AgdaSpace{}%
\AgdaOperator{\AgdaInductiveConstructor{⊗}}\AgdaSpace{}%
\AgdaBound{c₁}\AgdaSymbol{)}\AgdaSpace{}%
\AgdaOperator{\AgdaInductiveConstructor{⊕}}\AgdaSpace{}%
\AgdaSymbol{(}\AgdaInductiveConstructor{id⟷₁}\AgdaSpace{}%
\AgdaOperator{\AgdaInductiveConstructor{⊗}}\AgdaSpace{}%
\AgdaBound{c₂}\AgdaSymbol{))}\AgdaSpace{}%
\AgdaOperator{\AgdaInductiveConstructor{◎}}\AgdaSpace{}%
\AgdaInductiveConstructor{factor}\<%
\end{code}}

\newcommand{\resetnormtwo}{%
\begin{code}%
\>[0]\AgdaFunction{reversibleOrNorm}\AgdaSpace{}%
\AgdaSymbol{:}\AgdaSpace{}%
\AgdaFunction{𝟠+}\AgdaSpace{}%
\AgdaOperator{\AgdaDatatype{⟷₁₊}}\AgdaSpace{}%
\AgdaFunction{𝟠+}\<%
\\
\>[0]\AgdaFunction{reversibleOrNorm}\AgdaSpace{}%
\AgdaSymbol{=}%
\>[20]\AgdaSymbol{(}\AgdaInductiveConstructor{id⟷₁}\AgdaSpace{}%
\AgdaOperator{\AgdaInductiveConstructor{⊕}}\AgdaSpace{}%
\AgdaInductiveConstructor{id⟷₁}\AgdaSpace{}%
\AgdaOperator{\AgdaInductiveConstructor{⊕}}\AgdaSpace{}%
\AgdaInductiveConstructor{assocl₊}\AgdaSpace{}%
\AgdaOperator{\AgdaInductiveConstructor{◎}}\AgdaSpace{}%
\AgdaSymbol{(}\AgdaInductiveConstructor{swap₊}\AgdaSpace{}%
\AgdaOperator{\AgdaInductiveConstructor{⊕}}\AgdaSpace{}%
\AgdaInductiveConstructor{id⟷₁}\AgdaSymbol{)}\AgdaSpace{}%
\AgdaOperator{\AgdaInductiveConstructor{◎}}\AgdaSpace{}%
\AgdaInductiveConstructor{assocr₊}\AgdaSymbol{)}\AgdaSpace{}%
\AgdaOperator{\AgdaInductiveConstructor{◎}}\<%
\\
\>[20]\AgdaSymbol{(}\AgdaInductiveConstructor{id⟷₁}\AgdaSpace{}%
\AgdaOperator{\AgdaInductiveConstructor{⊕}}\AgdaSpace{}%
\AgdaInductiveConstructor{assocl₊}\AgdaSpace{}%
\AgdaOperator{\AgdaInductiveConstructor{◎}}\AgdaSpace{}%
\AgdaSymbol{(}\AgdaInductiveConstructor{swap₊}\AgdaSpace{}%
\AgdaOperator{\AgdaInductiveConstructor{⊕}}\AgdaSpace{}%
\AgdaInductiveConstructor{id⟷₁}\AgdaSymbol{)}\AgdaSpace{}%
\AgdaOperator{\AgdaInductiveConstructor{◎}}\AgdaSpace{}%
\AgdaInductiveConstructor{assocr₊}\AgdaSymbol{)}\AgdaSpace{}%
\AgdaOperator{\AgdaInductiveConstructor{◎}}\<%
\\
\>[20]\AgdaSymbol{(}\AgdaInductiveConstructor{assocl₊}\AgdaSpace{}%
\AgdaOperator{\AgdaInductiveConstructor{◎}}\AgdaSpace{}%
\AgdaSymbol{(}\AgdaInductiveConstructor{swap₊}\AgdaSpace{}%
\AgdaOperator{\AgdaInductiveConstructor{⊕}}\AgdaSpace{}%
\AgdaInductiveConstructor{id⟷₁}\AgdaSymbol{)}\AgdaSpace{}%
\AgdaOperator{\AgdaInductiveConstructor{◎}}\AgdaSpace{}%
\AgdaInductiveConstructor{assocr₊}\AgdaSymbol{)}\AgdaSpace{}%
\AgdaOperator{\AgdaInductiveConstructor{◎}}\<%
\\
\>[20]\AgdaSymbol{(}\AgdaInductiveConstructor{id⟷₁}\AgdaSpace{}%
\AgdaOperator{\AgdaInductiveConstructor{⊕}}\AgdaSpace{}%
\AgdaInductiveConstructor{id⟷₁}\AgdaSpace{}%
\AgdaOperator{\AgdaInductiveConstructor{⊕}}\AgdaSpace{}%
\AgdaInductiveConstructor{id⟷₁}\AgdaSpace{}%
\AgdaOperator{\AgdaInductiveConstructor{⊕}}\AgdaSpace{}%
\AgdaInductiveConstructor{assocl₊}\AgdaSpace{}%
\AgdaOperator{\AgdaInductiveConstructor{◎}}\AgdaSpace{}%
\AgdaSymbol{(}\AgdaInductiveConstructor{swap₊}\AgdaSpace{}%
\AgdaOperator{\AgdaInductiveConstructor{⊕}}\AgdaSpace{}%
\AgdaInductiveConstructor{id⟷₁}\AgdaSymbol{)}\AgdaSpace{}%
\AgdaOperator{\AgdaInductiveConstructor{◎}}\AgdaSpace{}%
\AgdaInductiveConstructor{assocr₊}\AgdaSymbol{)}\AgdaSpace{}%
\AgdaOperator{\AgdaInductiveConstructor{◎}}\<%
\\
\>[20]\AgdaSymbol{(}\AgdaInductiveConstructor{id⟷₁}\AgdaSpace{}%
\AgdaOperator{\AgdaInductiveConstructor{⊕}}\AgdaSpace{}%
\AgdaInductiveConstructor{id⟷₁}\AgdaSpace{}%
\AgdaOperator{\AgdaInductiveConstructor{⊕}}\AgdaSpace{}%
\AgdaInductiveConstructor{assocl₊}\AgdaSpace{}%
\AgdaOperator{\AgdaInductiveConstructor{◎}}\AgdaSpace{}%
\AgdaSymbol{(}\AgdaInductiveConstructor{swap₊}\AgdaSpace{}%
\AgdaOperator{\AgdaInductiveConstructor{⊕}}\AgdaSpace{}%
\AgdaInductiveConstructor{id⟷₁}\AgdaSymbol{)}\AgdaSpace{}%
\AgdaOperator{\AgdaInductiveConstructor{◎}}\AgdaSpace{}%
\AgdaInductiveConstructor{assocr₊}\AgdaSymbol{)}\AgdaSpace{}%
\AgdaOperator{\AgdaInductiveConstructor{◎}}\<%
\\
\>[20]\AgdaSymbol{(}\AgdaInductiveConstructor{id⟷₁}\AgdaSpace{}%
\AgdaOperator{\AgdaInductiveConstructor{⊕}}\AgdaSpace{}%
\AgdaInductiveConstructor{assocl₊}\AgdaSpace{}%
\AgdaOperator{\AgdaInductiveConstructor{◎}}\AgdaSpace{}%
\AgdaSymbol{(}\AgdaInductiveConstructor{swap₊}\AgdaSpace{}%
\AgdaOperator{\AgdaInductiveConstructor{⊕}}\AgdaSpace{}%
\AgdaInductiveConstructor{id⟷₁}\AgdaSymbol{)}\AgdaSpace{}%
\AgdaOperator{\AgdaInductiveConstructor{◎}}\AgdaSpace{}%
\AgdaInductiveConstructor{assocr₊}\AgdaSymbol{)}\AgdaSpace{}%
\AgdaOperator{\AgdaInductiveConstructor{◎}}\<%
\\
\>[20]\AgdaSymbol{(}\AgdaInductiveConstructor{assocl₊}\AgdaSpace{}%
\AgdaOperator{\AgdaInductiveConstructor{◎}}\AgdaSpace{}%
\AgdaSymbol{(}\AgdaInductiveConstructor{swap₊}\AgdaSpace{}%
\AgdaOperator{\AgdaInductiveConstructor{⊕}}\AgdaSpace{}%
\AgdaInductiveConstructor{id⟷₁}\AgdaSymbol{)}\AgdaSpace{}%
\AgdaOperator{\AgdaInductiveConstructor{◎}}\AgdaSpace{}%
\AgdaInductiveConstructor{assocr₊}\AgdaSymbol{)}\AgdaSpace{}%
\AgdaOperator{\AgdaInductiveConstructor{◎}}\<%
\\
\>[20]\AgdaSymbol{(}\AgdaInductiveConstructor{id⟷₁}\AgdaSpace{}%
\AgdaOperator{\AgdaInductiveConstructor{⊕}}\AgdaSpace{}%
\AgdaInductiveConstructor{id⟷₁}\AgdaSpace{}%
\AgdaOperator{\AgdaInductiveConstructor{⊕}}\AgdaSpace{}%
\AgdaInductiveConstructor{id⟷₁}\AgdaSpace{}%
\AgdaOperator{\AgdaInductiveConstructor{⊕}}\AgdaSpace{}%
\AgdaInductiveConstructor{id⟷₁}\AgdaSpace{}%
\AgdaOperator{\AgdaInductiveConstructor{⊕}}\AgdaSpace{}%
\AgdaInductiveConstructor{assocl₊}\AgdaSpace{}%
\AgdaOperator{\AgdaInductiveConstructor{◎}}\AgdaSpace{}%
\AgdaSymbol{(}\AgdaInductiveConstructor{swap₊}\AgdaSpace{}%
\AgdaOperator{\AgdaInductiveConstructor{⊕}}\AgdaSpace{}%
\AgdaInductiveConstructor{id⟷₁}\AgdaSymbol{)}\AgdaSpace{}%
\AgdaOperator{\AgdaInductiveConstructor{◎}}\AgdaSpace{}%
\AgdaInductiveConstructor{assocr₊}\AgdaSymbol{)}\AgdaSpace{}%
\AgdaOperator{\AgdaInductiveConstructor{◎}}\<%
\\
\>[20]\AgdaSymbol{(}\AgdaInductiveConstructor{id⟷₁}\AgdaSpace{}%
\AgdaOperator{\AgdaInductiveConstructor{⊕}}\AgdaSpace{}%
\AgdaInductiveConstructor{id⟷₁}\AgdaSpace{}%
\AgdaOperator{\AgdaInductiveConstructor{⊕}}\AgdaSpace{}%
\AgdaInductiveConstructor{id⟷₁}\AgdaSpace{}%
\AgdaOperator{\AgdaInductiveConstructor{⊕}}\AgdaSpace{}%
\AgdaInductiveConstructor{assocl₊}\AgdaSpace{}%
\AgdaOperator{\AgdaInductiveConstructor{◎}}\AgdaSpace{}%
\AgdaSymbol{(}\AgdaInductiveConstructor{swap₊}\AgdaSpace{}%
\AgdaOperator{\AgdaInductiveConstructor{⊕}}\AgdaSpace{}%
\AgdaInductiveConstructor{id⟷₁}\AgdaSymbol{)}\AgdaSpace{}%
\AgdaOperator{\AgdaInductiveConstructor{◎}}\AgdaSpace{}%
\AgdaInductiveConstructor{assocr₊}\AgdaSymbol{)}\AgdaSpace{}%
\AgdaOperator{\AgdaInductiveConstructor{◎}}\<%
\\
\>[20]\AgdaSymbol{(}\AgdaInductiveConstructor{id⟷₁}\AgdaSpace{}%
\AgdaOperator{\AgdaInductiveConstructor{⊕}}\AgdaSpace{}%
\AgdaInductiveConstructor{id⟷₁}\AgdaSpace{}%
\AgdaOperator{\AgdaInductiveConstructor{⊕}}\AgdaSpace{}%
\AgdaInductiveConstructor{assocl₊}\AgdaSpace{}%
\AgdaOperator{\AgdaInductiveConstructor{◎}}\AgdaSpace{}%
\AgdaSymbol{(}\AgdaInductiveConstructor{swap₊}\AgdaSpace{}%
\AgdaOperator{\AgdaInductiveConstructor{⊕}}\AgdaSpace{}%
\AgdaInductiveConstructor{id⟷₁}\AgdaSymbol{)}\AgdaSpace{}%
\AgdaOperator{\AgdaInductiveConstructor{◎}}\AgdaSpace{}%
\AgdaInductiveConstructor{assocr₊}\AgdaSymbol{)}\AgdaSpace{}%
\AgdaOperator{\AgdaInductiveConstructor{◎}}\<%
\\
\>[20]\AgdaSymbol{(}\AgdaInductiveConstructor{id⟷₁}\AgdaSpace{}%
\AgdaOperator{\AgdaInductiveConstructor{⊕}}\AgdaSpace{}%
\AgdaInductiveConstructor{assocl₊}\AgdaSpace{}%
\AgdaOperator{\AgdaInductiveConstructor{◎}}\AgdaSpace{}%
\AgdaSymbol{(}\AgdaInductiveConstructor{swap₊}\AgdaSpace{}%
\AgdaOperator{\AgdaInductiveConstructor{⊕}}\AgdaSpace{}%
\AgdaInductiveConstructor{id⟷₁}\AgdaSymbol{)}\AgdaSpace{}%
\AgdaOperator{\AgdaInductiveConstructor{◎}}\AgdaSpace{}%
\AgdaInductiveConstructor{assocr₊}\AgdaSymbol{)}\AgdaSpace{}%
\AgdaOperator{\AgdaInductiveConstructor{◎}}\<%
\\
\>[20]\AgdaSymbol{(}\AgdaInductiveConstructor{id⟷₁}\AgdaSpace{}%
\AgdaOperator{\AgdaInductiveConstructor{⊕}}\AgdaSpace{}%
\AgdaInductiveConstructor{id⟷₁}\AgdaSpace{}%
\AgdaOperator{\AgdaInductiveConstructor{⊕}}\AgdaSpace{}%
\AgdaInductiveConstructor{id⟷₁}\AgdaSpace{}%
\AgdaOperator{\AgdaInductiveConstructor{⊕}}\AgdaSpace{}%
\AgdaInductiveConstructor{id⟷₁}\AgdaSpace{}%
\AgdaOperator{\AgdaInductiveConstructor{⊕}}\AgdaSpace{}%
\AgdaInductiveConstructor{id⟷₁}\AgdaSpace{}%
\AgdaOperator{\AgdaInductiveConstructor{⊕}}\AgdaSpace{}%
\AgdaInductiveConstructor{assocl₊}\AgdaSpace{}%
\AgdaOperator{\AgdaInductiveConstructor{◎}}\AgdaSpace{}%
\AgdaSymbol{(}\AgdaInductiveConstructor{swap₊}\AgdaSpace{}%
\AgdaOperator{\AgdaInductiveConstructor{⊕}}\AgdaSpace{}%
\AgdaInductiveConstructor{id⟷₁}\AgdaSymbol{)}\AgdaSpace{}%
\AgdaOperator{\AgdaInductiveConstructor{◎}}\AgdaSpace{}%
\AgdaInductiveConstructor{assocr₊}\AgdaSymbol{)}\AgdaSpace{}%
\AgdaOperator{\AgdaInductiveConstructor{◎}}\<%
\\
\>[20]\AgdaSymbol{(}\AgdaInductiveConstructor{id⟷₁}\AgdaSpace{}%
\AgdaOperator{\AgdaInductiveConstructor{⊕}}\AgdaSpace{}%
\AgdaInductiveConstructor{id⟷₁}\AgdaSpace{}%
\AgdaOperator{\AgdaInductiveConstructor{⊕}}\AgdaSpace{}%
\AgdaInductiveConstructor{id⟷₁}\AgdaSpace{}%
\AgdaOperator{\AgdaInductiveConstructor{⊕}}\AgdaSpace{}%
\AgdaInductiveConstructor{id⟷₁}\AgdaSpace{}%
\AgdaOperator{\AgdaInductiveConstructor{⊕}}\AgdaSpace{}%
\AgdaInductiveConstructor{assocl₊}\AgdaSpace{}%
\AgdaOperator{\AgdaInductiveConstructor{◎}}\AgdaSpace{}%
\AgdaSymbol{(}\AgdaInductiveConstructor{swap₊}\AgdaSpace{}%
\AgdaOperator{\AgdaInductiveConstructor{⊕}}\AgdaSpace{}%
\AgdaInductiveConstructor{id⟷₁}\AgdaSymbol{)}\AgdaSpace{}%
\AgdaOperator{\AgdaInductiveConstructor{◎}}\AgdaSpace{}%
\AgdaInductiveConstructor{assocr₊}\AgdaSymbol{)}\AgdaSpace{}%
\AgdaOperator{\AgdaInductiveConstructor{◎}}\<%
\\
\>[20]\AgdaSymbol{(}\AgdaInductiveConstructor{id⟷₁}\AgdaSpace{}%
\AgdaOperator{\AgdaInductiveConstructor{⊕}}\AgdaSpace{}%
\AgdaInductiveConstructor{id⟷₁}\AgdaSpace{}%
\AgdaOperator{\AgdaInductiveConstructor{⊕}}\AgdaSpace{}%
\AgdaInductiveConstructor{id⟷₁}\AgdaSpace{}%
\AgdaOperator{\AgdaInductiveConstructor{⊕}}\AgdaSpace{}%
\AgdaInductiveConstructor{assocl₊}\AgdaSpace{}%
\AgdaOperator{\AgdaInductiveConstructor{◎}}\AgdaSpace{}%
\AgdaSymbol{(}\AgdaInductiveConstructor{swap₊}\AgdaSpace{}%
\AgdaOperator{\AgdaInductiveConstructor{⊕}}\AgdaSpace{}%
\AgdaInductiveConstructor{id⟷₁}\AgdaSymbol{)}\AgdaSpace{}%
\AgdaOperator{\AgdaInductiveConstructor{◎}}\AgdaSpace{}%
\AgdaInductiveConstructor{assocr₊}\AgdaSymbol{)}\AgdaSpace{}%
\AgdaOperator{\AgdaInductiveConstructor{◎}}\<%
\\
\>[20]\AgdaSymbol{(}\AgdaInductiveConstructor{id⟷₁}\AgdaSpace{}%
\AgdaOperator{\AgdaInductiveConstructor{⊕}}\AgdaSpace{}%
\AgdaInductiveConstructor{id⟷₁}\AgdaSpace{}%
\AgdaOperator{\AgdaInductiveConstructor{⊕}}\AgdaSpace{}%
\AgdaInductiveConstructor{assocl₊}\AgdaSpace{}%
\AgdaOperator{\AgdaInductiveConstructor{◎}}\AgdaSpace{}%
\AgdaSymbol{(}\AgdaInductiveConstructor{swap₊}\AgdaSpace{}%
\AgdaOperator{\AgdaInductiveConstructor{⊕}}\AgdaSpace{}%
\AgdaInductiveConstructor{id⟷₁}\AgdaSymbol{)}\AgdaSpace{}%
\AgdaOperator{\AgdaInductiveConstructor{◎}}\AgdaSpace{}%
\AgdaInductiveConstructor{assocr₊}\AgdaSymbol{)}\AgdaSpace{}%
\AgdaOperator{\AgdaInductiveConstructor{◎}}\AgdaSpace{}%
\AgdaInductiveConstructor{id⟷₁}\<%
\end{code}}

\newcommand{\resetperm}{%
\begin{code}%
\>[0]\AgdaFunction{reversibleOrPerm}\AgdaSpace{}%
\AgdaSymbol{:}\AgdaSpace{}%
\AgdaFunction{Aut}\AgdaSpace{}%
\AgdaSymbol{(}\AgdaFunction{Fin}\AgdaSpace{}%
\AgdaNumber{8}\AgdaSymbol{)}\<%
\\
\>[0]\AgdaFunction{reversibleOrPerm}\AgdaSpace{}%
\AgdaSymbol{=}\AgdaSpace{}%
\AgdaFunction{equiv}\AgdaSpace{}%
\AgdaFunction{f}\AgdaSpace{}%
\AgdaFunction{f}\AgdaSpace{}%
\AgdaFunction{f-f}\AgdaSpace{}%
\AgdaFunction{f-f}\<%
\\
\>[0][@{}l@{\AgdaIndent{0}}]%
\>[2]\AgdaKeyword{where}%
\>[377I]\AgdaFunction{f}\AgdaSpace{}%
\AgdaSymbol{:}\AgdaSpace{}%
\AgdaFunction{Fin}\AgdaSpace{}%
\AgdaNumber{8}\AgdaSpace{}%
\AgdaSymbol{→}\AgdaSpace{}%
\AgdaFunction{Fin}\AgdaSpace{}%
\AgdaNumber{8}\<%
\\
\>[.][@{}l@{}]\<[377I]%
\>[8]\AgdaFunction{f}\AgdaSpace{}%
\AgdaSymbol{=}\AgdaSpace{}%
\AgdaFunction{lookup}\AgdaSpace{}%
\AgdaSymbol{(}\AgdaNumber{0}\AgdaSpace{}%
\AgdaOperator{\AgdaInductiveConstructor{::}}\AgdaSpace{}%
\AgdaNumber{5}\AgdaSpace{}%
\AgdaOperator{\AgdaInductiveConstructor{::}}\AgdaSpace{}%
\AgdaNumber{6}\AgdaSpace{}%
\AgdaOperator{\AgdaInductiveConstructor{::}}\AgdaSpace{}%
\AgdaNumber{7}\AgdaSpace{}%
\AgdaOperator{\AgdaInductiveConstructor{::}}\AgdaSpace{}%
\AgdaNumber{4}\AgdaSpace{}%
\AgdaOperator{\AgdaInductiveConstructor{::}}\AgdaSpace{}%
\AgdaNumber{1}\AgdaSpace{}%
\AgdaOperator{\AgdaInductiveConstructor{::}}\AgdaSpace{}%
\AgdaNumber{2}\AgdaSpace{}%
\AgdaOperator{\AgdaInductiveConstructor{::}}\AgdaSpace{}%
\AgdaNumber{3}\AgdaSpace{}%
\AgdaOperator{\AgdaInductiveConstructor{::}}\AgdaSpace{}%
\AgdaInductiveConstructor{nil}\AgdaSymbol{)}\<%
\\
\>[8]\AgdaFunction{f-f}\AgdaSpace{}%
\AgdaSymbol{:}\AgdaSpace{}%
\AgdaSymbol{(}\AgdaBound{x}\AgdaSpace{}%
\AgdaSymbol{:}\AgdaSpace{}%
\AgdaFunction{Fin}\AgdaSpace{}%
\AgdaNumber{8}\AgdaSymbol{)}\AgdaSpace{}%
\AgdaSymbol{→}\AgdaSpace{}%
\AgdaFunction{f}\AgdaSpace{}%
\AgdaSymbol{(}\AgdaFunction{f}\AgdaSpace{}%
\AgdaBound{x}\AgdaSymbol{)}\AgdaSpace{}%
\AgdaOperator{\AgdaDatatype{==}}\AgdaSpace{}%
\AgdaBound{x}\<%
\\
\>[8]\AgdaComment{-- elided}\<%
\end{code}}
\begin{code}[hide]%
\>[8]\AgdaFunction{f-f}\AgdaSpace{}%
\AgdaSymbol{(}\AgdaNumber{0}\AgdaSpace{}%
\AgdaOperator{\AgdaInductiveConstructor{,}}\AgdaSpace{}%
\AgdaBound{ϕ}\AgdaSymbol{)}\AgdaSpace{}%
\AgdaSymbol{=}\AgdaSpace{}%
\AgdaFunction{fin=}\AgdaSpace{}%
\AgdaInductiveConstructor{idp}\<%
\\
\>[8]\AgdaFunction{f-f}\AgdaSpace{}%
\AgdaSymbol{(}\AgdaNumber{1}\AgdaSpace{}%
\AgdaOperator{\AgdaInductiveConstructor{,}}\AgdaSpace{}%
\AgdaBound{ϕ}\AgdaSymbol{)}\AgdaSpace{}%
\AgdaSymbol{=}\AgdaSpace{}%
\AgdaFunction{fin=}\AgdaSpace{}%
\AgdaInductiveConstructor{idp}\<%
\\
\>[8]\AgdaFunction{f-f}\AgdaSpace{}%
\AgdaSymbol{(}\AgdaNumber{2}\AgdaSpace{}%
\AgdaOperator{\AgdaInductiveConstructor{,}}\AgdaSpace{}%
\AgdaBound{ϕ}\AgdaSymbol{)}\AgdaSpace{}%
\AgdaSymbol{=}\AgdaSpace{}%
\AgdaFunction{fin=}\AgdaSpace{}%
\AgdaInductiveConstructor{idp}\<%
\\
\>[8]\AgdaFunction{f-f}\AgdaSpace{}%
\AgdaSymbol{(}\AgdaNumber{3}\AgdaSpace{}%
\AgdaOperator{\AgdaInductiveConstructor{,}}\AgdaSpace{}%
\AgdaBound{ϕ}\AgdaSymbol{)}\AgdaSpace{}%
\AgdaSymbol{=}\AgdaSpace{}%
\AgdaFunction{fin=}\AgdaSpace{}%
\AgdaInductiveConstructor{idp}\<%
\\
\>[8]\AgdaFunction{f-f}\AgdaSpace{}%
\AgdaSymbol{(}\AgdaNumber{4}\AgdaSpace{}%
\AgdaOperator{\AgdaInductiveConstructor{,}}\AgdaSpace{}%
\AgdaBound{ϕ}\AgdaSymbol{)}\AgdaSpace{}%
\AgdaSymbol{=}\AgdaSpace{}%
\AgdaFunction{fin=}\AgdaSpace{}%
\AgdaInductiveConstructor{idp}\<%
\\
\>[8]\AgdaFunction{f-f}\AgdaSpace{}%
\AgdaSymbol{(}\AgdaNumber{5}\AgdaSpace{}%
\AgdaOperator{\AgdaInductiveConstructor{,}}\AgdaSpace{}%
\AgdaBound{ϕ}\AgdaSymbol{)}\AgdaSpace{}%
\AgdaSymbol{=}\AgdaSpace{}%
\AgdaFunction{fin=}\AgdaSpace{}%
\AgdaInductiveConstructor{idp}\<%
\\
\>[8]\AgdaFunction{f-f}\AgdaSpace{}%
\AgdaSymbol{(}\AgdaNumber{6}\AgdaSpace{}%
\AgdaOperator{\AgdaInductiveConstructor{,}}\AgdaSpace{}%
\AgdaBound{ϕ}\AgdaSymbol{)}\AgdaSpace{}%
\AgdaSymbol{=}\AgdaSpace{}%
\AgdaFunction{fin=}\AgdaSpace{}%
\AgdaInductiveConstructor{idp}\<%
\\
\>[8]\AgdaFunction{f-f}\AgdaSpace{}%
\AgdaSymbol{(}\AgdaNumber{7}\AgdaSpace{}%
\AgdaOperator{\AgdaInductiveConstructor{,}}\AgdaSpace{}%
\AgdaBound{ϕ}\AgdaSymbol{)}\AgdaSpace{}%
\AgdaSymbol{=}\AgdaSpace{}%
\AgdaFunction{fin=}\AgdaSpace{}%
\AgdaInductiveConstructor{idp}\<%
\\
\>[8]\AgdaFunction{f-f}\AgdaSpace{}%
\AgdaSymbol{(}\AgdaBound{n}\AgdaSpace{}%
\AgdaOperator{\AgdaInductiveConstructor{,}}\AgdaSpace{}%
\AgdaInductiveConstructor{N.ltSR}\AgdaSpace{}%
\AgdaSymbol{(}\AgdaInductiveConstructor{N.ltSR}\AgdaSpace{}%
\AgdaSymbol{(}\AgdaInductiveConstructor{N.ltSR}\AgdaSpace{}%
\AgdaSymbol{(}\AgdaInductiveConstructor{N.ltSR}\AgdaSpace{}%
\AgdaSymbol{(}\AgdaInductiveConstructor{N.ltSR}\AgdaSpace{}%
\AgdaSymbol{(}\AgdaInductiveConstructor{N.ltSR}\AgdaSpace{}%
\AgdaSymbol{(}\AgdaInductiveConstructor{N.ltSR}\AgdaSpace{}%
\AgdaSymbol{(}\AgdaInductiveConstructor{N.ltSR}\AgdaSpace{}%
\AgdaSymbol{()))))))))}\<%
\\
\\[\AgdaEmptyExtraSkip]%
\>[0]\AgdaFunction{ccx}\AgdaSpace{}%
\AgdaSymbol{=}\AgdaSpace{}%
\AgdaFunction{toffoli}\AgdaSpace{}%
\AgdaNumber{3}\<%
\\
\>[0]\AgdaFunction{cx}\AgdaSpace{}%
\AgdaSymbol{=}\AgdaSpace{}%
\AgdaFunction{cnot}\<%
\\
\>[0]\AgdaFunction{x}\AgdaSpace{}%
\AgdaSymbol{=}\AgdaSpace{}%
\AgdaFunction{not}\<%
\\
\\[\AgdaEmptyExtraSkip]%
\>[0]\AgdaFunction{A[BC]-C[BA]}\AgdaSpace{}%
\AgdaSymbol{:}\AgdaSpace{}%
\AgdaSymbol{\{}\AgdaBound{A}\AgdaSpace{}%
\AgdaBound{B}\AgdaSpace{}%
\AgdaBound{C}\AgdaSpace{}%
\AgdaSymbol{:}\AgdaSpace{}%
\AgdaDatatype{U}\AgdaSymbol{\}}\AgdaSpace{}%
\AgdaSymbol{→}\AgdaSpace{}%
\AgdaBound{A}\AgdaSpace{}%
\AgdaOperator{\AgdaInductiveConstructor{×}}\AgdaSpace{}%
\AgdaSymbol{(}\AgdaBound{B}\AgdaSpace{}%
\AgdaOperator{\AgdaInductiveConstructor{×}}\AgdaSpace{}%
\AgdaBound{C}\AgdaSymbol{)}\AgdaSpace{}%
\AgdaOperator{\AgdaDatatype{⟷₁}}\AgdaSpace{}%
\AgdaBound{C}\AgdaSpace{}%
\AgdaOperator{\AgdaInductiveConstructor{×}}\AgdaSpace{}%
\AgdaSymbol{(}\AgdaBound{B}\AgdaSpace{}%
\AgdaOperator{\AgdaInductiveConstructor{×}}\AgdaSpace{}%
\AgdaBound{A}\AgdaSymbol{)}\<%
\\
\>[0]\AgdaFunction{A[BC]-C[BA]}\AgdaSpace{}%
\AgdaSymbol{=}\AgdaSpace{}%
\AgdaInductiveConstructor{swap⋆}\AgdaSpace{}%
\AgdaOperator{\AgdaInductiveConstructor{◎}}\AgdaSpace{}%
\AgdaSymbol{(}\AgdaInductiveConstructor{swap⋆}\AgdaSpace{}%
\AgdaOperator{\AgdaInductiveConstructor{⊗}}\AgdaSpace{}%
\AgdaInductiveConstructor{id⟷₁}\AgdaSymbol{)}\AgdaSpace{}%
\AgdaOperator{\AgdaInductiveConstructor{◎}}\AgdaSpace{}%
\AgdaInductiveConstructor{assocr⋆}\<%
\\
\\[\AgdaEmptyExtraSkip]%
\>[0]\AgdaFunction{A[BC]-B[AC]}\AgdaSpace{}%
\AgdaSymbol{:}\AgdaSpace{}%
\AgdaSymbol{\{}\AgdaBound{t₁}\AgdaSpace{}%
\AgdaBound{t₂}\AgdaSpace{}%
\AgdaBound{t₃}\AgdaSpace{}%
\AgdaSymbol{:}\AgdaSpace{}%
\AgdaDatatype{Pi.U}\AgdaSymbol{\}}\AgdaSpace{}%
\AgdaSymbol{→}\AgdaSpace{}%
\AgdaBound{t₁}\AgdaSpace{}%
\AgdaOperator{\AgdaInductiveConstructor{Pi.×}}\AgdaSpace{}%
\AgdaSymbol{(}\AgdaBound{t₂}\AgdaSpace{}%
\AgdaOperator{\AgdaInductiveConstructor{Pi.×}}\AgdaSpace{}%
\AgdaBound{t₃}\AgdaSymbol{)}\AgdaSpace{}%
\AgdaOperator{\AgdaDatatype{Pi.⟷₁}}\AgdaSpace{}%
\AgdaBound{t₂}\AgdaSpace{}%
\AgdaOperator{\AgdaInductiveConstructor{Pi.×}}\AgdaSpace{}%
\AgdaSymbol{(}\AgdaBound{t₁}\AgdaSpace{}%
\AgdaOperator{\AgdaInductiveConstructor{Pi.×}}\AgdaSpace{}%
\AgdaBound{t₃}\AgdaSymbol{)}\<%
\\
\>[0]\AgdaFunction{A[BC]-B[AC]}\AgdaSpace{}%
\AgdaSymbol{=}\AgdaSpace{}%
\AgdaInductiveConstructor{assocl⋆}\AgdaSpace{}%
\AgdaOperator{\AgdaInductiveConstructor{◎}}\AgdaSpace{}%
\AgdaSymbol{(}\AgdaInductiveConstructor{swap⋆}\AgdaSpace{}%
\AgdaOperator{\AgdaInductiveConstructor{⊗}}\AgdaSpace{}%
\AgdaInductiveConstructor{id⟷₁}\AgdaSymbol{)}\AgdaSpace{}%
\AgdaOperator{\AgdaInductiveConstructor{◎}}\AgdaSpace{}%
\AgdaInductiveConstructor{assocr⋆}\<%
\\
\\[\AgdaEmptyExtraSkip]%
\>[0]\AgdaFunction{A[BC]-B[CA]}\AgdaSpace{}%
\AgdaSymbol{:}\AgdaSpace{}%
\AgdaSymbol{\{}\AgdaBound{t₁}\AgdaSpace{}%
\AgdaBound{t₂}\AgdaSpace{}%
\AgdaBound{t₃}\AgdaSpace{}%
\AgdaSymbol{:}\AgdaSpace{}%
\AgdaDatatype{Pi.U}\AgdaSymbol{\}}\AgdaSpace{}%
\AgdaSymbol{→}\AgdaSpace{}%
\AgdaBound{t₁}\AgdaSpace{}%
\AgdaOperator{\AgdaInductiveConstructor{Pi.×}}\AgdaSpace{}%
\AgdaSymbol{(}\AgdaBound{t₂}\AgdaSpace{}%
\AgdaOperator{\AgdaInductiveConstructor{Pi.×}}\AgdaSpace{}%
\AgdaBound{t₃}\AgdaSymbol{)}\AgdaSpace{}%
\AgdaOperator{\AgdaDatatype{Pi.⟷₁}}\AgdaSpace{}%
\AgdaBound{t₂}\AgdaSpace{}%
\AgdaOperator{\AgdaInductiveConstructor{Pi.×}}\AgdaSpace{}%
\AgdaSymbol{(}\AgdaBound{t₃}\AgdaSpace{}%
\AgdaOperator{\AgdaInductiveConstructor{Pi.×}}\AgdaSpace{}%
\AgdaBound{t₁}\AgdaSymbol{)}\<%
\\
\>[0]\AgdaFunction{A[BC]-B[CA]}\AgdaSpace{}%
\AgdaSymbol{=}\AgdaSpace{}%
\AgdaInductiveConstructor{swap⋆}\AgdaSpace{}%
\AgdaOperator{\AgdaInductiveConstructor{◎}}\AgdaSpace{}%
\AgdaInductiveConstructor{assocr⋆}\<%
\\
\\[\AgdaEmptyExtraSkip]%
\>[0]\AgdaFunction{B[CA]-A[BC]}\AgdaSpace{}%
\AgdaSymbol{:}\AgdaSpace{}%
\AgdaSymbol{\{}\AgdaBound{t₁}\AgdaSpace{}%
\AgdaBound{t₂}\AgdaSpace{}%
\AgdaBound{t₃}\AgdaSpace{}%
\AgdaSymbol{:}\AgdaSpace{}%
\AgdaDatatype{Pi.U}\AgdaSymbol{\}}\AgdaSpace{}%
\AgdaSymbol{→}\AgdaSpace{}%
\AgdaBound{t₂}\AgdaSpace{}%
\AgdaOperator{\AgdaInductiveConstructor{Pi.×}}\AgdaSpace{}%
\AgdaSymbol{(}\AgdaBound{t₃}\AgdaSpace{}%
\AgdaOperator{\AgdaInductiveConstructor{Pi.×}}\AgdaSpace{}%
\AgdaBound{t₁}\AgdaSymbol{)}\AgdaSpace{}%
\AgdaOperator{\AgdaDatatype{Pi.⟷₁}}\AgdaSpace{}%
\AgdaBound{t₁}\AgdaSpace{}%
\AgdaOperator{\AgdaInductiveConstructor{Pi.×}}\AgdaSpace{}%
\AgdaSymbol{(}\AgdaBound{t₂}\AgdaSpace{}%
\AgdaOperator{\AgdaInductiveConstructor{Pi.×}}\AgdaSpace{}%
\AgdaBound{t₃}\AgdaSymbol{)}\<%
\\
\>[0]\AgdaFunction{B[CA]-A[BC]}\AgdaSpace{}%
\AgdaSymbol{=}\AgdaSpace{}%
\AgdaInductiveConstructor{assocl⋆}\AgdaSpace{}%
\AgdaOperator{\AgdaInductiveConstructor{◎}}\AgdaSpace{}%
\AgdaInductiveConstructor{swap⋆}\<%
\end{code}

\newcommand{\adder}{%
\begin{code}%
\>[0]\AgdaFunction{C[BA]-[CA]B}\AgdaSpace{}%
\AgdaSymbol{:}\AgdaSpace{}%
\AgdaGeneralizable{C}\AgdaSpace{}%
\AgdaOperator{\AgdaInductiveConstructor{×}}\AgdaSpace{}%
\AgdaSymbol{(}\AgdaGeneralizable{B}\AgdaSpace{}%
\AgdaOperator{\AgdaInductiveConstructor{×}}\AgdaSpace{}%
\AgdaGeneralizable{A}\AgdaSymbol{)}\AgdaSpace{}%
\AgdaOperator{\AgdaDatatype{⟷₁}}\AgdaSpace{}%
\AgdaSymbol{(}\AgdaGeneralizable{C}\AgdaSpace{}%
\AgdaOperator{\AgdaInductiveConstructor{×}}\AgdaSpace{}%
\AgdaGeneralizable{A}\AgdaSymbol{)}\AgdaSpace{}%
\AgdaOperator{\AgdaInductiveConstructor{×}}\AgdaSpace{}%
\AgdaGeneralizable{B}\<%
\\
\>[0]\AgdaFunction{C[BA]-[CA]B}\AgdaSpace{}%
\AgdaSymbol{=}\AgdaSpace{}%
\AgdaSymbol{(}\AgdaInductiveConstructor{id⟷₁}\AgdaSpace{}%
\AgdaOperator{\AgdaInductiveConstructor{⊗}}\AgdaSpace{}%
\AgdaInductiveConstructor{swap⋆}\AgdaSymbol{)}\AgdaSpace{}%
\AgdaOperator{\AgdaInductiveConstructor{◎}}\AgdaSpace{}%
\AgdaInductiveConstructor{assocl⋆}\<%
\end{code}}
\begin{code}[hide]%
\>[0]\AgdaFunction{[CA]B-A[BC]}\AgdaSpace{}%
\AgdaSymbol{:}\AgdaSpace{}%
\AgdaSymbol{\{}\AgdaBound{A}\AgdaSpace{}%
\AgdaBound{B}\AgdaSpace{}%
\AgdaBound{C}\AgdaSpace{}%
\AgdaSymbol{:}\AgdaSpace{}%
\AgdaDatatype{U}\AgdaSymbol{\}}\AgdaSpace{}%
\AgdaSymbol{→}\AgdaSpace{}%
\AgdaSymbol{(}\AgdaBound{C}\AgdaSpace{}%
\AgdaOperator{\AgdaInductiveConstructor{×}}\AgdaSpace{}%
\AgdaBound{A}\AgdaSymbol{)}\AgdaSpace{}%
\AgdaOperator{\AgdaInductiveConstructor{×}}\AgdaSpace{}%
\AgdaBound{B}\AgdaSpace{}%
\AgdaOperator{\AgdaDatatype{⟷₁}}\AgdaSpace{}%
\AgdaBound{A}\AgdaSpace{}%
\AgdaOperator{\AgdaInductiveConstructor{×}}\AgdaSpace{}%
\AgdaSymbol{(}\AgdaBound{B}\AgdaSpace{}%
\AgdaOperator{\AgdaInductiveConstructor{×}}\AgdaSpace{}%
\AgdaBound{C}\AgdaSymbol{)}\<%
\\
\>[0]\AgdaFunction{[CA]B-A[BC]}\AgdaSpace{}%
\AgdaSymbol{=}\AgdaSpace{}%
\AgdaFunction{!⟷₁}\AgdaSpace{}%
\AgdaFunction{C[BA]-[CA]B}\AgdaSpace{}%
\AgdaOperator{\AgdaInductiveConstructor{◎}}\AgdaSpace{}%
\AgdaFunction{!⟷₁}\AgdaSpace{}%
\AgdaFunction{A[BC]-C[BA]}\<%
\end{code}
\newcommand{\addertwo}{%
\begin{code}%
\>[0]\AgdaFunction{reversibleOr1}\AgdaSpace{}%
\AgdaSymbol{:}\AgdaSpace{}%
\AgdaFunction{𝔹}\AgdaSpace{}%
\AgdaNumber{3}\AgdaSpace{}%
\AgdaOperator{\AgdaDatatype{⟷₁}}\AgdaSpace{}%
\AgdaFunction{𝔹}\AgdaSpace{}%
\AgdaNumber{3}\<%
\\
\>[0]\AgdaFunction{reversibleOr1}\AgdaSpace{}%
\AgdaSymbol{=}\AgdaSpace{}%
\AgdaFunction{A[BC]-C[BA]}\AgdaSpace{}%
\AgdaOperator{\AgdaInductiveConstructor{◎}}\AgdaSpace{}%
\AgdaFunction{ccx}\AgdaSpace{}%
\AgdaOperator{\AgdaInductiveConstructor{◎}}\AgdaSpace{}%
\AgdaSymbol{(}\AgdaInductiveConstructor{id⟷₁}\AgdaSpace{}%
\AgdaOperator{\AgdaInductiveConstructor{⊗}}\AgdaSpace{}%
\AgdaFunction{cx}\AgdaSymbol{)}\AgdaSpace{}%
\AgdaOperator{\AgdaInductiveConstructor{◎}}\AgdaSpace{}%
\AgdaFunction{C[BA]-[CA]B}\AgdaSpace{}%
\AgdaOperator{\AgdaInductiveConstructor{◎}}\AgdaSpace{}%
\AgdaSymbol{(}\AgdaFunction{cx}\AgdaSpace{}%
\AgdaOperator{\AgdaInductiveConstructor{⊗}}\AgdaSpace{}%
\AgdaInductiveConstructor{id⟷₁}\AgdaSymbol{)}\AgdaSpace{}%
\AgdaOperator{\AgdaInductiveConstructor{◎}}\AgdaSpace{}%
\AgdaFunction{[CA]B-A[BC]}\<%
\end{code}}

\newcommand{\resettwo}{%
\begin{code}%
\>[0]\AgdaFunction{reversibleOr2}\AgdaSpace{}%
\AgdaSymbol{:}\AgdaSpace{}%
\AgdaFunction{𝔹}\AgdaSpace{}%
\AgdaNumber{3}\AgdaSpace{}%
\AgdaOperator{\AgdaDatatype{⟷₁}}\AgdaSpace{}%
\AgdaFunction{𝔹}\AgdaSpace{}%
\AgdaNumber{3}\<%
\\
\>[0]\AgdaFunction{reversibleOr2}\AgdaSpace{}%
\AgdaSymbol{=}\AgdaSpace{}%
\AgdaFunction{A[BC]-B[CA]}\AgdaSpace{}%
\AgdaOperator{\AgdaInductiveConstructor{◎}}\AgdaSpace{}%
\AgdaFunction{cif}\AgdaSpace{}%
\AgdaSymbol{(}\AgdaFunction{x}\AgdaSpace{}%
\AgdaOperator{\AgdaInductiveConstructor{⊗}}\AgdaSpace{}%
\AgdaInductiveConstructor{id⟷₁}\AgdaSymbol{)}\AgdaSpace{}%
\AgdaFunction{cx}\AgdaSpace{}%
\AgdaOperator{\AgdaInductiveConstructor{◎}}\AgdaSpace{}%
\AgdaFunction{B[CA]-A[BC]}\<%
\end{code}}

\begin{code}[hide]%
\>[0]\AgdaKeyword{open}\AgdaSpace{}%
\AgdaKeyword{import}\AgdaSpace{}%
\AgdaModule{Pi.Equiv.Equiv1NormHelpers}\<%
\\
\>[0]\AgdaFunction{step1}\AgdaSpace{}%
\AgdaSymbol{=}\AgdaSpace{}%
\AgdaFunction{pi\textasciicircum{}2list}\AgdaSpace{}%
\AgdaSymbol{(}\AgdaFunction{eval₁}\AgdaSpace{}%
\AgdaSymbol{(}\AgdaFunction{swaplr1}\AgdaSpace{}%
\AgdaSymbol{\{}\AgdaInductiveConstructor{I}\AgdaSymbol{\}}\AgdaSpace{}%
\AgdaSymbol{\{}\AgdaInductiveConstructor{I}\AgdaSymbol{\}}\AgdaSpace{}%
\AgdaSymbol{\{}\AgdaInductiveConstructor{I}\AgdaSymbol{\}))}\<%
\end{code}

\begin{code}[hide]%
\>[0]\AgdaKeyword{data}\AgdaSpace{}%
\AgdaOperator{\AgdaDatatype{\AgdaUnderscore{}⟷₂\AgdaUnderscore{}}}\AgdaSpace{}%
\AgdaSymbol{:}\AgdaSpace{}%
\AgdaSymbol{\{}\AgdaBound{X}\AgdaSpace{}%
\AgdaSymbol{:}\AgdaSpace{}%
\AgdaDatatype{U}\AgdaSymbol{\}}\AgdaSpace{}%
\AgdaSymbol{\{}\AgdaBound{Y}\AgdaSpace{}%
\AgdaSymbol{:}\AgdaSpace{}%
\AgdaDatatype{U}\AgdaSymbol{\}}\AgdaSpace{}%
\AgdaSymbol{→}\AgdaSpace{}%
\AgdaBound{X}\AgdaSpace{}%
\AgdaOperator{\AgdaDatatype{⟷₁}}\AgdaSpace{}%
\AgdaBound{Y}\AgdaSpace{}%
\AgdaSymbol{→}\AgdaSpace{}%
\AgdaBound{X}\AgdaSpace{}%
\AgdaOperator{\AgdaDatatype{⟷₁}}\AgdaSpace{}%
\AgdaBound{Y}\AgdaSpace{}%
\AgdaSymbol{→}\AgdaSpace{}%
\AgdaPrimitiveType{Set}\AgdaSpace{}%
\AgdaKeyword{where}\<%
\end{code}

\newcommand{\leveltwoblockone}{%
\begin{code}%
\>[0][@{}l@{\AgdaIndent{1}}]%
\>[2]\AgdaInductiveConstructor{assoc◎l}%
\>[12]\AgdaSymbol{:}\AgdaSpace{}%
\AgdaSymbol{\{}\AgdaBound{c₁}\AgdaSpace{}%
\AgdaSymbol{:}\AgdaSpace{}%
\AgdaGeneralizable{A}\AgdaSpace{}%
\AgdaOperator{\AgdaDatatype{⟷₁}}\AgdaSpace{}%
\AgdaGeneralizable{B}\AgdaSymbol{\}}\AgdaSpace{}%
\AgdaSymbol{\{}\AgdaBound{c₂}\AgdaSpace{}%
\AgdaSymbol{:}\AgdaSpace{}%
\AgdaGeneralizable{B}\AgdaSpace{}%
\AgdaOperator{\AgdaDatatype{⟷₁}}\AgdaSpace{}%
\AgdaGeneralizable{C}\AgdaSymbol{\}}\AgdaSpace{}%
\AgdaSymbol{\{}\AgdaBound{c₃}\AgdaSpace{}%
\AgdaSymbol{:}\AgdaSpace{}%
\AgdaGeneralizable{C}\AgdaSpace{}%
\AgdaOperator{\AgdaDatatype{⟷₁}}\AgdaSpace{}%
\AgdaGeneralizable{D}\AgdaSymbol{\}}\AgdaSpace{}%
\AgdaSymbol{→}\AgdaSpace{}%
\AgdaSymbol{(}\AgdaBound{c₁}\AgdaSpace{}%
\AgdaOperator{\AgdaInductiveConstructor{◎}}\AgdaSpace{}%
\AgdaSymbol{(}\AgdaBound{c₂}\AgdaSpace{}%
\AgdaOperator{\AgdaInductiveConstructor{◎}}\AgdaSpace{}%
\AgdaBound{c₃}\AgdaSymbol{))}\AgdaSpace{}%
\AgdaOperator{\AgdaDatatype{⟷₂}}\AgdaSpace{}%
\AgdaSymbol{((}\AgdaBound{c₁}\AgdaSpace{}%
\AgdaOperator{\AgdaInductiveConstructor{◎}}\AgdaSpace{}%
\AgdaBound{c₂}\AgdaSymbol{)}\AgdaSpace{}%
\AgdaOperator{\AgdaInductiveConstructor{◎}}\AgdaSpace{}%
\AgdaBound{c₃}\AgdaSymbol{)}\<%
\\
\>[2]\AgdaInductiveConstructor{assoc◎r}%
\>[12]\AgdaSymbol{:}\AgdaSpace{}%
\AgdaSymbol{\{}\AgdaBound{c₁}\AgdaSpace{}%
\AgdaSymbol{:}\AgdaSpace{}%
\AgdaGeneralizable{A}\AgdaSpace{}%
\AgdaOperator{\AgdaDatatype{⟷₁}}\AgdaSpace{}%
\AgdaGeneralizable{B}\AgdaSymbol{\}}\AgdaSpace{}%
\AgdaSymbol{\{}\AgdaBound{c₂}\AgdaSpace{}%
\AgdaSymbol{:}\AgdaSpace{}%
\AgdaGeneralizable{B}\AgdaSpace{}%
\AgdaOperator{\AgdaDatatype{⟷₁}}\AgdaSpace{}%
\AgdaGeneralizable{C}\AgdaSymbol{\}}\AgdaSpace{}%
\AgdaSymbol{\{}\AgdaBound{c₃}\AgdaSpace{}%
\AgdaSymbol{:}\AgdaSpace{}%
\AgdaGeneralizable{C}\AgdaSpace{}%
\AgdaOperator{\AgdaDatatype{⟷₁}}\AgdaSpace{}%
\AgdaGeneralizable{D}\AgdaSymbol{\}}\AgdaSpace{}%
\AgdaSymbol{→}\AgdaSpace{}%
\AgdaSymbol{((}\AgdaBound{c₁}\AgdaSpace{}%
\AgdaOperator{\AgdaInductiveConstructor{◎}}\AgdaSpace{}%
\AgdaBound{c₂}\AgdaSymbol{)}\AgdaSpace{}%
\AgdaOperator{\AgdaInductiveConstructor{◎}}\AgdaSpace{}%
\AgdaBound{c₃}\AgdaSymbol{)}\AgdaSpace{}%
\AgdaOperator{\AgdaDatatype{⟷₂}}\AgdaSpace{}%
\AgdaSymbol{(}\AgdaBound{c₁}\AgdaSpace{}%
\AgdaOperator{\AgdaInductiveConstructor{◎}}\AgdaSpace{}%
\AgdaSymbol{(}\AgdaBound{c₂}\AgdaSpace{}%
\AgdaOperator{\AgdaInductiveConstructor{◎}}\AgdaSpace{}%
\AgdaBound{c₃}\AgdaSymbol{))}\<%
\\
\>[2]\AgdaInductiveConstructor{assocl₊l}%
\>[12]\AgdaSymbol{:}\AgdaSpace{}%
\AgdaSymbol{\{}\AgdaBound{c₁}\AgdaSpace{}%
\AgdaSymbol{:}\AgdaSpace{}%
\AgdaGeneralizable{A}\AgdaSpace{}%
\AgdaOperator{\AgdaDatatype{⟷₁}}\AgdaSpace{}%
\AgdaGeneralizable{B}\AgdaSymbol{\}}\AgdaSpace{}%
\AgdaSymbol{\{}\AgdaBound{c₂}\AgdaSpace{}%
\AgdaSymbol{:}\AgdaSpace{}%
\AgdaGeneralizable{C}\AgdaSpace{}%
\AgdaOperator{\AgdaDatatype{⟷₁}}\AgdaSpace{}%
\AgdaGeneralizable{D}\AgdaSymbol{\}}\AgdaSpace{}%
\AgdaSymbol{\{}\AgdaBound{c₃}\AgdaSpace{}%
\AgdaSymbol{:}\AgdaSpace{}%
\AgdaGeneralizable{E}\AgdaSpace{}%
\AgdaOperator{\AgdaDatatype{⟷₁}}\AgdaSpace{}%
\AgdaGeneralizable{F}\AgdaSymbol{\}}\AgdaSpace{}%
\AgdaSymbol{→}\AgdaSpace{}%
\AgdaSymbol{((}\AgdaBound{c₁}\AgdaSpace{}%
\AgdaOperator{\AgdaInductiveConstructor{⊕}}\AgdaSpace{}%
\AgdaSymbol{(}\AgdaBound{c₂}\AgdaSpace{}%
\AgdaOperator{\AgdaInductiveConstructor{⊕}}\AgdaSpace{}%
\AgdaBound{c₃}\AgdaSymbol{))}\AgdaSpace{}%
\AgdaOperator{\AgdaInductiveConstructor{◎}}\AgdaSpace{}%
\AgdaInductiveConstructor{assocl₊}\AgdaSymbol{)}\AgdaSpace{}%
\AgdaOperator{\AgdaDatatype{⟷₂}}\AgdaSpace{}%
\AgdaSymbol{(}\AgdaInductiveConstructor{assocl₊}\AgdaSpace{}%
\AgdaOperator{\AgdaInductiveConstructor{◎}}\AgdaSpace{}%
\AgdaSymbol{((}\AgdaBound{c₁}\AgdaSpace{}%
\AgdaOperator{\AgdaInductiveConstructor{⊕}}\AgdaSpace{}%
\AgdaBound{c₂}\AgdaSymbol{)}\AgdaSpace{}%
\AgdaOperator{\AgdaInductiveConstructor{⊕}}\AgdaSpace{}%
\AgdaBound{c₃}\AgdaSymbol{))}\<%
\\
\>[2]\AgdaInductiveConstructor{assocl₊r}%
\>[12]\AgdaSymbol{:}\AgdaSpace{}%
\AgdaSymbol{\{}\AgdaBound{c₁}\AgdaSpace{}%
\AgdaSymbol{:}\AgdaSpace{}%
\AgdaGeneralizable{A}\AgdaSpace{}%
\AgdaOperator{\AgdaDatatype{⟷₁}}\AgdaSpace{}%
\AgdaGeneralizable{B}\AgdaSymbol{\}}\AgdaSpace{}%
\AgdaSymbol{\{}\AgdaBound{c₂}\AgdaSpace{}%
\AgdaSymbol{:}\AgdaSpace{}%
\AgdaGeneralizable{C}\AgdaSpace{}%
\AgdaOperator{\AgdaDatatype{⟷₁}}\AgdaSpace{}%
\AgdaGeneralizable{D}\AgdaSymbol{\}}\AgdaSpace{}%
\AgdaSymbol{\{}\AgdaBound{c₃}\AgdaSpace{}%
\AgdaSymbol{:}\AgdaSpace{}%
\AgdaGeneralizable{E}\AgdaSpace{}%
\AgdaOperator{\AgdaDatatype{⟷₁}}\AgdaSpace{}%
\AgdaGeneralizable{F}\AgdaSymbol{\}}\AgdaSpace{}%
\AgdaSymbol{→}\AgdaSpace{}%
\AgdaSymbol{(}\AgdaInductiveConstructor{assocl₊}\AgdaSpace{}%
\AgdaOperator{\AgdaInductiveConstructor{◎}}\AgdaSpace{}%
\AgdaSymbol{((}\AgdaBound{c₁}\AgdaSpace{}%
\AgdaOperator{\AgdaInductiveConstructor{⊕}}\AgdaSpace{}%
\AgdaBound{c₂}\AgdaSymbol{)}\AgdaSpace{}%
\AgdaOperator{\AgdaInductiveConstructor{⊕}}\AgdaSpace{}%
\AgdaBound{c₃}\AgdaSymbol{))}\AgdaSpace{}%
\AgdaOperator{\AgdaDatatype{⟷₂}}\AgdaSpace{}%
\AgdaSymbol{((}\AgdaBound{c₁}\AgdaSpace{}%
\AgdaOperator{\AgdaInductiveConstructor{⊕}}\AgdaSpace{}%
\AgdaSymbol{(}\AgdaBound{c₂}\AgdaSpace{}%
\AgdaOperator{\AgdaInductiveConstructor{⊕}}\AgdaSpace{}%
\AgdaBound{c₃}\AgdaSymbol{))}\AgdaSpace{}%
\AgdaOperator{\AgdaInductiveConstructor{◎}}\AgdaSpace{}%
\AgdaInductiveConstructor{assocl₊}\AgdaSymbol{)}\<%
\\
\>[2]\AgdaInductiveConstructor{assocr₊r}%
\>[12]\AgdaSymbol{:}\AgdaSpace{}%
\AgdaSymbol{\{}\AgdaBound{c₁}\AgdaSpace{}%
\AgdaSymbol{:}\AgdaSpace{}%
\AgdaGeneralizable{A}\AgdaSpace{}%
\AgdaOperator{\AgdaDatatype{⟷₁}}\AgdaSpace{}%
\AgdaGeneralizable{B}\AgdaSymbol{\}}\AgdaSpace{}%
\AgdaSymbol{\{}\AgdaBound{c₂}\AgdaSpace{}%
\AgdaSymbol{:}\AgdaSpace{}%
\AgdaGeneralizable{C}\AgdaSpace{}%
\AgdaOperator{\AgdaDatatype{⟷₁}}\AgdaSpace{}%
\AgdaGeneralizable{D}\AgdaSymbol{\}}\AgdaSpace{}%
\AgdaSymbol{\{}\AgdaBound{c₃}\AgdaSpace{}%
\AgdaSymbol{:}\AgdaSpace{}%
\AgdaGeneralizable{E}\AgdaSpace{}%
\AgdaOperator{\AgdaDatatype{⟷₁}}\AgdaSpace{}%
\AgdaGeneralizable{F}\AgdaSymbol{\}}\AgdaSpace{}%
\AgdaSymbol{→}\AgdaSpace{}%
\AgdaSymbol{(((}\AgdaBound{c₁}\AgdaSpace{}%
\AgdaOperator{\AgdaInductiveConstructor{⊕}}\AgdaSpace{}%
\AgdaBound{c₂}\AgdaSymbol{)}\AgdaSpace{}%
\AgdaOperator{\AgdaInductiveConstructor{⊕}}\AgdaSpace{}%
\AgdaBound{c₃}\AgdaSymbol{)}\AgdaSpace{}%
\AgdaOperator{\AgdaInductiveConstructor{◎}}\AgdaSpace{}%
\AgdaInductiveConstructor{assocr₊}\AgdaSymbol{)}\AgdaSpace{}%
\AgdaOperator{\AgdaDatatype{⟷₂}}\AgdaSpace{}%
\AgdaSymbol{(}\AgdaInductiveConstructor{assocr₊}\AgdaSpace{}%
\AgdaOperator{\AgdaInductiveConstructor{◎}}\AgdaSpace{}%
\AgdaSymbol{(}\AgdaBound{c₁}\AgdaSpace{}%
\AgdaOperator{\AgdaInductiveConstructor{⊕}}\AgdaSpace{}%
\AgdaSymbol{(}\AgdaBound{c₂}\AgdaSpace{}%
\AgdaOperator{\AgdaInductiveConstructor{⊕}}\AgdaSpace{}%
\AgdaBound{c₃}\AgdaSymbol{)))}\<%
\\
\>[2]\AgdaInductiveConstructor{assocr₊l}%
\>[12]\AgdaSymbol{:}\AgdaSpace{}%
\AgdaSymbol{\{}\AgdaBound{c₁}\AgdaSpace{}%
\AgdaSymbol{:}\AgdaSpace{}%
\AgdaGeneralizable{A}\AgdaSpace{}%
\AgdaOperator{\AgdaDatatype{⟷₁}}\AgdaSpace{}%
\AgdaGeneralizable{B}\AgdaSymbol{\}}\AgdaSpace{}%
\AgdaSymbol{\{}\AgdaBound{c₂}\AgdaSpace{}%
\AgdaSymbol{:}\AgdaSpace{}%
\AgdaGeneralizable{C}\AgdaSpace{}%
\AgdaOperator{\AgdaDatatype{⟷₁}}\AgdaSpace{}%
\AgdaGeneralizable{D}\AgdaSymbol{\}}\AgdaSpace{}%
\AgdaSymbol{\{}\AgdaBound{c₃}\AgdaSpace{}%
\AgdaSymbol{:}\AgdaSpace{}%
\AgdaGeneralizable{E}\AgdaSpace{}%
\AgdaOperator{\AgdaDatatype{⟷₁}}\AgdaSpace{}%
\AgdaGeneralizable{F}\AgdaSymbol{\}}\AgdaSpace{}%
\AgdaSymbol{→}\AgdaSpace{}%
\AgdaSymbol{(}\AgdaInductiveConstructor{assocr₊}\AgdaSpace{}%
\AgdaOperator{\AgdaInductiveConstructor{◎}}\AgdaSpace{}%
\AgdaSymbol{(}\AgdaBound{c₁}\AgdaSpace{}%
\AgdaOperator{\AgdaInductiveConstructor{⊕}}\AgdaSpace{}%
\AgdaSymbol{(}\AgdaBound{c₂}\AgdaSpace{}%
\AgdaOperator{\AgdaInductiveConstructor{⊕}}\AgdaSpace{}%
\AgdaBound{c₃}\AgdaSymbol{)))}\AgdaSpace{}%
\AgdaOperator{\AgdaDatatype{⟷₂}}\AgdaSpace{}%
\AgdaSymbol{(((}\AgdaBound{c₁}\AgdaSpace{}%
\AgdaOperator{\AgdaInductiveConstructor{⊕}}\AgdaSpace{}%
\AgdaBound{c₂}\AgdaSymbol{)}\AgdaSpace{}%
\AgdaOperator{\AgdaInductiveConstructor{⊕}}\AgdaSpace{}%
\AgdaBound{c₃}\AgdaSymbol{)}\AgdaSpace{}%
\AgdaOperator{\AgdaInductiveConstructor{◎}}\AgdaSpace{}%
\AgdaInductiveConstructor{assocr₊}\AgdaSymbol{)}\<%
\end{code}}

\newcommand{\leveltwoblocktwo}{%
\begin{code}%
\>[2]\AgdaInductiveConstructor{idl◎l}%
\>[10]\AgdaSymbol{:}\AgdaSpace{}%
\AgdaSymbol{\{}\AgdaBound{c}\AgdaSpace{}%
\AgdaSymbol{:}\AgdaSpace{}%
\AgdaGeneralizable{A}\AgdaSpace{}%
\AgdaOperator{\AgdaDatatype{⟷₁}}\AgdaSpace{}%
\AgdaGeneralizable{B}\AgdaSymbol{\}}\AgdaSpace{}%
\AgdaSymbol{→}\AgdaSpace{}%
\AgdaSymbol{(}\AgdaInductiveConstructor{id⟷₁}\AgdaSpace{}%
\AgdaOperator{\AgdaInductiveConstructor{◎}}\AgdaSpace{}%
\AgdaBound{c}\AgdaSymbol{)}\AgdaSpace{}%
\AgdaOperator{\AgdaDatatype{⟷₂}}\AgdaSpace{}%
\AgdaBound{c}\<%
\\
\>[2]\AgdaInductiveConstructor{idl◎r}%
\>[10]\AgdaSymbol{:}\AgdaSpace{}%
\AgdaSymbol{\{}\AgdaBound{c}\AgdaSpace{}%
\AgdaSymbol{:}\AgdaSpace{}%
\AgdaGeneralizable{A}\AgdaSpace{}%
\AgdaOperator{\AgdaDatatype{⟷₁}}\AgdaSpace{}%
\AgdaGeneralizable{B}\AgdaSymbol{\}}\AgdaSpace{}%
\AgdaSymbol{→}\AgdaSpace{}%
\AgdaBound{c}\AgdaSpace{}%
\AgdaOperator{\AgdaDatatype{⟷₂}}\AgdaSpace{}%
\AgdaInductiveConstructor{id⟷₁}\AgdaSpace{}%
\AgdaOperator{\AgdaInductiveConstructor{◎}}\AgdaSpace{}%
\AgdaBound{c}\<%
\\
\>[2]\AgdaInductiveConstructor{idr◎l}%
\>[10]\AgdaSymbol{:}\AgdaSpace{}%
\AgdaSymbol{\{}\AgdaBound{c}\AgdaSpace{}%
\AgdaSymbol{:}\AgdaSpace{}%
\AgdaGeneralizable{A}\AgdaSpace{}%
\AgdaOperator{\AgdaDatatype{⟷₁}}\AgdaSpace{}%
\AgdaGeneralizable{B}\AgdaSymbol{\}}\AgdaSpace{}%
\AgdaSymbol{→}\AgdaSpace{}%
\AgdaSymbol{(}\AgdaBound{c}\AgdaSpace{}%
\AgdaOperator{\AgdaInductiveConstructor{◎}}\AgdaSpace{}%
\AgdaInductiveConstructor{id⟷₁}\AgdaSymbol{)}\AgdaSpace{}%
\AgdaOperator{\AgdaDatatype{⟷₂}}\AgdaSpace{}%
\AgdaBound{c}\<%
\\
\>[2]\AgdaInductiveConstructor{idr◎r}%
\>[10]\AgdaSymbol{:}\AgdaSpace{}%
\AgdaSymbol{\{}\AgdaBound{c}\AgdaSpace{}%
\AgdaSymbol{:}\AgdaSpace{}%
\AgdaGeneralizable{A}\AgdaSpace{}%
\AgdaOperator{\AgdaDatatype{⟷₁}}\AgdaSpace{}%
\AgdaGeneralizable{B}\AgdaSymbol{\}}\AgdaSpace{}%
\AgdaSymbol{→}\AgdaSpace{}%
\AgdaBound{c}\AgdaSpace{}%
\AgdaOperator{\AgdaDatatype{⟷₂}}\AgdaSpace{}%
\AgdaSymbol{(}\AgdaBound{c}\AgdaSpace{}%
\AgdaOperator{\AgdaInductiveConstructor{◎}}\AgdaSpace{}%
\AgdaInductiveConstructor{id⟷₁}\AgdaSymbol{)}\<%
\\
\\[\AgdaEmptyExtraSkip]%
\>[2]\AgdaInductiveConstructor{linv◎l}%
\>[10]\AgdaSymbol{:}\AgdaSpace{}%
\AgdaSymbol{\{}\AgdaBound{c}\AgdaSpace{}%
\AgdaSymbol{:}\AgdaSpace{}%
\AgdaGeneralizable{A}\AgdaSpace{}%
\AgdaOperator{\AgdaDatatype{⟷₁}}\AgdaSpace{}%
\AgdaGeneralizable{B}\AgdaSymbol{\}}\AgdaSpace{}%
\AgdaSymbol{→}\AgdaSpace{}%
\AgdaSymbol{(}\AgdaBound{c}\AgdaSpace{}%
\AgdaOperator{\AgdaInductiveConstructor{◎}}\AgdaSpace{}%
\AgdaFunction{!⟷₁}\AgdaSpace{}%
\AgdaBound{c}\AgdaSymbol{)}\AgdaSpace{}%
\AgdaOperator{\AgdaDatatype{⟷₂}}\AgdaSpace{}%
\AgdaInductiveConstructor{id⟷₁}\<%
\\
\>[2]\AgdaInductiveConstructor{linv◎r}%
\>[10]\AgdaSymbol{:}\AgdaSpace{}%
\AgdaSymbol{\{}\AgdaBound{c}\AgdaSpace{}%
\AgdaSymbol{:}\AgdaSpace{}%
\AgdaGeneralizable{A}\AgdaSpace{}%
\AgdaOperator{\AgdaDatatype{⟷₁}}\AgdaSpace{}%
\AgdaGeneralizable{B}\AgdaSymbol{\}}\AgdaSpace{}%
\AgdaSymbol{→}\AgdaSpace{}%
\AgdaInductiveConstructor{id⟷₁}\AgdaSpace{}%
\AgdaOperator{\AgdaDatatype{⟷₂}}\AgdaSpace{}%
\AgdaSymbol{(}\AgdaBound{c}\AgdaSpace{}%
\AgdaOperator{\AgdaInductiveConstructor{◎}}\AgdaSpace{}%
\AgdaFunction{!⟷₁}\AgdaSpace{}%
\AgdaBound{c}\AgdaSymbol{)}\<%
\\
\>[2]\AgdaInductiveConstructor{rinv◎l}%
\>[10]\AgdaSymbol{:}\AgdaSpace{}%
\AgdaSymbol{\{}\AgdaBound{c}\AgdaSpace{}%
\AgdaSymbol{:}\AgdaSpace{}%
\AgdaGeneralizable{A}\AgdaSpace{}%
\AgdaOperator{\AgdaDatatype{⟷₁}}\AgdaSpace{}%
\AgdaGeneralizable{B}\AgdaSymbol{\}}\AgdaSpace{}%
\AgdaSymbol{→}\AgdaSpace{}%
\AgdaSymbol{(}\AgdaFunction{!⟷₁}\AgdaSpace{}%
\AgdaBound{c}\AgdaSpace{}%
\AgdaOperator{\AgdaInductiveConstructor{◎}}\AgdaSpace{}%
\AgdaBound{c}\AgdaSymbol{)}\AgdaSpace{}%
\AgdaOperator{\AgdaDatatype{⟷₂}}\AgdaSpace{}%
\AgdaInductiveConstructor{id⟷₁}\<%
\\
\>[2]\AgdaInductiveConstructor{rinv◎r}%
\>[10]\AgdaSymbol{:}\AgdaSpace{}%
\AgdaSymbol{\{}\AgdaBound{c}\AgdaSpace{}%
\AgdaSymbol{:}\AgdaSpace{}%
\AgdaGeneralizable{A}\AgdaSpace{}%
\AgdaOperator{\AgdaDatatype{⟷₁}}\AgdaSpace{}%
\AgdaGeneralizable{B}\AgdaSymbol{\}}\AgdaSpace{}%
\AgdaSymbol{→}\AgdaSpace{}%
\AgdaInductiveConstructor{id⟷₁}\AgdaSpace{}%
\AgdaOperator{\AgdaDatatype{⟷₂}}\AgdaSpace{}%
\AgdaSymbol{(}\AgdaFunction{!⟷₁}\AgdaSpace{}%
\AgdaBound{c}\AgdaSpace{}%
\AgdaOperator{\AgdaInductiveConstructor{◎}}\AgdaSpace{}%
\AgdaBound{c}\AgdaSymbol{)}\<%
\\
\\[\AgdaEmptyExtraSkip]%
\>[2]\AgdaInductiveConstructor{unite₊l⟷₂l}%
\>[14]\AgdaSymbol{:}\AgdaSpace{}%
\AgdaSymbol{\{}\AgdaBound{c₁}\AgdaSpace{}%
\AgdaSymbol{:}\AgdaSpace{}%
\AgdaInductiveConstructor{O}\AgdaSpace{}%
\AgdaOperator{\AgdaDatatype{⟷₁}}\AgdaSpace{}%
\AgdaInductiveConstructor{O}\AgdaSymbol{\}}\AgdaSpace{}%
\AgdaSymbol{\{}\AgdaBound{c₂}\AgdaSpace{}%
\AgdaSymbol{:}\AgdaSpace{}%
\AgdaGeneralizable{A}\AgdaSpace{}%
\AgdaOperator{\AgdaDatatype{⟷₁}}\AgdaSpace{}%
\AgdaGeneralizable{B}\AgdaSymbol{\}}\AgdaSpace{}%
\AgdaSymbol{→}\AgdaSpace{}%
\AgdaSymbol{(}\AgdaInductiveConstructor{unite₊l}\AgdaSpace{}%
\AgdaOperator{\AgdaInductiveConstructor{◎}}\AgdaSpace{}%
\AgdaBound{c₂}\AgdaSymbol{)}\AgdaSpace{}%
\AgdaOperator{\AgdaDatatype{⟷₂}}\AgdaSpace{}%
\AgdaSymbol{((}\AgdaBound{c₁}\AgdaSpace{}%
\AgdaOperator{\AgdaInductiveConstructor{⊕}}\AgdaSpace{}%
\AgdaBound{c₂}\AgdaSymbol{)}\AgdaSpace{}%
\AgdaOperator{\AgdaInductiveConstructor{◎}}\AgdaSpace{}%
\AgdaInductiveConstructor{unite₊l}\AgdaSymbol{)}\<%
\\
\>[2]\AgdaInductiveConstructor{unite₊l⟷₂r}%
\>[14]\AgdaSymbol{:}\AgdaSpace{}%
\AgdaSymbol{\{}\AgdaBound{c₁}\AgdaSpace{}%
\AgdaSymbol{:}\AgdaSpace{}%
\AgdaInductiveConstructor{O}\AgdaSpace{}%
\AgdaOperator{\AgdaDatatype{⟷₁}}\AgdaSpace{}%
\AgdaInductiveConstructor{O}\AgdaSymbol{\}}\AgdaSpace{}%
\AgdaSymbol{\{}\AgdaBound{c₂}\AgdaSpace{}%
\AgdaSymbol{:}\AgdaSpace{}%
\AgdaGeneralizable{A}\AgdaSpace{}%
\AgdaOperator{\AgdaDatatype{⟷₁}}\AgdaSpace{}%
\AgdaGeneralizable{B}\AgdaSymbol{\}}\AgdaSpace{}%
\AgdaSymbol{→}\AgdaSpace{}%
\AgdaSymbol{((}\AgdaBound{c₁}\AgdaSpace{}%
\AgdaOperator{\AgdaInductiveConstructor{⊕}}\AgdaSpace{}%
\AgdaBound{c₂}\AgdaSymbol{)}\AgdaSpace{}%
\AgdaOperator{\AgdaInductiveConstructor{◎}}\AgdaSpace{}%
\AgdaInductiveConstructor{unite₊l}\AgdaSymbol{)}\AgdaSpace{}%
\AgdaOperator{\AgdaDatatype{⟷₂}}\AgdaSpace{}%
\AgdaSymbol{(}\AgdaInductiveConstructor{unite₊l}\AgdaSpace{}%
\AgdaOperator{\AgdaInductiveConstructor{◎}}\AgdaSpace{}%
\AgdaBound{c₂}\AgdaSymbol{)}\<%
\\
\>[2]\AgdaInductiveConstructor{uniti₊l⟷₂l}%
\>[14]\AgdaSymbol{:}\AgdaSpace{}%
\AgdaSymbol{\{}\AgdaBound{c₁}\AgdaSpace{}%
\AgdaSymbol{:}\AgdaSpace{}%
\AgdaInductiveConstructor{O}\AgdaSpace{}%
\AgdaOperator{\AgdaDatatype{⟷₁}}\AgdaSpace{}%
\AgdaInductiveConstructor{O}\AgdaSymbol{\}}\AgdaSpace{}%
\AgdaSymbol{\{}\AgdaBound{c₂}\AgdaSpace{}%
\AgdaSymbol{:}\AgdaSpace{}%
\AgdaGeneralizable{A}\AgdaSpace{}%
\AgdaOperator{\AgdaDatatype{⟷₁}}\AgdaSpace{}%
\AgdaGeneralizable{B}\AgdaSymbol{\}}\AgdaSpace{}%
\AgdaSymbol{→}\AgdaSpace{}%
\AgdaSymbol{(}\AgdaInductiveConstructor{uniti₊l}\AgdaSpace{}%
\AgdaOperator{\AgdaInductiveConstructor{◎}}\AgdaSpace{}%
\AgdaSymbol{(}\AgdaBound{c₁}\AgdaSpace{}%
\AgdaOperator{\AgdaInductiveConstructor{⊕}}\AgdaSpace{}%
\AgdaBound{c₂}\AgdaSymbol{))}\AgdaSpace{}%
\AgdaOperator{\AgdaDatatype{⟷₂}}\AgdaSpace{}%
\AgdaSymbol{(}\AgdaBound{c₂}\AgdaSpace{}%
\AgdaOperator{\AgdaInductiveConstructor{◎}}\AgdaSpace{}%
\AgdaInductiveConstructor{uniti₊l}\AgdaSymbol{)}\<%
\\
\>[2]\AgdaInductiveConstructor{uniti₊l⟷₂r}%
\>[14]\AgdaSymbol{:}\AgdaSpace{}%
\AgdaSymbol{\{}\AgdaBound{c₁}\AgdaSpace{}%
\AgdaSymbol{:}\AgdaSpace{}%
\AgdaInductiveConstructor{O}\AgdaSpace{}%
\AgdaOperator{\AgdaDatatype{⟷₁}}\AgdaSpace{}%
\AgdaInductiveConstructor{O}\AgdaSymbol{\}}\AgdaSpace{}%
\AgdaSymbol{\{}\AgdaBound{c₂}\AgdaSpace{}%
\AgdaSymbol{:}\AgdaSpace{}%
\AgdaGeneralizable{A}\AgdaSpace{}%
\AgdaOperator{\AgdaDatatype{⟷₁}}\AgdaSpace{}%
\AgdaGeneralizable{B}\AgdaSymbol{\}}\AgdaSpace{}%
\AgdaSymbol{→}\AgdaSpace{}%
\AgdaSymbol{(}\AgdaBound{c₂}\AgdaSpace{}%
\AgdaOperator{\AgdaInductiveConstructor{◎}}\AgdaSpace{}%
\AgdaInductiveConstructor{uniti₊l}\AgdaSymbol{)}\AgdaSpace{}%
\AgdaOperator{\AgdaDatatype{⟷₂}}\AgdaSpace{}%
\AgdaSymbol{(}\AgdaInductiveConstructor{uniti₊l}\AgdaSpace{}%
\AgdaOperator{\AgdaInductiveConstructor{◎}}\AgdaSpace{}%
\AgdaSymbol{(}\AgdaBound{c₁}\AgdaSpace{}%
\AgdaOperator{\AgdaInductiveConstructor{⊕}}\AgdaSpace{}%
\AgdaBound{c₂}\AgdaSymbol{))}\<%
\\
\>[2]\AgdaInductiveConstructor{swapl₊⟷₂}%
\>[14]\AgdaSymbol{:}\AgdaSpace{}%
\AgdaSymbol{\{}\AgdaBound{c₁}\AgdaSpace{}%
\AgdaSymbol{:}\AgdaSpace{}%
\AgdaGeneralizable{A}\AgdaSpace{}%
\AgdaOperator{\AgdaDatatype{⟷₁}}\AgdaSpace{}%
\AgdaGeneralizable{B}\AgdaSymbol{\}}\AgdaSpace{}%
\AgdaSymbol{\{}\AgdaBound{c₂}\AgdaSpace{}%
\AgdaSymbol{:}\AgdaSpace{}%
\AgdaGeneralizable{C}\AgdaSpace{}%
\AgdaOperator{\AgdaDatatype{⟷₁}}\AgdaSpace{}%
\AgdaGeneralizable{D}\AgdaSymbol{\}}\AgdaSpace{}%
\AgdaSymbol{→}\AgdaSpace{}%
\AgdaSymbol{(}\AgdaInductiveConstructor{swap₊}\AgdaSpace{}%
\AgdaOperator{\AgdaInductiveConstructor{◎}}\AgdaSpace{}%
\AgdaSymbol{(}\AgdaBound{c₁}\AgdaSpace{}%
\AgdaOperator{\AgdaInductiveConstructor{⊕}}\AgdaSpace{}%
\AgdaBound{c₂}\AgdaSymbol{))}\AgdaSpace{}%
\AgdaOperator{\AgdaDatatype{⟷₂}}\AgdaSpace{}%
\AgdaSymbol{((}\AgdaBound{c₂}\AgdaSpace{}%
\AgdaOperator{\AgdaInductiveConstructor{⊕}}\AgdaSpace{}%
\AgdaBound{c₁}\AgdaSymbol{)}\AgdaSpace{}%
\AgdaOperator{\AgdaInductiveConstructor{◎}}\AgdaSpace{}%
\AgdaInductiveConstructor{swap₊}\AgdaSymbol{)}\<%
\\
\>[2]\AgdaInductiveConstructor{swapr₊⟷₂}%
\>[14]\AgdaSymbol{:}\AgdaSpace{}%
\AgdaSymbol{\{}\AgdaBound{c₁}\AgdaSpace{}%
\AgdaSymbol{:}\AgdaSpace{}%
\AgdaGeneralizable{A}\AgdaSpace{}%
\AgdaOperator{\AgdaDatatype{⟷₁}}\AgdaSpace{}%
\AgdaGeneralizable{B}\AgdaSymbol{\}}\AgdaSpace{}%
\AgdaSymbol{\{}\AgdaBound{c₂}\AgdaSpace{}%
\AgdaSymbol{:}\AgdaSpace{}%
\AgdaGeneralizable{C}\AgdaSpace{}%
\AgdaOperator{\AgdaDatatype{⟷₁}}\AgdaSpace{}%
\AgdaGeneralizable{D}\AgdaSymbol{\}}\AgdaSpace{}%
\AgdaSymbol{→}\AgdaSpace{}%
\AgdaSymbol{((}\AgdaBound{c₂}\AgdaSpace{}%
\AgdaOperator{\AgdaInductiveConstructor{⊕}}\AgdaSpace{}%
\AgdaBound{c₁}\AgdaSymbol{)}\AgdaSpace{}%
\AgdaOperator{\AgdaInductiveConstructor{◎}}\AgdaSpace{}%
\AgdaInductiveConstructor{swap₊}\AgdaSymbol{)}\AgdaSpace{}%
\AgdaOperator{\AgdaDatatype{⟷₂}}\AgdaSpace{}%
\AgdaSymbol{(}\AgdaInductiveConstructor{swap₊}\AgdaSpace{}%
\AgdaOperator{\AgdaInductiveConstructor{◎}}\AgdaSpace{}%
\AgdaSymbol{(}\AgdaBound{c₁}\AgdaSpace{}%
\AgdaOperator{\AgdaInductiveConstructor{⊕}}\AgdaSpace{}%
\AgdaBound{c₂}\AgdaSymbol{))}\<%
\\
\\[\AgdaEmptyExtraSkip]%
\>[2]\AgdaInductiveConstructor{id⟷₂}%
\>[11]\AgdaSymbol{:}\AgdaSpace{}%
\AgdaSymbol{\{}\AgdaBound{c}\AgdaSpace{}%
\AgdaSymbol{:}\AgdaSpace{}%
\AgdaGeneralizable{A}\AgdaSpace{}%
\AgdaOperator{\AgdaDatatype{⟷₁}}\AgdaSpace{}%
\AgdaGeneralizable{B}\AgdaSymbol{\}}\AgdaSpace{}%
\AgdaSymbol{→}\AgdaSpace{}%
\AgdaBound{c}\AgdaSpace{}%
\AgdaOperator{\AgdaDatatype{⟷₂}}\AgdaSpace{}%
\AgdaBound{c}\<%
\\
\>[2]\AgdaOperator{\AgdaInductiveConstructor{\AgdaUnderscore{}■\AgdaUnderscore{}}}%
\>[11]\AgdaSymbol{:}\AgdaSpace{}%
\AgdaSymbol{\{}\AgdaBound{c₁}\AgdaSpace{}%
\AgdaBound{c₂}\AgdaSpace{}%
\AgdaBound{c₃}\AgdaSpace{}%
\AgdaSymbol{:}\AgdaSpace{}%
\AgdaGeneralizable{A}\AgdaSpace{}%
\AgdaOperator{\AgdaDatatype{⟷₁}}\AgdaSpace{}%
\AgdaGeneralizable{B}\AgdaSymbol{\}}\AgdaSpace{}%
\AgdaSymbol{→}\AgdaSpace{}%
\AgdaSymbol{(}\AgdaBound{c₁}\AgdaSpace{}%
\AgdaOperator{\AgdaDatatype{⟷₂}}\AgdaSpace{}%
\AgdaBound{c₂}\AgdaSymbol{)}\AgdaSpace{}%
\AgdaSymbol{→}\AgdaSpace{}%
\AgdaSymbol{(}\AgdaBound{c₂}\AgdaSpace{}%
\AgdaOperator{\AgdaDatatype{⟷₂}}\AgdaSpace{}%
\AgdaBound{c₃}\AgdaSymbol{)}\AgdaSpace{}%
\AgdaSymbol{→}\AgdaSpace{}%
\AgdaSymbol{(}\AgdaBound{c₁}\AgdaSpace{}%
\AgdaOperator{\AgdaDatatype{⟷₂}}\AgdaSpace{}%
\AgdaBound{c₃}\AgdaSymbol{)}\<%
\\
\>[2]\AgdaOperator{\AgdaInductiveConstructor{\AgdaUnderscore{}⊡\AgdaUnderscore{}}}%
\>[11]\AgdaSymbol{:}\AgdaSpace{}%
\AgdaSymbol{\{}\AgdaBound{c₁}\AgdaSpace{}%
\AgdaSymbol{:}\AgdaSpace{}%
\AgdaGeneralizable{A}\AgdaSpace{}%
\AgdaOperator{\AgdaDatatype{⟷₁}}\AgdaSpace{}%
\AgdaGeneralizable{B}\AgdaSymbol{\}}\AgdaSpace{}%
\AgdaSymbol{\{}\AgdaBound{c₂}\AgdaSpace{}%
\AgdaSymbol{:}\AgdaSpace{}%
\AgdaGeneralizable{B}\AgdaSpace{}%
\AgdaOperator{\AgdaDatatype{⟷₁}}\AgdaSpace{}%
\AgdaGeneralizable{C}\AgdaSymbol{\}}\AgdaSpace{}%
\AgdaSymbol{\{}\AgdaBound{c₃}\AgdaSpace{}%
\AgdaSymbol{:}\AgdaSpace{}%
\AgdaGeneralizable{A}\AgdaSpace{}%
\AgdaOperator{\AgdaDatatype{⟷₁}}\AgdaSpace{}%
\AgdaGeneralizable{B}\AgdaSymbol{\}}\AgdaSpace{}%
\AgdaSymbol{\{}\AgdaBound{c₄}\AgdaSpace{}%
\AgdaSymbol{:}\AgdaSpace{}%
\AgdaGeneralizable{B}\AgdaSpace{}%
\AgdaOperator{\AgdaDatatype{⟷₁}}\AgdaSpace{}%
\AgdaGeneralizable{C}\AgdaSymbol{\}}\AgdaSpace{}%
\AgdaSymbol{→}\AgdaSpace{}%
\AgdaSymbol{(}\AgdaBound{c₁}\AgdaSpace{}%
\AgdaOperator{\AgdaDatatype{⟷₂}}\AgdaSpace{}%
\AgdaBound{c₃}\AgdaSymbol{)}\AgdaSpace{}%
\AgdaSymbol{→}\AgdaSpace{}%
\AgdaSymbol{(}\AgdaBound{c₂}\AgdaSpace{}%
\AgdaOperator{\AgdaDatatype{⟷₂}}\AgdaSpace{}%
\AgdaBound{c₄}\AgdaSymbol{)}\AgdaSpace{}%
\AgdaSymbol{→}\AgdaSpace{}%
\AgdaSymbol{(}\AgdaBound{c₁}\AgdaSpace{}%
\AgdaOperator{\AgdaInductiveConstructor{◎}}\AgdaSpace{}%
\AgdaBound{c₂}\AgdaSymbol{)}\AgdaSpace{}%
\AgdaOperator{\AgdaDatatype{⟷₂}}\AgdaSpace{}%
\AgdaSymbol{(}\AgdaBound{c₃}\AgdaSpace{}%
\AgdaOperator{\AgdaInductiveConstructor{◎}}\AgdaSpace{}%
\AgdaBound{c₄}\AgdaSymbol{)}\<%
\\
\>[2]\AgdaInductiveConstructor{resp⊕⟷₂}%
\>[11]\AgdaSymbol{:}\AgdaSpace{}%
\AgdaSymbol{\{}\AgdaBound{c₁}\AgdaSpace{}%
\AgdaSymbol{:}\AgdaSpace{}%
\AgdaGeneralizable{A}\AgdaSpace{}%
\AgdaOperator{\AgdaDatatype{⟷₁}}\AgdaSpace{}%
\AgdaGeneralizable{B}\AgdaSymbol{\}}\AgdaSpace{}%
\AgdaSymbol{\{}\AgdaBound{c₂}\AgdaSpace{}%
\AgdaSymbol{:}\AgdaSpace{}%
\AgdaGeneralizable{C}\AgdaSpace{}%
\AgdaOperator{\AgdaDatatype{⟷₁}}\AgdaSpace{}%
\AgdaGeneralizable{D}\AgdaSymbol{\}}\AgdaSpace{}%
\AgdaSymbol{\{}\AgdaBound{c₃}\AgdaSpace{}%
\AgdaSymbol{:}\AgdaSpace{}%
\AgdaGeneralizable{A}\AgdaSpace{}%
\AgdaOperator{\AgdaDatatype{⟷₁}}\AgdaSpace{}%
\AgdaGeneralizable{B}\AgdaSymbol{\}}\AgdaSpace{}%
\AgdaSymbol{\{}\AgdaBound{c₄}\AgdaSpace{}%
\AgdaSymbol{:}\AgdaSpace{}%
\AgdaGeneralizable{C}\AgdaSpace{}%
\AgdaOperator{\AgdaDatatype{⟷₁}}\AgdaSpace{}%
\AgdaGeneralizable{D}\AgdaSymbol{\}}\AgdaSpace{}%
\AgdaSymbol{→}\AgdaSpace{}%
\AgdaSymbol{(}\AgdaBound{c₁}\AgdaSpace{}%
\AgdaOperator{\AgdaDatatype{⟷₂}}\AgdaSpace{}%
\AgdaBound{c₃}\AgdaSymbol{)}\AgdaSpace{}%
\AgdaSymbol{→}\AgdaSpace{}%
\AgdaSymbol{(}\AgdaBound{c₂}\AgdaSpace{}%
\AgdaOperator{\AgdaDatatype{⟷₂}}\AgdaSpace{}%
\AgdaBound{c₄}\AgdaSymbol{)}\AgdaSpace{}%
\AgdaSymbol{→}\AgdaSpace{}%
\AgdaSymbol{(}\AgdaBound{c₁}\AgdaSpace{}%
\AgdaOperator{\AgdaInductiveConstructor{⊕}}\AgdaSpace{}%
\AgdaBound{c₂}\AgdaSymbol{)}\AgdaSpace{}%
\AgdaOperator{\AgdaDatatype{⟷₂}}\AgdaSpace{}%
\AgdaSymbol{(}\AgdaBound{c₃}\AgdaSpace{}%
\AgdaOperator{\AgdaInductiveConstructor{⊕}}\AgdaSpace{}%
\AgdaBound{c₄}\AgdaSymbol{)}\<%
\\
\\[\AgdaEmptyExtraSkip]%
\>[2]\AgdaInductiveConstructor{id⟷₁⊕id⟷₁⟷₂}%
\>[15]\AgdaSymbol{:}\AgdaSpace{}%
\AgdaSymbol{(}\AgdaInductiveConstructor{id⟷₁}\AgdaSpace{}%
\AgdaSymbol{\{}\AgdaGeneralizable{A}\AgdaSymbol{\}}\AgdaSpace{}%
\AgdaOperator{\AgdaInductiveConstructor{⊕}}\AgdaSpace{}%
\AgdaInductiveConstructor{id⟷₁}\AgdaSpace{}%
\AgdaSymbol{\{}\AgdaGeneralizable{B}\AgdaSymbol{\})}\AgdaSpace{}%
\AgdaOperator{\AgdaDatatype{⟷₂}}\AgdaSpace{}%
\AgdaInductiveConstructor{id⟷₁}\<%
\\
\>[2]\AgdaInductiveConstructor{split⊕-id⟷₁}%
\>[15]\AgdaSymbol{:}\AgdaSpace{}%
\AgdaSymbol{(}\AgdaInductiveConstructor{id⟷₁}\AgdaSpace{}%
\AgdaSymbol{\{}\AgdaGeneralizable{A}\AgdaSpace{}%
\AgdaOperator{\AgdaInductiveConstructor{+}}\AgdaSpace{}%
\AgdaGeneralizable{B}\AgdaSymbol{\})}\AgdaSpace{}%
\AgdaOperator{\AgdaDatatype{⟷₂}}\AgdaSpace{}%
\AgdaSymbol{(}\AgdaInductiveConstructor{id⟷₁}\AgdaSpace{}%
\AgdaOperator{\AgdaInductiveConstructor{⊕}}\AgdaSpace{}%
\AgdaInductiveConstructor{id⟷₁}\AgdaSymbol{)}\<%
\\
\>[2]\AgdaInductiveConstructor{hom⊕◎⟷₂}%
\>[15]\AgdaSymbol{:}\AgdaSpace{}%
\AgdaSymbol{\{}\AgdaBound{c₁}\AgdaSpace{}%
\AgdaSymbol{:}\AgdaSpace{}%
\AgdaGeneralizable{E}\AgdaSpace{}%
\AgdaOperator{\AgdaDatatype{⟷₁}}\AgdaSpace{}%
\AgdaGeneralizable{A}\AgdaSymbol{\}}\AgdaSpace{}%
\AgdaSymbol{\{}\AgdaBound{c₂}\AgdaSpace{}%
\AgdaSymbol{:}\AgdaSpace{}%
\AgdaGeneralizable{F}\AgdaSpace{}%
\AgdaOperator{\AgdaDatatype{⟷₁}}\AgdaSpace{}%
\AgdaGeneralizable{B}\AgdaSymbol{\}}\AgdaSpace{}%
\AgdaSymbol{\{}\AgdaBound{c₃}\AgdaSpace{}%
\AgdaSymbol{:}\AgdaSpace{}%
\AgdaGeneralizable{A}\AgdaSpace{}%
\AgdaOperator{\AgdaDatatype{⟷₁}}\AgdaSpace{}%
\AgdaGeneralizable{C}\AgdaSymbol{\}}\AgdaSpace{}%
\AgdaSymbol{\{}\AgdaBound{c₄}\AgdaSpace{}%
\AgdaSymbol{:}\AgdaSpace{}%
\AgdaGeneralizable{B}\AgdaSpace{}%
\AgdaOperator{\AgdaDatatype{⟷₁}}\AgdaSpace{}%
\AgdaGeneralizable{D}\AgdaSymbol{\}}\AgdaSpace{}%
\AgdaSymbol{→}\AgdaSpace{}%
\AgdaSymbol{((}\AgdaBound{c₁}\AgdaSpace{}%
\AgdaOperator{\AgdaInductiveConstructor{◎}}\AgdaSpace{}%
\AgdaBound{c₃}\AgdaSymbol{)}\AgdaSpace{}%
\AgdaOperator{\AgdaInductiveConstructor{⊕}}\AgdaSpace{}%
\AgdaSymbol{(}\AgdaBound{c₂}\AgdaSpace{}%
\AgdaOperator{\AgdaInductiveConstructor{◎}}\AgdaSpace{}%
\AgdaBound{c₄}\AgdaSymbol{))}\AgdaSpace{}%
\AgdaOperator{\AgdaDatatype{⟷₂}}\AgdaSpace{}%
\AgdaSymbol{((}\AgdaBound{c₁}\AgdaSpace{}%
\AgdaOperator{\AgdaInductiveConstructor{⊕}}\AgdaSpace{}%
\AgdaBound{c₂}\AgdaSymbol{)}\AgdaSpace{}%
\AgdaOperator{\AgdaInductiveConstructor{◎}}\AgdaSpace{}%
\AgdaSymbol{(}\AgdaBound{c₃}\AgdaSpace{}%
\AgdaOperator{\AgdaInductiveConstructor{⊕}}\AgdaSpace{}%
\AgdaBound{c₄}\AgdaSymbol{))}\<%
\\
\>[2]\AgdaInductiveConstructor{hom◎⊕⟷₂}%
\>[15]\AgdaSymbol{:}\AgdaSpace{}%
\AgdaSymbol{\{}\AgdaBound{c₁}\AgdaSpace{}%
\AgdaSymbol{:}\AgdaSpace{}%
\AgdaGeneralizable{E}\AgdaSpace{}%
\AgdaOperator{\AgdaDatatype{⟷₁}}\AgdaSpace{}%
\AgdaGeneralizable{A}\AgdaSymbol{\}}\AgdaSpace{}%
\AgdaSymbol{\{}\AgdaBound{c₂}\AgdaSpace{}%
\AgdaSymbol{:}\AgdaSpace{}%
\AgdaGeneralizable{F}\AgdaSpace{}%
\AgdaOperator{\AgdaDatatype{⟷₁}}\AgdaSpace{}%
\AgdaGeneralizable{B}\AgdaSymbol{\}}\AgdaSpace{}%
\AgdaSymbol{\{}\AgdaBound{c₃}\AgdaSpace{}%
\AgdaSymbol{:}\AgdaSpace{}%
\AgdaGeneralizable{A}\AgdaSpace{}%
\AgdaOperator{\AgdaDatatype{⟷₁}}\AgdaSpace{}%
\AgdaGeneralizable{C}\AgdaSymbol{\}}\AgdaSpace{}%
\AgdaSymbol{\{}\AgdaBound{c₄}\AgdaSpace{}%
\AgdaSymbol{:}\AgdaSpace{}%
\AgdaGeneralizable{B}\AgdaSpace{}%
\AgdaOperator{\AgdaDatatype{⟷₁}}\AgdaSpace{}%
\AgdaGeneralizable{D}\AgdaSymbol{\}}\AgdaSpace{}%
\AgdaSymbol{→}\AgdaSpace{}%
\AgdaSymbol{((}\AgdaBound{c₁}\AgdaSpace{}%
\AgdaOperator{\AgdaInductiveConstructor{⊕}}\AgdaSpace{}%
\AgdaBound{c₂}\AgdaSymbol{)}\AgdaSpace{}%
\AgdaOperator{\AgdaInductiveConstructor{◎}}\AgdaSpace{}%
\AgdaSymbol{(}\AgdaBound{c₃}\AgdaSpace{}%
\AgdaOperator{\AgdaInductiveConstructor{⊕}}\AgdaSpace{}%
\AgdaBound{c₄}\AgdaSymbol{))}\AgdaSpace{}%
\AgdaOperator{\AgdaDatatype{⟷₂}}\AgdaSpace{}%
\AgdaSymbol{((}\AgdaBound{c₁}\AgdaSpace{}%
\AgdaOperator{\AgdaInductiveConstructor{◎}}\AgdaSpace{}%
\AgdaBound{c₃}\AgdaSymbol{)}\AgdaSpace{}%
\AgdaOperator{\AgdaInductiveConstructor{⊕}}\AgdaSpace{}%
\AgdaSymbol{(}\AgdaBound{c₂}\AgdaSpace{}%
\AgdaOperator{\AgdaInductiveConstructor{◎}}\AgdaSpace{}%
\AgdaBound{c₄}\AgdaSymbol{))}\<%
\\
\\[\AgdaEmptyExtraSkip]%
\>[2]\AgdaInductiveConstructor{triangle₊l}%
\>[14]\AgdaSymbol{:}\AgdaSpace{}%
\AgdaSymbol{(}\AgdaFunction{unite₊r}\AgdaSpace{}%
\AgdaSymbol{\{}\AgdaGeneralizable{A}\AgdaSymbol{\}}\AgdaSpace{}%
\AgdaOperator{\AgdaInductiveConstructor{⊕}}\AgdaSpace{}%
\AgdaInductiveConstructor{id⟷₁}\AgdaSpace{}%
\AgdaSymbol{\{}\AgdaGeneralizable{B}\AgdaSymbol{\})}\AgdaSpace{}%
\AgdaOperator{\AgdaDatatype{⟷₂}}\AgdaSpace{}%
\AgdaInductiveConstructor{assocr₊}\AgdaSpace{}%
\AgdaOperator{\AgdaInductiveConstructor{◎}}\AgdaSpace{}%
\AgdaSymbol{(}\AgdaInductiveConstructor{id⟷₁}\AgdaSpace{}%
\AgdaOperator{\AgdaInductiveConstructor{⊕}}\AgdaSpace{}%
\AgdaInductiveConstructor{unite₊l}\AgdaSymbol{)}\<%
\\
\>[2]\AgdaInductiveConstructor{triangle₊r}%
\>[14]\AgdaSymbol{:}\AgdaSpace{}%
\AgdaInductiveConstructor{assocr₊}\AgdaSpace{}%
\AgdaOperator{\AgdaInductiveConstructor{◎}}\AgdaSpace{}%
\AgdaSymbol{(}\AgdaInductiveConstructor{id⟷₁}\AgdaSpace{}%
\AgdaSymbol{\{}\AgdaGeneralizable{A}\AgdaSymbol{\}}\AgdaSpace{}%
\AgdaOperator{\AgdaInductiveConstructor{⊕}}\AgdaSpace{}%
\AgdaInductiveConstructor{unite₊l}\AgdaSpace{}%
\AgdaSymbol{\{}\AgdaGeneralizable{B}\AgdaSymbol{\})}\AgdaSpace{}%
\AgdaOperator{\AgdaDatatype{⟷₂}}\AgdaSpace{}%
\AgdaFunction{unite₊r}\AgdaSpace{}%
\AgdaOperator{\AgdaInductiveConstructor{⊕}}\AgdaSpace{}%
\AgdaInductiveConstructor{id⟷₁}\<%
\\
\>[2]\AgdaInductiveConstructor{pentagon₊l}%
\>[14]\AgdaSymbol{:}\AgdaSpace{}%
\AgdaInductiveConstructor{assocr₊}\AgdaSpace{}%
\AgdaOperator{\AgdaInductiveConstructor{◎}}\AgdaSpace{}%
\AgdaSymbol{(}\AgdaInductiveConstructor{assocr₊}\AgdaSpace{}%
\AgdaSymbol{\{}\AgdaGeneralizable{A}\AgdaSymbol{\}}\AgdaSpace{}%
\AgdaSymbol{\{}\AgdaGeneralizable{B}\AgdaSymbol{\}}\AgdaSpace{}%
\AgdaSymbol{\{}\AgdaGeneralizable{C}\AgdaSpace{}%
\AgdaOperator{\AgdaInductiveConstructor{+}}\AgdaSpace{}%
\AgdaGeneralizable{D}\AgdaSymbol{\})}\AgdaSpace{}%
\AgdaOperator{\AgdaDatatype{⟷₂}}\AgdaSpace{}%
\AgdaSymbol{((}\AgdaInductiveConstructor{assocr₊}\AgdaSpace{}%
\AgdaOperator{\AgdaInductiveConstructor{⊕}}\AgdaSpace{}%
\AgdaInductiveConstructor{id⟷₁}\AgdaSymbol{)}\AgdaSpace{}%
\AgdaOperator{\AgdaInductiveConstructor{◎}}\AgdaSpace{}%
\AgdaInductiveConstructor{assocr₊}\AgdaSymbol{)}\AgdaSpace{}%
\AgdaOperator{\AgdaInductiveConstructor{◎}}\AgdaSpace{}%
\AgdaSymbol{(}\AgdaInductiveConstructor{id⟷₁}\AgdaSpace{}%
\AgdaOperator{\AgdaInductiveConstructor{⊕}}\AgdaSpace{}%
\AgdaInductiveConstructor{assocr₊}\AgdaSymbol{)}\<%
\\
\>[2]\AgdaInductiveConstructor{pentagon₊r}%
\>[14]\AgdaSymbol{:}\AgdaSpace{}%
\AgdaSymbol{((}\AgdaInductiveConstructor{assocr₊}\AgdaSpace{}%
\AgdaSymbol{\{}\AgdaGeneralizable{A}\AgdaSymbol{\}}\AgdaSpace{}%
\AgdaSymbol{\{}\AgdaGeneralizable{B}\AgdaSymbol{\}}\AgdaSpace{}%
\AgdaSymbol{\{}\AgdaGeneralizable{C}\AgdaSymbol{\}}\AgdaSpace{}%
\AgdaOperator{\AgdaInductiveConstructor{⊕}}\AgdaSpace{}%
\AgdaInductiveConstructor{id⟷₁}\AgdaSpace{}%
\AgdaSymbol{\{}\AgdaGeneralizable{D}\AgdaSymbol{\})}\AgdaSpace{}%
\AgdaOperator{\AgdaInductiveConstructor{◎}}\AgdaSpace{}%
\AgdaInductiveConstructor{assocr₊}\AgdaSymbol{)}\AgdaSpace{}%
\AgdaOperator{\AgdaInductiveConstructor{◎}}\AgdaSpace{}%
\AgdaSymbol{(}\AgdaInductiveConstructor{id⟷₁}\AgdaSpace{}%
\AgdaOperator{\AgdaInductiveConstructor{⊕}}\AgdaSpace{}%
\AgdaInductiveConstructor{assocr₊}\AgdaSymbol{)}\AgdaSpace{}%
\AgdaOperator{\AgdaDatatype{⟷₂}}\AgdaSpace{}%
\AgdaInductiveConstructor{assocr₊}\AgdaSpace{}%
\AgdaOperator{\AgdaInductiveConstructor{◎}}\AgdaSpace{}%
\AgdaInductiveConstructor{assocr₊}\<%
\\
\\[\AgdaEmptyExtraSkip]%
\>[2]\AgdaInductiveConstructor{unite₊l-coh-l}%
\>[17]\AgdaSymbol{:}\AgdaSpace{}%
\AgdaInductiveConstructor{unite₊l}\AgdaSpace{}%
\AgdaSymbol{\{}\AgdaGeneralizable{A}\AgdaSymbol{\}}\AgdaSpace{}%
\AgdaOperator{\AgdaDatatype{⟷₂}}\AgdaSpace{}%
\AgdaInductiveConstructor{swap₊}\AgdaSpace{}%
\AgdaOperator{\AgdaInductiveConstructor{◎}}\AgdaSpace{}%
\AgdaFunction{unite₊r}\<%
\\
\>[2]\AgdaInductiveConstructor{unite₊l-coh-r}%
\>[17]\AgdaSymbol{:}\AgdaSpace{}%
\AgdaInductiveConstructor{swap₊}\AgdaSpace{}%
\AgdaOperator{\AgdaInductiveConstructor{◎}}\AgdaSpace{}%
\AgdaFunction{unite₊r}\AgdaSpace{}%
\AgdaOperator{\AgdaDatatype{⟷₂}}\AgdaSpace{}%
\AgdaInductiveConstructor{unite₊l}\AgdaSpace{}%
\AgdaSymbol{\{}\AgdaGeneralizable{A}\AgdaSymbol{\}}\<%
\\
\>[2]\AgdaInductiveConstructor{hexagonr₊l}%
\>[17]\AgdaSymbol{:}\AgdaSpace{}%
\AgdaSymbol{(}\AgdaInductiveConstructor{assocr₊}\AgdaSpace{}%
\AgdaOperator{\AgdaInductiveConstructor{◎}}\AgdaSpace{}%
\AgdaInductiveConstructor{swap₊}\AgdaSymbol{)}\AgdaSpace{}%
\AgdaOperator{\AgdaInductiveConstructor{◎}}\AgdaSpace{}%
\AgdaInductiveConstructor{assocr₊}\AgdaSpace{}%
\AgdaSymbol{\{}\AgdaGeneralizable{A}\AgdaSymbol{\}}\AgdaSpace{}%
\AgdaSymbol{\{}\AgdaGeneralizable{B}\AgdaSymbol{\}}\AgdaSpace{}%
\AgdaSymbol{\{}\AgdaGeneralizable{C}\AgdaSymbol{\}}\AgdaSpace{}%
\AgdaOperator{\AgdaDatatype{⟷₂}}\AgdaSpace{}%
\AgdaSymbol{((}\AgdaInductiveConstructor{swap₊}\AgdaSpace{}%
\AgdaOperator{\AgdaInductiveConstructor{⊕}}\AgdaSpace{}%
\AgdaInductiveConstructor{id⟷₁}\AgdaSymbol{)}\AgdaSpace{}%
\AgdaOperator{\AgdaInductiveConstructor{◎}}\AgdaSpace{}%
\AgdaInductiveConstructor{assocr₊}\AgdaSymbol{)}\AgdaSpace{}%
\AgdaOperator{\AgdaInductiveConstructor{◎}}\AgdaSpace{}%
\AgdaSymbol{(}\AgdaInductiveConstructor{id⟷₁}\AgdaSpace{}%
\AgdaOperator{\AgdaInductiveConstructor{⊕}}\AgdaSpace{}%
\AgdaInductiveConstructor{swap₊}\AgdaSymbol{)}\<%
\\
\>[2]\AgdaInductiveConstructor{hexagonr₊r}%
\>[17]\AgdaSymbol{:}\AgdaSpace{}%
\AgdaSymbol{((}\AgdaInductiveConstructor{swap₊}\AgdaSpace{}%
\AgdaOperator{\AgdaInductiveConstructor{⊕}}\AgdaSpace{}%
\AgdaInductiveConstructor{id⟷₁}\AgdaSymbol{)}\AgdaSpace{}%
\AgdaOperator{\AgdaInductiveConstructor{◎}}\AgdaSpace{}%
\AgdaInductiveConstructor{assocr₊}\AgdaSymbol{)}\AgdaSpace{}%
\AgdaOperator{\AgdaInductiveConstructor{◎}}\AgdaSpace{}%
\AgdaSymbol{(}\AgdaInductiveConstructor{id⟷₁}\AgdaSpace{}%
\AgdaOperator{\AgdaInductiveConstructor{⊕}}\AgdaSpace{}%
\AgdaInductiveConstructor{swap₊}\AgdaSymbol{)}\AgdaSpace{}%
\AgdaOperator{\AgdaDatatype{⟷₂}}\AgdaSpace{}%
\AgdaSymbol{(}\AgdaInductiveConstructor{assocr₊}\AgdaSpace{}%
\AgdaOperator{\AgdaInductiveConstructor{◎}}\AgdaSpace{}%
\AgdaInductiveConstructor{swap₊}\AgdaSymbol{)}\AgdaSpace{}%
\AgdaOperator{\AgdaInductiveConstructor{◎}}\AgdaSpace{}%
\AgdaInductiveConstructor{assocr₊}\AgdaSpace{}%
\AgdaSymbol{\{}\AgdaGeneralizable{A}\AgdaSymbol{\}}\AgdaSpace{}%
\AgdaSymbol{\{}\AgdaGeneralizable{B}\AgdaSymbol{\}}\AgdaSpace{}%
\AgdaSymbol{\{}\AgdaGeneralizable{C}\AgdaSymbol{\}}\<%
\\
\>[2]\AgdaInductiveConstructor{hexagonl₊l}%
\>[17]\AgdaSymbol{:}\AgdaSpace{}%
\AgdaSymbol{(}\AgdaInductiveConstructor{assocl₊}\AgdaSpace{}%
\AgdaOperator{\AgdaInductiveConstructor{◎}}\AgdaSpace{}%
\AgdaInductiveConstructor{swap₊}\AgdaSymbol{)}\AgdaSpace{}%
\AgdaOperator{\AgdaInductiveConstructor{◎}}\AgdaSpace{}%
\AgdaInductiveConstructor{assocl₊}\AgdaSpace{}%
\AgdaSymbol{\{}\AgdaGeneralizable{A}\AgdaSymbol{\}}\AgdaSpace{}%
\AgdaSymbol{\{}\AgdaGeneralizable{B}\AgdaSymbol{\}}\AgdaSpace{}%
\AgdaSymbol{\{}\AgdaGeneralizable{C}\AgdaSymbol{\}}\AgdaSpace{}%
\AgdaOperator{\AgdaDatatype{⟷₂}}\AgdaSpace{}%
\AgdaSymbol{((}\AgdaInductiveConstructor{id⟷₁}\AgdaSpace{}%
\AgdaOperator{\AgdaInductiveConstructor{⊕}}\AgdaSpace{}%
\AgdaInductiveConstructor{swap₊}\AgdaSymbol{)}\AgdaSpace{}%
\AgdaOperator{\AgdaInductiveConstructor{◎}}\AgdaSpace{}%
\AgdaInductiveConstructor{assocl₊}\AgdaSymbol{)}\AgdaSpace{}%
\AgdaOperator{\AgdaInductiveConstructor{◎}}\AgdaSpace{}%
\AgdaSymbol{(}\AgdaInductiveConstructor{swap₊}\AgdaSpace{}%
\AgdaOperator{\AgdaInductiveConstructor{⊕}}\AgdaSpace{}%
\AgdaInductiveConstructor{id⟷₁}\AgdaSymbol{)}\<%
\\
\>[2]\AgdaInductiveConstructor{hexagonl₊r}%
\>[17]\AgdaSymbol{:}\AgdaSpace{}%
\AgdaSymbol{((}\AgdaInductiveConstructor{id⟷₁}\AgdaSpace{}%
\AgdaOperator{\AgdaInductiveConstructor{⊕}}\AgdaSpace{}%
\AgdaInductiveConstructor{swap₊}\AgdaSymbol{)}\AgdaSpace{}%
\AgdaOperator{\AgdaInductiveConstructor{◎}}\AgdaSpace{}%
\AgdaInductiveConstructor{assocl₊}\AgdaSymbol{)}\AgdaSpace{}%
\AgdaOperator{\AgdaInductiveConstructor{◎}}\AgdaSpace{}%
\AgdaSymbol{(}\AgdaInductiveConstructor{swap₊}\AgdaSpace{}%
\AgdaOperator{\AgdaInductiveConstructor{⊕}}\AgdaSpace{}%
\AgdaInductiveConstructor{id⟷₁}\AgdaSymbol{)}\AgdaSpace{}%
\AgdaOperator{\AgdaDatatype{⟷₂}}\AgdaSpace{}%
\AgdaSymbol{(}\AgdaInductiveConstructor{assocl₊}\AgdaSpace{}%
\AgdaOperator{\AgdaInductiveConstructor{◎}}\AgdaSpace{}%
\AgdaInductiveConstructor{swap₊}\AgdaSymbol{)}\AgdaSpace{}%
\AgdaOperator{\AgdaInductiveConstructor{◎}}\AgdaSpace{}%
\AgdaInductiveConstructor{assocl₊}\AgdaSpace{}%
\AgdaSymbol{\{}\AgdaGeneralizable{A}\AgdaSymbol{\}}\AgdaSpace{}%
\AgdaSymbol{\{}\AgdaGeneralizable{B}\AgdaSymbol{\}}\AgdaSpace{}%
\AgdaSymbol{\{}\AgdaGeneralizable{C}\AgdaSymbol{\}}\<%
\end{code}}

\begin{code}[hide]%
\>[0]\AgdaKeyword{postulate}\<%
\end{code}

\AtBeginDocument{%
  \providecommand\BibTeX{{%
    \normalfont B\kern-0.5em{\scshape i\kern-0.25em b}\kern-0.8em\TeX}}}

\acmJournal{PACMPL}
\acmVolume{1}
\acmNumber{POPL} 
\acmArticle{1}
\acmYear{2022}
\acmMonth{1}
\acmDOI{} 
\startPage{1}

\setcopyright{rightsretained}

\begin{document}


\title{Symmetries in Reversible Programming}
\subtitle{From Symmetric Rig Groupoids to Reversible Programming Languages}






\author{Vikraman Choudhury}
\orcid{0000-0003-2030-8056}
\affiliation{
  \department{Department of Computer Science}
  \institution{Indiana University}
  \city{Bloomington}
  \postcode{47408}
  \country{USA}
}
\email{vikraman@indiana.edu}
\affiliation{
  \department{Department of Computer Science and Technology}
  \institution{University of Cambridge}
  \city{Cambridge}
  \postcode{CB3 0FD}
  \country{UK}
}
\email{vc378@cam.ac.uk}
\author{Jacek Karwowski}
\affiliation{
  \institution{University of Warsaw}
  \city{Warsaw}
  \postcode{00-927}
  \country{Poland}
}
\email{jac.karwowski@gmail.com}
\author{Amr Sabry}
\orcid{0000-0002-1025-7331}
\affiliation{
  \department{Department of Computer Science}
  \institution{Indiana University}
  \city{Bloomington}
  \postcode{47408}
  \country{USA}
}





\begin{abstract}
The $\PiLang$ family of reversible programming languages for boolean circuits is presented as a syntax of combinators
witnessing type isomorphisms of algebraic datatypes. In this paper, we give a denotational semantics for this language,
using the language of weak groupoids \`{a} la Homotopy Type Theory, and show how to derive an equational theory for it,
presented by 2-combinators witnessing equivalences of reversible circuits.

We establish a correspondence between the syntactic groupoid of the language and a formally presented univalent
subuniverse of finite types. The correspondence relates 1-combinators to 1-paths, and 2-combinators to 2-paths in the
universe, which is shown to be sound and complete for both levels, establishing full abstraction and adequacy. We extend
the already established Curry-Howard correspondence for $\PiLang$ to a Curry-Howard-Lambek correspondence between
Reversible Logic, Reversible Programming Languages, and Symmetric Rig Groupoids, by showing that the syntax of $\PiLang$
is presented by the free symmetric rig groupoid, given by finite sets and permutations. Our proof uses techniques from
the theory of group presentations and rewriting systems to solve the word problem for symmetric groups.

Using the formalisation of our results, we show how to perform normalisation-by-evaluation, verification, and synthesis
of reversible logic gates, motivated by examples from quantum computing.
\end{abstract}



\maketitle


\renewcommand{\appendixsectionformat}[2]{
  {Supplementary material for Section~#1}
}

\section{Introduction}
\label{sec:introduction}

Consider two programs that implement the same permutation using different sequences of type isomorphisms:
\begin{align*}
      A + (B + C) \isoone
      (A + B) + C & \isoone
      C + (A + B) \isoone
      C + (B + A)
      \\
      A + (B + C) \isoone
      A + (C + B) \isoone
      (A + C) + B & \isoone
      (C + A) + B \isoone
      C + (A + B) \isoone
      C + (B + A)
\end{align*}

\noindent These permutations are written as sequences of primitive operations: associativity, symmetry, and
composition. Can we find necessary and sufficient equations to identify \emph{all} such equivalent sequences of type
isomorphisms?

Such sequences of type isomorphisms are pervasive in reversible boolean circuits, which are at the core of quantum
computing. Typically, they might be formalised as permutations $\{0,1\}^k \to \{0,1\}^k$ on bit strings of some length
$k$~\cite{aaronson_et_al:LIPIcs:2017:8173,1201583}. But from the perspective of programming languages, permutations and
reversible circuits can be more conveniently expressed as isomorphisms over algebraic datatypes as proposed in the
$\PiLang$ family of reversible languages~\cite*{jamesInformationEffects2012,theseus} whose syntax is inspired by the
sound and complete axiomatisation of type isomorphisms~\cite{fioreRemarksIsomorphismsTyped2002}, with respect to the
bicartesian structure, that is, coproducts and products. Since $\PiLang$ programs correspond to type isomorphisms of
finite types, our observation is that the syntax of $\PiLang$ is a presentation of the free symmetric rig groupoid (on
zero generators).

It is folklore that the groupoid of finite sets and permutations is the free symmetric rig groupoid on zero
generators~\cite{laplaza72,kelly74,baez2000finite}. Our main result formally establishes this correspondence by giving an
equational theory for $\PiLang$ that exactly includes all the necessary equations to decide equivalence of $\PiLang$
programs. As conjectured by \citet{caretteComputingSemiringsWeak2016}, these equations correspond to the \emph{coherence
  conditions of symmetric rig groupoids}, answering the conjecture in the positive.

\paragraph{Equivalence by Example} Without going into proofs, let us try to answer the question for the original
pair of examples.

\begin{figure}
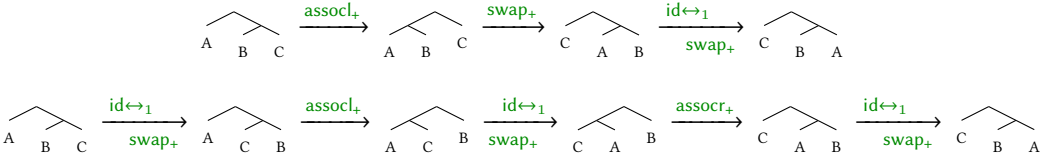

      \[
            \Tree [ {\tiny A} [ {\tiny B} {\tiny C} ] ] ~\xrightarrow{\assoclp}~
            \Tree [ [ {\tiny A} {\tiny B} ] {\tiny C} ] ~\xrightarrow{\swapp}~
            \Tree [ {\tiny C} [ {\tiny A} {\tiny B} ] ] ~\xrightarrow[\phantom{xx}\swapp]{\idc\phantom{xx}}~
            \Tree [ {\tiny C} [ {\tiny B} {\tiny A} ] ] ~
      \]

      \[
            \Tree [ {\tiny A} [ {\tiny B} {\tiny C} ] ] ~\xrightarrow[\phantom{xx}\swapp]{\idc\phantom{xx}}~
            \Tree [ {\tiny A} [ {\tiny C} {\tiny B} ] ] ~\xrightarrow{\assoclp}~
            \Tree [ [ {\tiny A} {\tiny C} ] {\tiny B} ] ~\xrightarrow[\swapp\phantom{x}]{\phantom{x}\idc}~
            \Tree [ [ {\tiny C} {\tiny A} ] {\tiny B} ] ~\xrightarrow{\assocrp}~
            \Tree [ {\tiny C} [ {\tiny A} {\tiny B} ] ] ~\xrightarrow[\phantom{xx}\swapp]{\idc\phantom{xx}}~
            \Tree [ {\tiny C} [ {\tiny B} {\tiny A} ] ] ~
      \]
      \label{fig:example-first-stage}
      \caption{Type isomorphisms as tree transformations}
\end{figure}

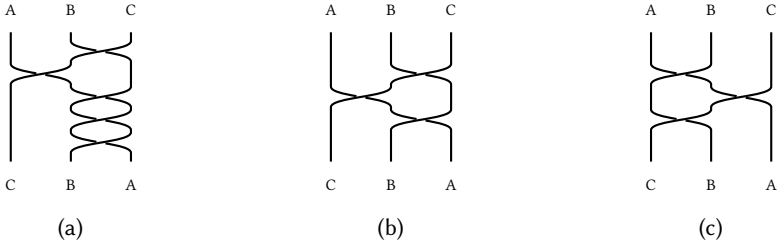
\begin{figure}
      \begin{subfigure}[b]{0.3\textwidth}
            \centering
            \begin{tikzpicture}
                  \def\nstrandsdf{3}
                  \pic[local bounding box=my braid,braid/.cd,
                        number of strands = \nstrandsdf,
                        width = 0.8cm,
                        height = 0.3cm,
                        border height = 0.1cm,
                        thick,
                        name prefix=braid]
                  {braid={s_2, s_2, s_2, s_1, s_2}};
                  \node at (braid-1-s)[yshift = 2.0cm] {\tiny A};
                  \node at (braid-2-s)[yshift = 2.0cm] {\tiny B};
                  \node at (braid-3-s)[yshift = 2.0cm] {\tiny C};

                  \node at (braid-1-e)[yshift = -2.0cm] {\tiny A};
                  \node at (braid-2-e)[yshift = -2.0cm] {\tiny B};
                  \node at (braid-3-e)[yshift = -2.0cm] {\tiny C};
            \end{tikzpicture}
            \caption{}
            \label{fig:cone-braid}
      \end{subfigure}
      \begin{subfigure}[b]{0.3\textwidth}
            \centering
            \begin{tikzpicture}
                  \def\nstrandsdf{3}
                  \pic[local bounding box=my braid,braid/.cd,
                        number of strands = \nstrandsdf,
                        width = 0.8cm,
                        height = 0.3cm,
                        border height = 0.1cm,
                        thick,
                        name prefix=braid]
                  {braid={1, s_2,, s_1, s_2, 1}};
                  \node at (braid-1-s)[yshift = 2.0cm] {\tiny A};
                  \node at (braid-2-s)[yshift = 2.0cm] {\tiny B};
                  \node at (braid-3-s)[yshift = 2.0cm] {\tiny C};

                  \node at (braid-1-e)[yshift = -2.0cm] {\tiny A};
                  \node at (braid-2-e)[yshift = -2.0cm] {\tiny B};
                  \node at (braid-3-e)[yshift = -2.0cm] {\tiny C};
            \end{tikzpicture}
            \caption{}
            \label{fig:ctwo-braid}
      \end{subfigure}
      \begin{subfigure}[b]{0.3\textwidth}
            \centering
            \begin{tikzpicture}
                  \def\nstrandsdf{3}
                  \pic[local bounding box=my braid,braid/.cd,
                        number of strands = \nstrandsdf,
                        width = 0.8cm,
                        height = 0.3cm,
                        border height = 0.1cm,
                        thick,
                        name prefix=braid]
                  {braid={1, s_1,, s_2, s_1, 1}};
                  \node at (braid-1-s)[yshift = 2.0cm] {\tiny A};
                  \node at (braid-2-s)[yshift = 2.0cm] {\tiny B};
                  \node at (braid-3-s)[yshift = 2.0cm] {\tiny C};

                  \node at (braid-1-e)[yshift = -2.0cm] {\tiny A};
                  \node at (braid-2-e)[yshift = -2.0cm] {\tiny B};
                  \node at (braid-3-e)[yshift = -2.0cm] {\tiny C};
            \end{tikzpicture}
            \caption{}
            \label{fig:norm-braid}
      \end{subfigure}

      \label{fig:example-braid}
      \caption{Type isomorphisms as braid diagrams}
\end{figure}

The first step is to realise that the use of associativity is uninteresting, and the only steps with interesting
computational content are the swaps. The swaps can be either big or small -- they can happen between leaf nodes, or
between bigger subtrees, respectively.

First, we normalise the types to a list-like representation $A + (B + (C + \zerot))$, where $\zerot$ is the empty type.
In this representation, each type is identified with its index, $A$ has index 0, $B$ has index 1, and $C$ has index 2.
Then, we need to compile each primitive isomorphism to a list of adjacent transpositions, and then compose by appending
the lists. A small swap is trivially implemented as a single adjacent transposition. To compile a big swap into adjacent
transpositions, we need to traverse the subtrees in-order and recursively swap elements from the left by transposing
them across the ones in the middle -- this generates a large number of adjacent transpositions. For the two programs, we
get the following two lists: $[1,0,1,1,1]$ (\cref{fig:cone-braid}) and $[1,0,1]$ (\cref{fig:ctwo-braid}), where the
number $k$ encodes a transposition of elements at indices $k$ and $k+1$.

Swapping is symmetric, that is, swapping two elements and immediately swapping them back produces the same permutation.
The first list contains such redundant operations, so to eliminate these, we have to normalise the lists, which we will
do by setting up reduction relations and an appropriate rewriting system. This system will normalise both lists to
$[0,1,0]$ (\cref{fig:norm-braid}), which is lexicographically the smallest list that corresponds to this permutation.
Since both programs have the same normal forms, they're equivalent!

The crux of designing an equational theory for $\PiLang$ relies on the choice of relations we use to design this
rewriting system. First, there should be enough equations relating the programs to their normalised forms, and second,
there should be enough equations corresponding to the reduction rules. We will show that these correspond to the
coherence conditions for symmetric rig groupoids.

From the normalised list $[0,1,0]$, we can construct a permutation on a list of 3 elements $[a,b,c]$ as follows. We think
of it as insertion-sorting the elements of the list. Starting from the list $[a, b, c]$, we first insert $b$ at the
right place -- by applying transposition $0$, we get the list $[b, a, c]$. Then, element $c$ is inserted in the right
place by applying transpositions $1$ and $0$ -- as a result, we get the desired permutation $[c, b, a]$. Notice that we
could specify a more compact way of describing this process, since the key information was only how many shifts we
needed to apply to an element. We could describe this procedure using a code $(0, 1, 2)$, which says how many inversions
to apply to elements $a$, $b$, and $c$, respectively.

Using this algorithm, we can turn a type isomorphism into a permutation of finite sets, relating the operational
semantics of $\PiLang$ as permutations of finite sets of bits, to our denotational semantics. This allows us to
establish full abstraction and adequacy.

\paragraph{Outline and Contributions} Our main result is a proof of the soundness and completeness of $\PiLang$ with
respect to its semantics in the weak symmetric rig groupoid of finite sets and permutations. We state our result as a
Curry-Howard-Lambek correspondence for Reversible Logic, Reversible Programming Languages, and Symmetric Rig Groupoids,
using which we can build a toolbox of technical devices for reasoning about reversible circuits.

\begin{itemize}[leftmargin=*]
      \item We start in~\Cref{sec:qiskit} by presenting a few reversible circuits in the popular IBM Qiskit framework to
            serve as running examples throughout the paper.
      \item In~\Cref{sec:pi}, we introduce the two-level language
            $\PiLang$~\cite{jamesInformationEffects2012,caretteComputingSemiringsWeak2016} and illustrate how to write
            reversible circuits and their equivalences using 1-combinators and 2-combinators respectively. We give a
            semantic account of the language by translating each level-1 program to a bijection between finite sets, and
            verifying that programs identified by level-2 constructs denote the same bijection.
      \item ~\Cref{sec:ufin} describes the construction of $\UFin$, the groupoid of finite sets and permutations, in
            Homotopy Type Theory. We define and characterise the notion of a univalent subuniverse, and construct
            $\UFin$ as a univalent subuniverse which classifies all finite types. We establish that paths in $\UFin$ are
            families of loops on finite sets of specified cardinality, given by $\Aut[\Fin[n]]$, which produces the
            permutation group on $\Fin[n]$.
      \item In~\Cref{sec:finite}, we proceed to give a presentation of the permutation group, as the symmetric group
            $\Sn$ with generators and relations, and solve its word problem. In particular, we present $\Sn$ as a
            Coxeter group, build a rewriting system based on Coxeter relations, and prove confluence and termination.
            Using our rewriting system, we establish that normal forms for words in $\Sn$ are equivalent to Lehmer
            codes~\cite{lehmerTeachingCombinatorialTricks1960}, which are a convenient and compact representation of
            permutations. Finally, we show that there is an equivalence between Lehmer codes and permutations
            $\Aut[\Fin[n]]$ given by the Lehmer encode-decode algorithm.
      \item In~\Cref{sec:equivalence}, we show how to interpret the language $\PiLang$ into the groupoid $\UFin$, in
            stages. First we define a subset $\PiPlusLang$ of the language which only includes the additive monoidal
            structure, and show how to translate $\PiLang$ programs to $\PiPlusLang$ programs. Then, we further define a
            normalised form for this language called $\PiHatLang$, which has normalised 1-combinators and 2-combinators
            corresponding to adjacent transpositions. We show that $\PiPlusLang$ can be translated to $\PiHatLang$ and
            back. Then, we show how to interpret this language $\PiHatLang$ into $\UFin$ -- the 1-combinators are
            translated into permutations via words in $\Sn$, and 2-combinators are interpreted as 2-paths in $\UFin$. We
            further show how to quote back a permutation in $\UFin$ into a 1-combinator using the normal forms for words
            in $\Sn$. The main result of this section is a symmetric monoidal equivalence between the syntactic
            groupoids of $\PiPlusLang$ and $\PiHatLang$, with $\UFin$. Finally, we also establish full abstraction and
            adequacy of this model with respect to the operational semantics.
      \item In~\Cref{sec:applications}, we show applications of our results to reversible circuits, using our
            formalisation. Our results are stated using HoTT~\cite{univalentfoundationsprogramHomotopyTypeTheory2013},
            and formalised using the HoTT-Agda library (around 7,500 lines of code). Using the formalisation, we are
            able to extract procedures for:
            \begin{enumerate*}
                  \item the synthesis of a reversible circuit from a permutation on a finite set,
                  \item the verification that a reversible circuit realises a given permutation on finite sets,
                  \item a normalisation-by-evaluation (NbE) procedure that reduces reversible circuits to canonical normal forms,
                  \item a sound and complete calculus for reasoning about reversible circuits and their equivalences, and
                  \item the transfer of theorems about permutations and reversible circuits from one representation to the other.
            \end{enumerate*}
\end{itemize}

\noindent
The proofs of some of our lemmas and propositions and theorems, as well as additional material, are given in the
supplementary appendices, and we refer to them in the text. Our accompanying Agda code contains the formalisation of the
full syntax, and most of our proofs.

\section{Reversible Circuits in Qiskit}
\label{sec:qiskit}
\label{sec:examples}

Classical reversible boolean circuits are at the core of most quantum algorithms and hence are supported by popular
platforms for quantum computing such as IBM Qiskit~\cite{aleksandrowiczQiskitOpensourceFramework2019}. Specifically, the
Qiskit framework provides the following universal set of gates for reversible computing: \textsf{not} (boolean negation,
called \verb|x|), \textsf{cnot} (conditional negation of the second input if the first is true; called \verb|cx|), and
\textsf{toffoli} (conditional negation of the third input if both the first two inputs are true; called \verb|ccx|)
gates. Additionally, Qiskit allows implicit re-shuffling of bits by allowing each operation to specify the indices of
its input bits.

For concreteness, we demonstrate two different circuits that implement the
following reversible function specification
$\mathit{reversibleOr}(h,b_1,b_2) ~=~ (h \,\underline{\vee}\, (b_1 \vee b_2),
~b_1, ~b_2)$ where $\vee$ is boolean disjunction and $\underline{\vee}$ is the
exclusive-or operation. The circuits are presented in both the textual interface
\verb|qasm| and the graphical interface:

\medskip
\begin{tabular}{c@{\qquad}c}
\begin{minipage}[t]{0.42\linewidth}
  \begin{verbatim}
  ccx q[1], q[2], q[0];
  cx  q[1], q[0];
  cx  q[2], q[0];
  \end{verbatim}
  \includegraphics[scale=0.7]{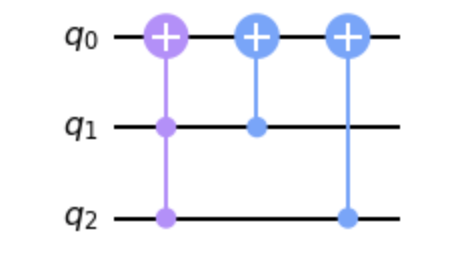}
  \end{minipage}
&
\begin{minipage}[t]{0.43\linewidth}
  \begin{verbatim}
  cx  q[1], q[0];
  x   q[1];
  ccx q[1], q[2], q[0];
  x   q[1];
  \end{verbatim}
  \includegraphics[scale=0.6]{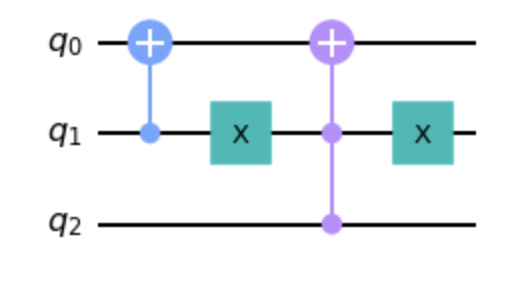}
  \end{minipage}
\end{tabular}

\medskip

There is a wealth of manual and algorithmic approaches for producing circuits such
as the two above~\cite{maslov:2003:rls:1087512,1201583}. The circuit on the left
was manually produced using a standard synthesis algorithm for reversible
circuits~\cite{10.1145/775832.775915}. The circuit on the right was produced
using an approach that analyzes the recursive structure of the circuit (and
would generalise to computing the disjunction of more than two inputs):

From the specification of the circuit, we expect input \verb|011| to be mapped
to \verb|111|. To gain some intuition, we trace the evaluation of each circuit for
input \verb|011|. In this context, the most significant bit is at index 0. In
the first circuit, the \verb|ccx| gate negates \verb|q[0]| since both \verb|q[1]| and
\verb|q[2]| are true producing \verb|111|; the following \verb|cx| gate produces
\verb|011|; finally the last \verb|cx| produces the result \verb|111|. For the
second circuit, the trace of the execution on the same input value \verb|011| goes through the stages \verb|111|, \verb|101|, \verb|101|, and finally \verb|111|.

Although rather trivial, the examples above illustrate the general idea. A more
interesting example would be the classical core of Shor's algorithm which
requires a circuit implementing $f(r) = a^{r} \mod N$ for fixed $a$ and $N$. The
specification of the circuit is relatively straightforward to calculate. Here it
is for $a=11$ and $N=15$:
\[\begin{array}{rcll}
g(r,h) &=& \left\{ \begin{array}{ll}
                     (r,h+1) & \mbox{when~$r$~even~and~$h$~even} \\
                     (r,h-1) & \mbox{when~$r$~even~and~$h$~odd} \\
                     (r,11-h) & \mbox{when~$r$~odd~and~$4 > h \geq 0$~or~$12 > h \geq 8$} \\
                     (r,19-h) & \mbox{when~$r$~odd~and~$8 > h \geq 4$~or~$16 > h \geq 12$}
                                \end{array}\right.
\end{array}\]

\noindent However, as explained in standard accounts of the algorithm (e.g., the
Qiskit implementation), producing an efficient modular exponentiation circuit
from this specification is not straightforward and is actually the bottleneck in
Shor’s algorithm. Typical derivations of the circuit start from elementary
gates, build a circuit for reversible disjunction (like the two circuits above),
reversible conjunction, a circuit for a half-adder, a circuit for computing the
carry, progressing to a circuit for modular addition, which is used to build a
circuit for modular multiplication, and then finally a circuit for modular
exponentiation taking care at each step to avoid the exponential blowup (e.g.,
by implementing exponentiation by squaring instead of repeated
multiplication)~\cite{shorefficient}.

\begin{figure}[t]
  {\scalebox{\scalef}{$%
        \begin{array}{rrcll}
          \idc :     & A                     & \isoone & A                            & : \idc      \\
          \\
          \identlp : & \zerot + A            & \isoone & A                            & : \identrp  \\
          \swapp :   & A + B                 & \isoone & B + A                        & : \swapp    \\
          \assoclp : & A + (B + C)           & \isoone & (A + B) + C                  & : \assocrp  \\ [1.5ex]
          \identlt : & \onet \times A        & \isoone & A                            & : \identrt  \\
          \swapt :   & A \times B            & \isoone & B \times A                   & : \swapt    \\
          \assoclt : & A \times (B \times C) & \isoone & (A \times B) \times C        & : \assocrt  \\ [1.5ex]
          \absorbr : & ~ \zerot \times A     & \isoone & \zerot ~                     & : \factorzl \\
          \dist :    & ~ (A + B) \times C    & \isoone & (A \times C) + (B \times C)~ & : \factor
        \end{array}$}}

  \medskip

  {\scalebox{\scalef}{%
      \Rule{}
      {\jdg{}{}{c_1 : A \isoone B} \quad \vdash c_2 : B \isoone C}
      {\jdg{}{}{c_1 \fatsemi c_2 : A \isoone C}}
      {}

      \Rule{}
      {\jdg{}{}{c_1 : A \isoone B} \quad \vdash c_2 : C \isoone D}
      {\jdg{}{}{c_1 \oplus c_2 : A + C \isoone B + D}}
      {}

      \Rule{}
      {\jdg{}{}{c_1 : A \isoone B} \quad \vdash c_2 : C \isoone D}
      {\jdg{}{}{c_1 \otimes c_2 : A \times C \isoone B \times D}}
      {}
    }}
  \caption{$\Pi$-terms, combinators, and their types.}
  \label{fig:pi-terms}
\end{figure}

\section{A Reversible Programming Language}
\label{sec:pi}
\label{sec:reversibleone}
\label{sec:reversibletwo}
\label{langeqeq}
\label{sec:informal}

The circuit model of reversible computation discussed in the previous section is a useful abstraction close to the
hardware platform. However, since its main data abstraction is a \emph{sequence of wires}, it only provides an
``assembly-level'' programming abstraction (e.g., \verb|qasm|). As motivated by \citet{LAFONT2003257}, a mathematical
model based on permutations of finite sets provides a richer algebraic structure, using which we describe a reversible
programming language in this section.

\subsection{The $\Pi$ Family of Languages}
\label{sec:langRev-examples}
\label{examples}

In reversible boolean circuits, the number of input bits matches the number of output bits. Thus, a key insight for a
programming language of reversible circuits is to ensure that each primitive operation preserves the number of bits,
which is just a natural number. The algebraic structure of natural numbers as the free commutative semiring (or,
commutative rig), with $(0,+)$ for addition, and $(1,\times)$ for multiplication then provides sequential, vertical, and
horizontal circuit composition operators.

These commutative rig identities can be used to design a logic for reversible
programming~\cite*{sparksSuperstructuralReversibleLogic2014}. To interpret natural number identities as reversible programs,
the logic needs to be equipped with values and types, and a notion of operational semantics and contextual equivalence,
giving a computational interpretation of the commutative rig structure~\cite{jamesInformationEffects2012}. On the
semantic side, the natural space to consider is the groupoidification of a commutative rig, that is, a symmetric rig
groupoid.

Putting these ideas together, the programming language $\PiLang$, whose syntax is given below, embodies the
computational content of isomorphisms of finite types, or permutations.

\medskip

{\scalebox{\scalef}{$%
      \begin{array}{lrcl}
        \textit{Value types}   & A,B,C,D & ::= & \zerot \alt \onet \alt A+B \alt A\times B        \\
        \textit{Values}        & v,w,x,y & ::= & \Acon{tt} \alt \inlv{v} \alt \inrv{v} \alt (v,w) \\
        \textit{Program types} &         &     & A \isoone B                                      \\
        \textit{Programs}      & c       & ::= & (\textrm{See Fig.~\ref{fig:pi-terms}})
      \end{array}$}}

\medskip\noindent The language of types is built from the empty type ($\zerot$), the unit type
($\onet$) containing just one value~$\Acon{tt}$, the sum type ($+$) containing values of the form $\inlv{v}$ and
$\inrv{v}$, and the product type ($\times$) containing pairs of values $(v_1,v_2)$.
%

To see how this language expresses reversible circuits, we present a few examples in the Agda embedding of the
language. The Agda syntax is almost identical to the syntax in Fig.~\ref{fig:pi-terms} except that sequential
composition is written using $\circledcirc$. First it is possible to directly mimic the \verb|qasm|-perspective by
defining types that describe sequences of booleans. We use the type $\mathbb{2} = \onet + \onet$ to represent booleans
with $\inlv{\Acon{tt}}$ representing \textsf{true} and $\inrv{\Acon{tt}}$ representing $\textsf{false}$. Boolean
negation (the \verb|x|-gate) is straightforward to define using the primitive combinator $\swapp$. We can represent
$n$-bit words using an n-ary product of boolean values, thus the type $\mathbb{2} \times (\mathbb{2} \times \mathbb{2})$
(abbreviated $\mathbb{B}~3$) corresponds to a collection of wires that can transmit three bits.
%
%
To express the \verb|cx| and \verb|ccx| gates, we need to encode a notion of conditional expression. Such conditionals
turn out to be expressible using the distributivity and factoring identities of rigs as shown below:

\medskip

\cif{}

\noindent The input value of type $\mathbb{2} \times A$ is processed by the distribute operator \ensuremath{\dist},
which converts it into a value of type $(\onet \times A) + (\onet \times A)$. In the left branch, which corresponds to
the case when the boolean is \textsf{true}, the combinator~\ensuremath{c_1} is applied to the value of
type~\ensuremath{A}. The right branch, which corresponds to the boolean being \textsf{false}, passes the value of type $A$
through the combinator \ensuremath{c_2}.  The inverse of \ensuremath{\dist}, namely \ensuremath{\factor} is applied to
get the final result. Using this conditional operator, \verb|cx| is defined as $\Afun{cif}~\verb|x|~\idc$ and
\verb|ccx| is defined as $\Afun{cif}~\verb|cx|~\idc$. With these conventions, the first circuit in the previous section
is transcribed as follows:

\medskip

\adder{}

\addertwo{}

\noindent where we clearly see the sequences of the three operations \verb|ccx|, \verb|cx|, and \verb|cx| but, instead
of using the indices in the sequence of wires to identify the relevant parameters, here we use structural isomorphisms
to re-shuffle the types. We only show one of these re-shuffling isomorphisms and elide the others. For the second circuit, instead of transcribing it directly, we express it using
a slightly more abstract notation:

\medskip

\resettwo{}

\noindent Like the original circuit, we examine the bit at index 1 (corresponding to the component $B$ in a tuple
$(A,(B,C))$): if the bit is true, we perform an \verb|x| operation on component $A$, and otherwise we perform a
\verb|cx| operation on $(C,A)$. The two uses of \verb|x| gates in the circuit are now unnecessary as they were only needed
to encode a two-way conditional expression using a sequence of one-way conditional expressions (the only ones available in
the linear circuit model).

All of this is only half the story, however. A sound semantics for $\PiLang$ in weak rig
groupoids was established by \citet{caretteComputingSemiringsWeak2016}, and conjectured to be complete. For this
semantics, \emph{coherence conditions} for symmetric rig groupoids that identify different syntactic representations of the same
permutation~\cite{laplaza72,caretteComputingSemiringsWeak2016}, were collected  in a
second level of $\PiLang$ syntax as level-2 combinators.  Each level-2 combinator is of the form $c_1 \isotwo c_2$ for appropriate
$c_1$ and $c_2$ of the same level-1 type $A \isoone B$ and asserts that $c_1$ and $c_2$ denote the same bijection. For
example, we have the following level-2 combinators dealing with associativity:

\medskip

\leveltwoblockone{}

\noindent The full set of level-2 combinators is large; the remaining combinators are listed in~\cref{app:leveltwo}.

\begin{toappendix}
  \subsection{2-combinators}

  The additional level-2 combinators:

  \label{app:leveltwo}

  \medskip
  \leveltwoblocktwo{}
\end{toappendix}





\subsection{Semantics}
\label{subsec:denotational}

\noindent Below we present a simple denotational semantics for our language, using finite types and type isomorphisms.
Each $\PiLang$ type $A$ is mapped to a (finite) set $\denot{A}$ and each combinator $c : A \isoone B$ is mapped to a
(bijective) function $\denot{A} \to \denot{B}$. We describe this semantics by writing an interpreter in Agda. First, we
state the semantics for types, where $\bot$ is the empty set, $\top$ is the singleton set, $\sqcup$ is the disjoint
union of sets, and $\times$ is the cartesian product of sets.


\begin{center}
  {\scalebox{\scalef}{$%
        \begin{array}{rcl}
          \denot{\zerot}     & = & \bot                       \\
          \denot{\onet}      & = & \top                       \\
          \denot{A + B}      & = & \denot{A} \sqcup \denot{B} \\
          \denot{A \times B} & = & \denot{A} \times \denot{B}
        \end{array}$}}
\end{center}

For combinators, we show explicitly how values reduce along each combinator, similar to a big-step operational
semantics~\cite{chenComputationalInterpretationCompact2021,theseus}.


\begin{multicols}{2}
  {\scalebox{\scalef}{$%
        \begin{array}[t]{rlcl}
          \denot{\identlp} & (\inl{v})         & = & v               \\
          \denot{\identrp} & v                 & = & \inl{v}         \\
          \denot{\swapp}   & (\inl{v})         & = & \inr{v}         \\
          \denot{\swapp}   & (\inr{v})         & = & \inl{v}         \\
          \denot{\assoclp} & (\inl{v})         & = & \inl{(\inl{v})} \\
          \denot{\assoclp} & (\inr{(\inl{v})}) & = & \inl{(\inr{v})} \\
          \denot{\assoclp} & (\inr{(\inr{v})}) & = & \inr{v}         \\
          \denot{\assocrp} & (\inl{(\inl{v})}) & = & \inl{v}         \\
          \denot{\assocrp} & (\inl{(\inr{v})}) & = & \inr{(\inl{v})} \\
          \denot{\assocrp} & (\inr{v})         & = & \inr{(\inr{v})}
        \end{array}$}}

  {\scalebox{\scalef}{$%
        \begin{array}[t]{rlcl}
          \denot{\identlt} & (\ttt , v)          & = & v                   \\
          \denot{\identrt} & v                   & = & (\ttt , v)          \\
          \denot{\swapt}   & (v_1 , v_2)         & = & (v_2 , v_1)         \\
          \denot{\assoclt} & (v_1 , (v_2 , v_3)) & = & ((v_1 , v_2) , v_3) \\
          \denot{\assocrt} & ((v_1 , v_2) , v_3) & = & (v_1 , (v_2 , v_3)) \\
          \denot{\dist}    & (\inl{v_1} , v_3)   & = & \inl{(v_1 , v_3)}   \\
          \denot{\dist}    & (\inr{v_2 , v_3})   & = & \inr{(v_2 , v_3)}   \\
          \denot{\factor}  & (\inl{(v_1 , v_3)}) & = & (\inl{v_1} , v_3)   \\
          \denot{\factor}  & (\inr{(v_2 , v_3)}) & = & (\inr{v_2} , v_3)   \\
          \denot{\idc}     & v                   & = & v
        \end{array}$}}
\end{multicols}

\begin{center}
  {\scalebox{\scalef}{$%
        \begin{array}{rlcl}
          \denot{(c_1 \fatsemi c_2)} & v           & = & (\denot{c_2} \circ \denot{c_1}) v   \\
          \denot{(c_1 \oplus c_2)}   & (\inl{v})   & = & \inl{(\denot{c_1}~v)}               \\
          \denot{(c_1 \oplus c_2)}   & (\inr{v})   & = & \inr{(\denot{c_2}~v)}               \\
          \denot{(c_1 \otimes c_2)}  & (v_1 , v_2) & = & (\denot{c_1} v_1 , \denot{c_2} v_2)
        \end{array}
      $}}
\end{center}

\begin{theoremrep}\label{thm:semone}
  The semantics is sound in the following sense:
  \begin{itemize}
    \item For every level-1 combinator $c : A \isoone B$, we have that $\denot{c}$ is a bijection between $\denot{A}$ and $\denot{B}$.
    \item For every pair of combinators $c_1$ and $c_2$ of the same type $A \isoone B$, if there exists a level-2
          combinator $\alpha$ such that $\alpha : c_1 \isotwo c_2$, then $\denot{c_1} = \denot{c_2}$ using
          extensional equivalence of functions.
  \end{itemize}
\end{theoremrep}
\begin{proof}
  For every primitive combinator $c$ listed on one side of Fig.~\ref{fig:pi-terms}, let $!c$ be the combinator listed on
  the other side. Thus $! \assoclp$ is $\assocrp$ and $! \swapp$ is $\swapp$ itself. Then we have that $\denot{c}$ and
  $\denot{!c}$ form an equivalence. For the level-2 combinator \Afun{idr◎l}, we check
  $\denot{\AgdaBound{c}~\AgdaOperator{\AgdaInductiveConstructor{◎}}~\AgdaInductiveConstructor{id⟷₁}}
    = \mathit{id} \circ \denot{c} = \denot{c}$. The other cases are only slightly more involved calculations.
\end{proof}

\section{The Groupoid of Finite Types}
\label{sec:ufin}


In categorical language, the setting for the semantics in the previous section is the category of finite sets and
functions $\SetFin$. However, as $\PiLang$ only refers to bijective functions, a more precise setting is the groupoid
$\BFin = \term{core}(\SetFin)$ of finite sets and bijections. $\SetFin$ has finite coproducts $(\emptyt, \sqcup)$ and
finite products $(\unit, \times)$ and in $\BFin$ these restrict to additive and multiplicative symmetric monoidal
structures, respectively, making $\BFin$ a symmetric rig groupoid -- the \emph{vertical categorification} of the
commutative rig of natural numbers $\Nat$~\cite{baezHIGHERDIMENSIONALALGEBRA2010}.

The semantics interprets types of $\PiLang$ as objects in $\BFin$, 1-combinators as isomorphisms, and for every pair of 1-combinators
related by a 2-combinator, their interpretations in~$\BFin$ are equal.  The groupoid $\BFin$ is strict, since the
collection of isomorphisms is a set, that is, a discrete category. There is no explicit witness for the equality of two
isomorphisms, since we can decide by evaluating two bijections whether they are equal. To be able to establish
completeness for $\PiLang$, we want a witness for this equality, so that we can quote back to the syntax and produce a
2-combinator witnessing the equality of the corresponding 1-combinators.




We observe that the implicit equalities between the isomorphisms are pointwise equalities of functions, that is,
homotopies. We therefore \emph{weaken} the groupoid $\BFin$, exposing these homotopies, by using higher invertible cells.
We work in HoTT (Univalent Foundations) as it provides a proof-relevant, constructive metatheory to get a handle
on these equalities and provides a rich internal language for describing weak groupoids, using the ``types are weak
$\infty$-groupoids'' correspondence.





Every type in HoTT is a weak $\infty$-groupoid whose points are the terms of the type, and the (iterated) identity type
gives the (higher) morphisms. The groupoid we are interested in has types as points, type equivalences for 1-cells, and
higher homotopies for higher cells. (See~\cref{app:grpdexample} for an example on a 3-element set.) This is the
groupoid structure for the universe type $\UU$, since the identity type on types can be characterised as type
equivalences (by \emph{univalence}). But, we only want to carve out a \emph{subuniverse} of \emph{finite types}, still
satisfying univalence, to get the groupoid structure. In this section, we formally define \emph{univalent subuniverses},
and proceed to construct the particular instance for finite types, $\UFin$ (\cref{def:ufin}).

\begin{toappendix}
  \label{app:grpdexample}
  We give an example of the groupoid structure on a 3-element set.

  \[
    \begin{tikzcd}
      \Fin[3]
      \arrow[""{name=0, anchor=center, inner sep=0}, "{f_{3}}", no head, loop, distance=4em, in=115, out=65]
      \arrow[""{name=0, anchor=center, inner sep=0}, "{f_{2}}", no head, loop, distance=8em, in=125, out=55]
      \arrow[""{name=1, anchor=center, inner sep=0}, "{f_{1}}"', no head, loop, distance=12em, in=135, out=45]
      \arrow["", "{h}", shorten <=3pt, shorten >=3pt, Rightarrow, no head, from=0, to=1]
    \end{tikzcd}
  \]

  \noindent We have $\Fin[3] = \Set{0,1,2} \eqv \unit \sqcup (\unit \sqcup \unit)$ which fixes a particular enumeration of the
  elements. Suppose we have a set $X = (\unit \sqcup \unit) \sqcup \unit$, it has the same cardinality as $\Fin[3]$, so it
  is represented by the same 0-cell. But, $X$ can be made equivalent to $\Fin[3]$ in many different ways since there are
  many bijections between them. One bijection is
  $\Set{\inl(\inl(\ttt)) \mapsto 0, \inl(\inr(\ttt)) \mapsto 1, \inr(\ttt) \mapsto 2}$ which can be written in two
  different ways by composing more primitive operations, $f_{1} = \assocrp$, or
  $f_{2} = \swapp \compc \assoclp \compc \swapp$. Another bijection is
  $\Set{\inl(\inl(\ttt)) \mapsto 1, \inl(\inr(\ttt)) \mapsto 2, \inr(\ttt) \mapsto 0}$ which is given by $f_{3} = \ldots$.
  Since $f_{1}$ and $f_{2}$ produce the same enumeration of the elements of $X$, they are identified by a homotopy $h$
  which is encoded in the 2-cell between them.

  At level 0, all we know is that if $X : \UFin[3]$, then X is merely equal to $\Fin[3]$, that is
  $\Trunc[-1]{X \id \Fin[3]}$, and we don't have access to the bijection. At level 1, if we know that both $X$ and $Y$ are
  \emph{equal} in $\UFin[3]$, then we can extract an equivalence between them, that is, $(X \id Y) \to (X \eqv Y)$.
  $\UFin[3]$ being a univalent subuniverse asserts that there are as many elements (upto higher homotopy) in $X \id Y$ as
  there are $X \eqv Y$.
\end{toappendix}

\subsection{The Type Theory}
\label{subsec:type-theory}


We use the type theory of the HoTT book~\cite{univalentfoundationsprogramHomotopyTypeTheory2013}, that is, we use
intensional Martin-L\"{o}f Type Theory, with a (univalent) universe $\UU$, and a few Higher Inductive Types (HITs) for
propositional and set truncation, and set-quotients. All arguments will hold in a Cubical Type
Theory~\cite*{cohenCubicalTypeTheory2018,angiuliComputationalSemanticsCartesianCubical2019,vezzosiCubicalAgdaDependently2019}
as well.  In~\cref{app:identitytypes},~\cref{app:homotopytypes}, and~\cref{app:hits}, we review the basics of identity
types, homotopy types, and HITs and refer the reader to the book for more details.


\begin{toappendix}
  \subsection{Identity Types}
  \label{app:identitytypes}

  Given two terms $x:A$ and $y:A$, we write $x \id_{A} y$, or simply $x \id y$, for the identity type, which is the type
  of equalities or identifications between them. The identity type is generated by reflexivity $\refl_{x} : x \id_{A} x$,
  and the eliminator for the identity type is given by path induction or the $J$-rule (\cref{def:path-induction}). This
  construction can be iterated, giving the identity type between two terms of an identity type, repeating ad infinitum.
  Using the iterated identity type for morphisms, each type is equipped with the structure of a weak $\infty$-groupoid,
  where each morphism satisfies groupoid laws only upto a higher one. Given an arbitrary type (or groupoid) $A$, we list
  some laws that are provable using path induction.

  \begin{definition}[Path Induction]
    \label{def:path-induction}
    Given a type family $C : \dfun{x,y:A}{(x \id_A y)} \to \UU$, and a function $c : \dfun{x:A}{C(x,x,\refl_x)}$, there is
    a function $f : \dfun{x,y:A}{\dfun{p:x \id_A y}{C(x,y,p)}}$ such that $f(x,x,\refl_x) \defeq c(x)$.
  \end{definition}

  \begin{gather*}
    \begin{aligned}
      \term{\inv{\blank}}      & : (x \id_{A} y) \to (y \id_{A} x)                   \\
      \term{\blank\comp\blank} & : (x \id_{A} y) \to (y \id_{A} z) \to (x \id_{A} z)
    \end{aligned}
    \qquad
    \begin{aligned}
      \term{assoc} & : (p : x \id_{A} y)  (q : y \id_{A} z) (r : z \id_{A} w) \\
                   & \to (p \comp q) \comp r \id p \comp (q \comp r)          \\
      \term{invr}  & : (p : x \id_{A} y) \to p \comp \inv{p} \id \refl_{x}
    \end{aligned}
  \end{gather*}

  \noindent A homotopy between functions $f \htpy g$ is given by pointwise equality between them $\dfun{x:A}{f(x) \id_{B} g(x)}$.
  The identity type for functions is equivalent to homotopies between them ${(f \id_{A \to B} g)} \eqv {(f \htpy g)}$, by
  function extensionality. An equivalence between types $A \eqv B$ is given by a pair of functions between them which
  compose to the identity, $f \comp g \htpy \idfunc_{B}$ and $g \comp f \htpy \idfunc_{A}$, and this is equivalent to the
  identity type for the universe, $(A \id_{\UU} B) \eqv (A \eqv B)$, by univalence.

  Functions between types are functors between groupoids. Given a function $f : A \to B$, there is a functorial action
  on the paths given by $\term{ap}$. Type families, that is, types indexed by terms, are simply functions from a type to
  the universe, such as $A \to \UU$. For a type family $P : A \to \UU$ and a point $x : A$, the type $P(x)$ is called
  the fiber over $x$.  The $\term{transport}$ operation, named $\term{transport}/term{tr}$, lifts paths in the indexing
  type to functions between fibers.

  \begin{gather*}
    \term{ap}_{f} : \dfun{x,y:A}{x \id_{A} y \to f(x) \id_{B} f(y)}
    \qquad
    \term{transport}_{P} : \dfun{x,y:A}{x \id_{A} y \to P(x) \to P(y)}
  \end{gather*}

  The type $\dsum*{x:A}{P(x)}$ is the collection of all the fibers and is called the total space of $P$. The first
  projection ${\pi_1 : \dsum*{x:A}{P(x)} \to A}$ from the total space to the base space $A$ has the structure of a
  fibration, that is, there is a lifting operation (\cref{fig:lift}) which lifts paths in the base space to paths in the
  total space. Given a path $p : x \id_{A} y$ in the base space, and $u : P(x)$ a point in the fiber over $x$, we have:
  \[
    \term{lift}(u,p) : (x , u) \id_{\dsum{x:A}{P(x)}} (y , \transport{P}{p}{u})
  \]


  \begin{figure}
    \begin{center}
      \begin{tikzpicture}[yscale=.5,xscale=2]
        \draw (0,0) arc (-90:170:8ex) node[anchor=south east] {$A$} arc (170:270:8ex);
        \draw (0,6) arc (-90:170:10ex) node[anchor=south east] {$\dsum{x:A}{P(x)}$} arc (170:270:10ex);
        \draw[->] (0,5.8) -- node[auto] {$\fst$} (0,3.2);
        \node[circle,fill,inner sep=1pt,label=left:{$x$}] (b1) at (-.5,1.4) {};
        \node[circle,fill,inner sep=1pt,label=right:{$y$}] (b2) at (.5,1.4) {};
        \draw[decorate,decoration={snake,amplitude=1}] (b1) -- node[auto,swap] {$p$} (b2);
        \node[circle,fill,inner sep=1pt,label=left:{$\pair{x,u}$}] (b1) at (-.5,7.2) {};
        \node[circle,fill,inner sep=1pt,label=right:{$\pair{y,\transport*{P}{p}{u}}$}] (b2) at (.5,7.2) {};
        \draw[decorate,decoration={snake,amplitude=1}] (b1) -- node[auto] {$\term{lift}(u,p)$} (b2);
      \end{tikzpicture}
    \end{center}
    \caption{Lifting operation in $P$}
    \label{fig:lift}
  \end{figure}

\end{toappendix}

\subsection{Univalent Fibrations}

Functions between types are functors between groupoids, and type families (or functions to the universe) are indexed
families of groupoids. A type family $P : A \to \UU$ comes equipped with a functor $\pi_1 : \dsum*{x:A}{P(x)} \to A$
which has a lifting operation giving it the structure of a fibration. The $\term{transport}$ operation lifts paths in
the base space to functions between the fibers.  Using the groupoid structure of $A$, for any $x, y : A$ and a path 
$p : x \id_{A} y$, $\term{transport}(p)$ and $\term{transport}(\inv{p})$ form an equivalence.
\[
  \tptEqv{P} : (x \id_{A} y) \to (P(x) \eqv P(y))
\]

\noindent The type families (or fibrations) we are interested in are the ones where paths in the base space completely determine
the equivalences in the fibers -- these are called univalent
fibrations~\cite*{kapulkinUnivalenceSimplicialSets2018,kapulkinSimplicialModelUnivalent2021,christensenCharacterizationUnivalentFibrations2015}.

\begin{definition}[Univalent Fibration]
  $P$ is a univalent type family (or, $\fst : {\dsum*{x:A}{P(x)}} \to A$ is a univalent fibration) if $\tptEqv{P}$ is an
  equivalence.
\end{definition}

Univalent fibrations were introduced by~\citet*{kapulkinUnivalenceSimplicialSets2018}, to build a model of Voevodsky's
\emph{univalence} principle in simplicial sets. Indeed, univalence characterises paths in the universe as equivalences
between types, which follows from the canonical fibration $\idfunc : \UU \to \UU$ being univalent.

\begin{toappendix}

  \subsection{Homotopy Types}
  \label{app:homotopytypes}

  A type is \emph{contractible} (-2-type) if it has a unique element, that is, there is a center of contraction and every
  other point is equal to it. A type is a \emph{proposition} (-1-type) if its equality types are contractible, that is, it
  has at most one inhabitant. Iterating this, we can define sets or 0-types (whose equality types are propositions) and
  1-groupoids or 1-types (whose equality types are sets), and similarly, higher homotopy $n$-types.

  \begin{gather*}
    \begin{aligned}
      \isContr{A} & \defeq \dsum{x:A}{\dfun{y:A}{y \id x}} \\
      \isProp{A}  & \defeq \dfun{x,y:A}{\isContr{x \id y}}
    \end{aligned}
    \qquad
    \begin{aligned}
      \isSet{A} & \defeq \dfun{x,y:A}{\isProp{x \id y}} \\
      \isGpd{A} & \defeq \dfun{x,y:A}{\isSet{x \id y}}
    \end{aligned}
  \end{gather*}

  \subsection{Higher Inductive Types}
  \label{app:hits}

  Higher Inductive Types generalise Inductive Types, by allowing path constructors besides point constructors. While point
  constructors generate the elements of the type, path constructors generate equalities between points in the type. We
  describe a few basic HITs that we use.

  Given a type $A$, the propositional truncation $\Trunc[-1]{A}$, squashes the elements of $A$ turning it into a
  proposition. It is given by a HIT with a point constructor $\trunc{\blank} : A \to \Trunc{A}$, and a path constructor
  $\term{trunc}(x,y) : x \id_{\Trunc{A}} y$, which equates every pair of points in the truncation
  (see~\cref{def:prop-trunc}).

  \begin{definition}[Propositional Truncation]
    \label{def:prop-trunc}
    Given a type $A$, the propositional truncation $\Trunc[-1]{A}$, or simply $\Trunc{A}$, is a higher inductive type
    generated by the following constructors,
    \begin{itemize}
      \item an inclusion function $\trunc{\blank} : A \to \Trunc{A}$,
      \item for each $x, y : \Trunc{A}$, a path $\term{trunc}(x,y) : x \id_{\Trunc{A}} y$,
    \end{itemize}
    such that, given any type $B$ with
    \begin{itemize}
      \item a function $g : A \to B$,
      \item for each $x, y : B$, a path $\term{trunc*}(x,y) : x \id_{B} y$,
    \end{itemize}
    there is a unique function $f : \Trunc{A} \to B$ such that,
    \begin{itemize}
      \item $f(\trunc{a}) \equiv g(a)$
      \item for each $x, y : \Trunc{A}$, $\ap{f}{\term{trunc}(x,y)} \id_{B} \term{trunc*}(f(x),f(y))$.
    \end{itemize}
  \end{definition}

  \begin{definition}[${\fib_{f}} : B \to \UU$]
    \label{def:fib}
    The fiber of $f : A \to B$ at $b : B$ is
    \[
      \fib_{f}(b) \defeq \dsum{a:A}{f(a) \id_{B} b}.
    \]
  \end{definition}

  \begin{definition}[${\term{im}} : (f : A \to B) \to \UU$]
    \label{def:im}
    The image of $f$ is the (-1)-truncation of its fiber.
    \[
      \im{f} \defeq \dsum{b:B}{\Trunc[-1]{\fib_{f}(b)}}
    \]
  \end{definition}

  \begin{lemma}
    The following are equivalent.
    \begin{enumerate}
      \item $f : A \to B$ is an equivalence.
      \item $f$ has a left and right inverse.
      \item $f$ has contractible fibers.
    \end{enumerate}
  \end{lemma}

  Another HIT that we use is the set-quotient $\quot{A}{R}$ which takes an set $A$ and a relation $R : A \to A \to \UU$.
  It has an inclusion of points $\quotinc : A \to \quot{A}{R}$, and adds paths between related pairs of elements
  $\quotrel : R(x,y) \to \quotinc(x) \id_{\quot{A}{R}} \quotinc(y)$ (see~\cref{def:set-quot}).  We recall that the
  quotient is \emph{effective} if $R$ is a prop-valued equivalence relation, that is, $R(x,y)$ holds iff $(q(x)
  \id_{\quot{A}{R}} q(y))$.

  \begin{definition}[Set Quotient]
    \label{def:set-quot}
    Given a type $A$ which is a set, and a relation $R : A \to A \to \UU$, the set-quotient $\quot{A}{R}$ is the higher
    inductive type generated by
    \begin{itemize}
      \item an inclusion function $\quotinc : A \to \quot{A}{R}$,
      \item for each $x, y : A$ such that $R(x,y)$, a path $\quotinc(x) \id_{\quot{A}{R}} \quotinc(y)$,
      \item a set truncation, for each $x, y : \quot{A}{R}$ and $r, s : x \id_{\quot{A}{R}} y$, we have $r \id s$,
    \end{itemize}
    with an appropriate induction principle.
  \end{definition}

\end{toappendix}

\subsection{Univalent Subuniverses}

Starting from a univalent universe which classifies all types, we want to define a subuniverse which classifies only
certain types, for example, types that satisfy some desired property. We use a prop-valued type family, that is, a
predicate on the universe, which picks out only those types, and collect them into a univalent subuniverse. Being
univalent ensures that the equality type of the ambient universe is reflected in the subuniverse.

\begin{definition}[Universe]
  A universe \`{a} la Tarski is given by the following pieces of data,
  \begin{itemize}
    \item a code $U : \UU$,
    \item a decoding type family $\El : U \to \UU$.
  \end{itemize}
  If $\El$ is univalent, we call $(U,\El)$ a \emph{univalent} universe.
\end{definition}

\begin{propositionrep}[Univalent Subuniverse]
  \label{prop:univsub}
  A universe predicate is a type family $P : \UU \to \UU$ whose fibers are propositions, that is, $P(X)$ is a
  proposition for every $X$. Given such a predicate $P$, the fibration ${\fst : \dsum*{X:\UU}{P(X)} \to \UU}$ is
  univalent and generates a univalent subuniverse ${\UU_{P} \defeq (\dsum*{X:\UU}{P(X)}, \fst)}$.~\footnote{Univalent
    typoids~\cite{petrakisUnivalentTypoids2019a} are a different presentation of univalent subuniverses.}
\end{propositionrep}

\begin{proof}
  Suppose $(U, \El) \defeq (\dsum*{X:\UU}{P(X)}, \fst)$ is a subuniverse generated by a subtype $P : \UU \to \UU$. For
  any $X, Y : \UU$ such that $\phi : P(X)$ and $\psi : P(Y)$, we want to show that $\tptEqv{\fst} : (X,\phi) \id
    (Y,\psi) \to X \eqv Y$ is an equivalence. We construct $X \eqv Y \to (X,\phi) \id (Y,\psi)$ by $\ua$ and using the
  fact that $P(\blank)$ is a proposition. That it is an inverse follows by calculation using the appropriate computation
  rules.
\end{proof}

The types we are interested in are the finite types. In constructive mathematics, the notion of finiteness is
subtle~\cite{spiwackConstructivelyFinite2010}. We use the notion of Bishop-finiteness: a type is finite if it is merely
equivalent to a finite set (\cref*{def:finite-set,def:isfin}).

\begin{definition}[$\Fin$]
  \label{def:finite-set}
  The type family $\Fin : \Nat \to \UU$ is the type of finite sets indexed by their cardinality. It is defined
  equivalently in two different ways,
  \begin{gather*}
    \Fin[n] \defeq \dsum*{k:\Nat}{k < n}
    \qquad\qquad \text{or} \qquad\qquad
    \begin{aligned}
       & \Fin[0] \defeq \bot                      \\
       & \Fin[\suc[n]] \defeq \top \sqcup \Fin[n]
    \end{aligned}
  \end{gather*}
  Note that $\Fin[n]$ is a set, and we use both definitions interchangeably.
\end{definition}

\begin{definition}[$\isFin$]
  \label{def:isfin}
  We say that a type is finite if it is merely equal to $\Fin[n]$ for some $n : \Nat$.
  \[
    \isFin[X] \defeq \dsum*{n:\Nat}{\SubP{X}{\Fin[n]}}
  \]
  Note that the natural number $n$ need not be truncated, as justified below.
\end{definition}

\begin{lemmarep}
  \label{prop:isFin}
  For any type $X$, $\isFin[X]$ is a proposition.
\end{lemmarep}

\begin{proof}
  Suppose we have $(n,\phi) : \isFin[X]$ and $(m,\psi) : \isFin[X]$, we need to show that $(n,\phi) \id (m,\psi)$. It is
  enough to show that $n \id m$. Since $\Nat$ is a set, this is a proposition, so we can use the induction principle of
  propositional truncation to eliminate to $n \id m$, applying it on $\phi$ and $\psi$ respectively. This gives us the
  equalities $X \id \Fin[n]$ and $X \id \Fin[m]$, which gives us $\Fin[n] \id \Fin[m]$, from which $n \id m$ follows by
  applying the first projection.
\end{proof}

Since $\isFin$ is a predicate on the universe $\UU$, we easily get our univalent subuniverse $\UFin$.

\begin{definition}
  \label{def:ufin}
  The univalent subuniverse of \emph{all finite types} is given by
  $
    \UFin \defeq \dsum*{X:\UU}{\isFin[X]}.
  $
  We write $F_{n} \defeq (\Fin[n], n, \trunc{\refl})$, for the image of the inclusion of $\Fin[n]$.
\end{definition}

This definition of the groupoid of finite types has also been considered
in~\cite{yorgeyCombinatorialSpeciesLabelled2014}. While $\UFin$ has \emph{all} the finite types, we are also interested
in constructing a subuniverse of finite types of a specified cardinality. To do so, we will start with the subuniverse
$\BAut[T]$, for any type $T : \UU$. \footnote{Characterisations of univalent fibrations using the $\BAut$ construction
have been studied by~\citet{christensenCharacterizationUnivalentFibrations2015}.}

\begin{definition}[$\BAut$]
  The predicate $P(X) \defeq \Trunc[-1]{X \id T}$ picks out exactly those types that are merely equal to $T$, and this
  generates the subuniverse
  \[
    \BAut[T] \defeq \Sub{T}.
  \]
  We write $T_0 \defeq (T, \trunc{\refl_{T}})$ for the image of the inclusion of $T$ in $\BAut[T]$.
\end{definition}

Using $\BAut$, we can talk about types that are equivalent to a finite set of specified cardinality, for example, the
subuniverse of 2-element sets is given by $\BAut[\Bool]$. This has been used to construct the real projective spaces in
HoTT~\cite{buchholtzRealProjectiveSpaces2017}, and also to give the denotational semantics for a 1-bit reversible
programming language~\cite{caretteReversibleProgramsUnivalent2018}.

\begin{definition}[${\UFin[n]}$]
  For any $n : \Nat$, we define $\UFin[n] \defeq \BAut[\Fin[n]]$ to be the univalent subuniverse of $n$-element sets.
  Note that, $\UFin$ can be equivalently seen as the collection of all types of finite cardinality, that is,
  $\UFin \eqv \dsum*{n:\Nat}{\UFin[n]}$.
\end{definition}

Since $\BAut[T]$ is a univalent subuniverse, we can characterise its path space. The intuition is that $\BAut[T]$ only
has one point $T_0$, and 1-paths $T_0 \id T_0$, that is, loops, and higher paths between these loops. The
type of loops on $T_0$, $\loopspace[\BAut[T],T_{0}]$, is shown to be equivalent to $\Aut[T] \defeq T \eqv T$, which is
the group of automorphisms of $T$.

\begin{lemmarep}
  \leavevmode
  \begin{enumerate}
    \item If $T$ is an $n$-type, $\BAut[T]$ is an $(n+1)$-type.
    \item For any $T : \UU$, $\BAut[T]$ is 0-connected.
    \item For any $T : \UU$, \( \loopspace[\BAut[T],T_{0}] \eqv \Aut[T] \). \label{lem:loop-deloop}
  \end{enumerate}
\end{lemmarep}

\begin{proof}
  We need to show that the equality type of $\BAut[T]$ is an $n$-type. Assume $X, Y : \BAut[T]$. Since $\BAut[T]$ is a
  univalent subuniverse, we have $(X \id Y) \eqv (\fst(X) \eqv \fst(Y))$. Note that being an $n$-type is a proposition.
  Since $T$ is an $n$-type, and $\fst(X)$ and $\fst(Y)$ are merely equal to $T$, they're also $n$-types. It follows that
  $\fst(X) \eqv \fst(Y)$ is an $n$-type, and hence $X \id Y$ is an $n$-type.

  Since $\BAut[T]$ is a univalent universe, it follows that
  \[
    (T_{0} \id_{\BAut[T]} T_{0}) \eqv (\fst(T_{0}) \eqv \fst(T_{0})) \equiv (T \eqv T) \equiv \Aut[T].
  \]
\end{proof}

\begin{theorem}
  $\UFin[n]$ is a pointed, connected, 1-groupoid for every $n:\Nat$, and \( \loopspace[\UFin[n],F_{n}] \eqv
  \Aut[\Fin[n]] \). $\UFin$ is a 1-groupoid with connected components for every $n:\Nat$.
\end{theorem}

We have shown that loops in $\UFin$ exactly encode the automorphism group $\Aut[\Fin[n]]$ for every~$n$. This is a
general technique called \emph{delooping}, where a group can be identified with a 1-object groupoid, internally in HoTT.
This technique also allows defining higher groups~\cite{buchholtzHigherGroupsHomotopy2018}. The loopspace of a pointed
type automatically has the structure of a group, with $\refl_{\pt}$ for the neutral element, path composition for the
group multiplication, and path inverse for the group inverse operation. The group axioms are given by the higher paths
corresponding to groupoid laws.


\subsection{Rig structure}
\label{subsec:rig}

Similar to $\BFin$, the groupoid $\UFin$ has two symmetric monoidal structures, the additive and the multiplicative ones,
and the multiplicative tensor product distributes over the additive one. To construct these, we first state and prove
some equivalences on $\Fin$, and some general type isomorphisms. Then we simply lift these equivalences to $\UFin$, by
the univalence principle.

\begin{proposition}
  For any $n, m : \Nat$, and for any types $X, Y, Z$,
  \begin{gather*}
    \begin{aligned}
      \Fin[0]                & \eqv \bot                  \\
      \Fin[n] \sqcup \Fin[m] & \eqv \Fin[n + m]           \\
      \\
      \bot \sqcup X          & \eqv X                     \\
      X \sqcup \bot          & \eqv X                     \\
      (X \sqcup Y) \sqcup Z  & \eqv X \sqcup (Y \sqcup Z) \\
      X \sqcup Y             & \eqv Y \sqcup X            \\
      X \times \bot          & \eqv \bot
    \end{aligned}
    \qquad
    \begin{aligned}
      \Fin[1]                & \eqv \top                             \\
      \Fin[n] \times \Fin[m] & \eqv \Fin[n * m]                      \\
      \\
      \top \times X          & \eqv X                                \\
      X \times \top          & \eqv X                                \\
      (X \times Y) \times Z  & \eqv X \times (Y \times Z)            \\
      X \times Y             & \eqv Y \times X                       \\
      X \times (Y \sqcup Z)  & \eqv (X \times Y) \sqcup (X \times Z)
    \end{aligned}
  \end{gather*}
\end{proposition}

\begin{theorem}
  $\UFin$ has two symmetric monoidal structures, the additive and multiplicative ones, given by $(F_0, \sqcup)$ and
  $(F_1, \times)$, with corresponding natural isomorphisms $\lambda_{X}$, $\rho_{X}$, $\alpha_{X,Y,Z}$, and the braiding
  isomorphism $\mathcal{B}_{X,Y}$ upto 1-paths in $\UFin$. These isomorphisms satisfy the Mac Lane coherence conditions
  for symmetric monoidal categories~\cite{maclaneNaturalAssociativityCommutativity1963}, that is, the triangle,
  pentagon, and hexagon identities, and the symmetry of the braiding, upto 2-paths in $\UFin$. The multiplicative
  structure distributes over the additive structure and satisfies the Laplaza coherence conditions for rig
  categories~\cite{laplaza72}.
\end{theorem}

\begin{toappendix}
  \begin{definition}[Additive symmetric monoidal structure]
    \label{def:additive}
    \begin{align*}
      O                 & \defeq F_{0}                                       \\
      X \oplus Y        & \defeq X \sqcup Y                                  \\
      \lambda_{X}       & : O \oplus X \eqv X                                \\
      \rho_{X}          & : X \oplus O \eqv X                                \\
      \alpha_{X,Y,Z}    & : (X \oplus Y) \oplus Z \eqv X \oplus (Y \oplus Z) \\
      \mathcal{B}_{X,Y} & : X \oplus Y \eqv Y \oplus X
    \end{align*}
  \end{definition}
  \begin{proposition}
    \label{prop:additive}
    \[\begin{tikzcd}
        {(X \oplus I) \oplus Y} && {X \oplus (I \oplus Y)} \\
        {} & {X \oplus Y}
        \arrow["{\alpha_{X,I,Y}}", from=1-1, to=1-3]
        \arrow[""{name=0, anchor=center, inner sep=0}, "{\rho_{X} \oplus 1_{Y}}"', from=1-1, to=2-2]
        \arrow[""{name=1, anchor=center, inner sep=0}, "{1_{X} \oplus \lambda_{Y}}", from=1-3, to=2-2]
        \arrow["\id", Rightarrow, draw=none, from=0, to=1]
      \end{tikzcd}\]
    \[\begin{tikzcd}
        & {(W \oplus X) \oplus (Y \oplus Z)} \\
        {((W \oplus X) \oplus Y) \oplus Z} && {W \oplus (X \oplus (Y \oplus Z))} \\
        \\
        {(W \oplus (X \oplus Y)) \oplus Z} && {W \oplus ((X \oplus Y) \oplus Z)}
        \arrow["{\alpha_{W \oplus X, Y, Z}}", from=2-1, to=1-2]
        \arrow["{\alpha_{W,X,Y \oplus Z}}", from=1-2, to=2-3]
        \arrow["{1_{W} \oplus \alpha_{X,Y,Z}}"', from=4-3, to=2-3]
        \arrow["{\alpha_{W,X,Y} \oplus 1_{Z}}"', from=2-1, to=4-1]
        \arrow["{\alpha_{W,X \oplus Y,Z}}"', from=4-1, to=4-3]
        \arrow["\id"{description}, shift right=5, draw=none, from=2-1, to=2-3]
      \end{tikzcd}\]
    \[\begin{tikzcd}
        & {X \oplus (Y \oplus Z)} \\
        {(X \oplus Y) \oplus Z} && {(Y \oplus Z) \oplus X} \\
        {(Y \oplus X) \oplus Z} && {Y \oplus (Z \oplus X)} \\
        & {Y \oplus (X \oplus Z)}
        \arrow["{\alpha_{X,Y,Z}}", from=2-1, to=1-2]
        \arrow[""{name=0, anchor=center, inner sep=0}, "{\mathcal{B}_{X,Y} \oplus 1_{Z}}"', from=2-1, to=3-1]
        \arrow["{\alpha_{Y,X,Z}}"', from=3-1, to=4-2]
        \arrow["{1_{Y} \oplus \mathcal{B}_{X,Z}}"', from=4-2, to=3-3]
        \arrow[""{name=1, anchor=center, inner sep=0}, "{\alpha_{Y,Z,X}}", from=2-3, to=3-3]
        \arrow["{\mathcal{B}_{X,Y \oplus Z}}", from=1-2, to=2-3]
        \arrow["\id", Rightarrow, draw=none, from=0, to=1]
      \end{tikzcd}\]
    \[\begin{tikzcd}
        {X \oplus Y} && {X \oplus Y} \\
        & {Y \oplus X}
        \arrow["{1_{X \oplus Y}}", Rightarrow, no head, from=1-1, to=1-3]
        \arrow[""{name=0, anchor=center, inner sep=0}, "{\mathcal{B}_{X,Y}}"', from=1-1, to=2-2]
        \arrow[""{name=1, anchor=center, inner sep=0}, "{\mathcal{B}_{Y,X}}"', from=2-2, to=1-3]
        \arrow["\id", Rightarrow, draw=none, from=0, to=1]
      \end{tikzcd}\]
  \end{proposition}
\end{toappendix}

\begin{toappendix}
  \begin{definition}[Multiplicative symmetric monoidal structure]
    \label{def:multiplicative}
    \begin{align*}
      I                 & \defeq F_{1}                                           \\
      X \otimes Y       & \defeq X \times Y                                      \\
      \lambda_{X}       & : I \times X \eqv X                                    \\
      \rho_{X}          & : X \times I \eqv X                                    \\
      \alpha_{X,Y,Z}    & : (X \otimes Y) \otimes Z \eqv X \otimes (Y \otimes Z) \\
      \mathcal{B}_{X,Y} & : X \otimes Y \eqv Y \otimes X
    \end{align*}
  \end{definition}
  \begin{proposition}
    \label{prop:multiplicative}
    \[\begin{tikzcd}
        {(X \otimes I) \otimes Y} && {X \otimes (I \otimes Y)} \\
        {} & {X \otimes Y}
        \arrow["{\alpha_{X,I,Y}}", from=1-1, to=1-3]
        \arrow[""{name=0, anchor=center, inner sep=0}, "{\rho_{X} \otimes 1_{Y}}"', from=1-1, to=2-2]
        \arrow[""{name=1, anchor=center, inner sep=0}, "{1_{X} \otimes \lambda_{Y}}", from=1-3, to=2-2]
        \arrow["\id", Rightarrow, draw=none, from=0, to=1]
      \end{tikzcd}\]
    \[\begin{tikzcd}
        & {(W \otimes X) \otimes (Y \otimes Z)} \\
        {((W \otimes X) \otimes Y) \otimes Z} && {W \otimes (X \otimes (Y \otimes Z))} \\
        \\
        {(W \otimes (X \otimes Y)) \otimes Z} && {W \otimes ((X \otimes Y) \otimes Z)}
        \arrow["{\alpha_{W \otimes X, Y, Z}}", from=2-1, to=1-2]
        \arrow["{\alpha_{W,X,Y \otimes Z}}", from=1-2, to=2-3]
        \arrow["{1_{W} \otimes \alpha_{X,Y,Z}}"', from=4-3, to=2-3]
        \arrow["{\alpha_{W,X,Y} \otimes 1_{Z}}"', from=2-1, to=4-1]
        \arrow["{\alpha_{W,X \otimes Y,Z}}"', from=4-1, to=4-3]
        \arrow["\id"{description}, shift right=5, draw=none, from=2-1, to=2-3]
      \end{tikzcd}\]
    \[\begin{tikzcd}
        & {X \otimes (Y \otimes Z)} \\
        {(X \otimes Y) \otimes Z} && {(Y \otimes Z) \otimes X} \\
        {(Y \otimes X) \otimes Z} && {Y \otimes (Z \otimes X)} \\
        & {Y \otimes (X \otimes Z)}
        \arrow["{\alpha_{X,Y,Z}}", from=2-1, to=1-2]
        \arrow[""{name=0, anchor=center, inner sep=0}, "{\mathcal{B}_{X,Y} \otimes 1_{Z}}"', from=2-1, to=3-1]
        \arrow["{\alpha_{Y,X,Z}}"', from=3-1, to=4-2]
        \arrow["{1_{Y} \otimes \mathcal{B}_{X,Z}}"', from=4-2, to=3-3]
        \arrow[""{name=1, anchor=center, inner sep=0}, "{\alpha_{Y,Z,X}}", from=2-3, to=3-3]
        \arrow["{\mathcal{B}_{X,Y \otimes Z}}", from=1-2, to=2-3]
        \arrow["\id", Rightarrow, draw=none, from=0, to=1]
      \end{tikzcd}\]
    \[\begin{tikzcd}
        {X \otimes Y} && {X \otimes Y} \\
        & {Y \otimes X}
        \arrow["{1_{X \otimes Y}}", Rightarrow, no head, from=1-1, to=1-3]
        \arrow[""{name=0, anchor=center, inner sep=0}, "{\mathcal{B}_{X,Y}}"', from=1-1, to=2-2]
        \arrow[""{name=1, anchor=center, inner sep=0}, "{\mathcal{B}_{Y,X}}"', from=2-2, to=1-3]
        \arrow["\id", Rightarrow, draw=none, from=0, to=1]
      \end{tikzcd}\]
  \end{proposition}
\end{toappendix}

\begin{toappendix}
  \begin{proposition}[Distributivity]
    \label{prop:distributivity}
    \begin{gather*}
      \begin{aligned}
        \delta_{l} : X \otimes (Y \oplus Z) & \eqv (X \otimes Y) \oplus (X \otimes Z) \\
        \delta_{r} : (X \oplus Y) \otimes Z & \eqv (X \otimes Z) \oplus (Y \otimes Z)
      \end{aligned}
      \qquad
      \begin{aligned}
        a_{l} : X \otimes O \eqv O \\
        a_{r} : O \otimes X \eqv O
      \end{aligned}
    \end{gather*}
  \end{proposition}
\end{toappendix}

\section{The Group of Permutations}
\label{sec:finite}

In~\Cref{sec:ufin}, we established that paths in $\UFin$ are equivalent to families of loops on $\Fin[n]$ for every
$n:\Nat$, that is, automorphisms of finite sets of size $n$, with the loopspace encoding the automorphism group.  This
is also known to be the finite symmetric group $\Sn[n]$, making $\UFin$ the \emph{horizontal categorification} of $\Sn$
for every $n$. In this section, we will describe this group syntactically.

In order to study syntactic descriptions of permutations, we will hit the problem of deciding whether two descriptions
refer to the same permutation -- in group theory, this is the \emph{word problem} for $\Sn$. Putting it in this form
allows us to connect it to the broader scope of computational group theory and combinatorics -- we can borrow ideas such
as Coxeter relations and Lehmer codes.~\footnote{The concepts we use can be found in any standard textbook on group
theory, or see~\cite{symmetryBook2021} for a univalent point of view.}

Thus, the goal of this section is to reconcile two different approaches to defining the symmetric group -- as an
automorphism group, and as a group syntactically presented using \emph{generators} and \emph{relations}. The generators
of the group are similar to the primitive combinators in a (reversible) programming language -- the group structure
gives the composition and inverse operations, and the relations describe how these primitive combinators interact with
each other.

First, we will define the required notions of free groups and group presentations, and state some of their most
important properties. Then, we introduce our chosen \emph{Coxeter presentation} for $\Sn$. To solve the word problem for
$\Sn$, we will use a rewriting system, with a suitable, well behaved collection of reduction rules corresponding to the
Coxeter presentation equations. Finally, we describe the normal forms in this rewriting system, using Lehmer codes, and prove the
correspondence between them and the type $\Aut[\Fin[n]]$ of automorphisms on a finite set. The generators and relations
we use here will be used to quote back to 1 and 2-combinators in $\PiLang$ (see~\cref{sec:equivalence}).

\subsection{Presenting the permutation group}

One way of thinking about presentations of $\Sn$ is via sorting algorithms, which use different primitive operations. A
sorting algorithm has to calculate a permutation of a list or a finite set, which satisfies the invariant of being a
sorted sequence, which means, the primitive operations of a sorting algorithm are able to generate all the permutations
on a given list. So, a chosen set of reversible operations in a sorting algorithm can be a good candidate for the
generators of a permutation group. For example, we could generate the permutation group on $\Fin[n]$ by using generators
(primitive operations) that:

\begin{itemize}
  \item swap the $i$-th element with the $(i+1)$-th element, that is, adjacent swaps, or
  \item swap the $i$-th element with the $j$-th element, for arbitrary $i$-s and $j$-s, or
  \item swap the $i$-th element with an element at a fixed position, or
  \item reverses a prefix $\Fin[k]$ of $\Fin[n]$ for $k \leq n$, or
  \item cyclically shift any subset of $\Fin[n]$.
\end{itemize}

Bubble sort uses the primitive operation of adjacent swaps, insertion sort and selection sort use the primitive
operation of swapping the $i$-th element with the $j$-th element, cycle sort uses cyclical shifts of subsequences,
pancake sort uses reversals of the prefixes of the list, et cetera. The choice of generators for our presentation is
important for the following reasons.

\begin{itemize}
  \item It affects the difficulty of solving the word problem in $\Sn$ and formalising the proof of its correctness.
  \item The choice of generators dictates which words become normal forms in this presentation of $\Sn$. These normal
        forms dictate the shape of the synthesised and normalised boolean circuits, which is the application we have in
        mind.
  \item Finally, the generators have to closely match the $\PiLang$ combinators so that we can quote back a permutation
        to a $\PiLang$ program, for the proof of completeness.
\end{itemize}

We will show that it is possible to encode all $\PiLang$ combinators using adjacent transpositions
(in~\cref{sec:equivalence}). Group presentations are built by adding equations to a \emph{free group}.

\begin{toappendix}
  \subsection{Groups}
  \label{subsec:groups}

  From universal algebra, a group is simply a set with a 0-ary constant $e$ (the neutral element), a binary operation
  $\blank\mult\blank$ for group multiplication, and a unary inverse operation $\inv{\blank}$. The neutral element has to
  satisfy unit and inverse laws, and the multiplication has to be associative (see~\cref{def:group}).

  A very simple example of a group is $\mathbb{Z}$, where the neutral element is 0, the inverse of $k$ is $-k$, and the
  group multiplication is given by integer addition.

  \begin{definition}[Group]
    \label{def:group}
    In type theory, a group $G$ can be defined as a set $S$ with the following pieces of data:

    \begin{enumerate}
      \item a unit or neutral element $e : S$
      \item a multiplication function $m : S \times S \to S$ written as $(g_{1}, g_{2}) \mapsto g_{1} \mult g_{2}$, that satisfies
            \begin{enumerate}
              \item the unit laws, for all $g : S$, that \( g \mult e \id g \) and \( e \mult g \id g \)
              \item the associativity law, for all $g_{1}, g_{2}, g_{3} : S$, that \( g_{1} \mult (g_{2} \mult g_{3}) \id (g_{1} \mult g_{2}) \mult g_{3} \)
            \end{enumerate}
      \item an inversion function $i : S \to S$ written as $g \mapsto \inv{g}$, that satisfies
            \begin{enumerate}
              \item the inverse laws, for all $g : S$, that \( g \mult \inv{g} \id e \) and \( \inv{g} \mult g \id e \)
            \end{enumerate}
    \end{enumerate}
  \end{definition}
\end{toappendix}

\subsection{Free groups}

Usually, there are are many equations, besides the group axioms, that hold for the elements of a group. For example, in
the group $\Aut[\Bool]$, or $\mathbb{Z}_2$, we have an equation $1 + 1 = 0$, which is not a consequence of the group
axioms, but is specific to this particular group. A free group has the property that no other equations hold except the
ones directly implied by the group axioms. For example, the additive group of integers $\mathbb{Z}$ is the free group on
the singleton set.

\begin{toappendix}
  \begin{definition}[Free group]
    \label{def:free-group}
    Given a set $A$, the free group $F(A)$ on it is given by a higher inductive type with the following point and path
    constructors. Notice the similarity with the definition of a group structure (\cref{def:group}), but note that each
    operation here is a generator for $F(A)$.
    \begin{itemize}
      \item An inclusion function $\eta_{A} : A \to F(A)$
      \item A multiplication function $m : F(A) \times F(A) \to F(A)$
      \item An element $e : F(A)$
      \item An inverse function $i : F(A) \to F(A)$
    \end{itemize}
    \smallskip
    \begin{itemize}
      \item For every $x, y, z : F(A)$, a path $\term{assoc} : m(x, m(y, z)) \id m(m(x, y), z)$
      \item For every $x : F(A)$, paths $\term{unitr} : m(x, e) \id x$ and $\term{unitl} : m(e, x) \id x$
      \item For every $x : F(A)$, paths $\term{invr} : m(x, i(x)) \id e$ and $\term{invl} : m(i(x), x) \id e$
      \item A 0-truncation, for every $x, y : F(A)$ and $p, q : x \id y$, a 2-path $\term{trunc} : p \id q$
    \end{itemize}
  \end{definition}
\end{toappendix}

A group homomorphism between groups $G$ and $H$ is a function $f : G \to H$ between the underlying sets that preserves
the group structure. Giving a group homomorphism out of the free group is equivalent to giving a function out of the
generating set. This is the universal property of free groups, stemming from the free-forgetful adjunction between the
category of groups and sets.


\begin{propositionrep}[Universal Property of $F(A)$]
  \label{prop:free-groups}
  Given a group $G$ and a map $f : A \to G$, there is a unique group homomorphism $\extend{f} : \term{Hom}(F(A), G)$
  such that $\extend{f} \comp \eta_A \htpy f$. Equivalently, composition with $\eta_A$ gives an equivalence
  $\term{Hom}(F(A),G) \eqv A \to G$.
\end{propositionrep}

\begin{toappendix}
  Alternatively, the type of group homomorphisms $h : \term{Hom}(F(A),G)$ satisfying $h \comp \eta_A \htpy f$ is
  contractible.

  \[\begin{tikzcd}
      {F(A)} && G \\
      \\
      A
      \arrow[""{name=0, anchor=center, inner sep=0}, "{\eta_A}", from=3-1, to=1-1]
      \arrow[""{name=1, anchor=center, inner sep=0}, "f"', from=3-1, to=1-3]
      \arrow["{\extend{f}}", dashed, from=1-1, to=1-3]
      \arrow["\id", Rightarrow, draw=none, from=0, to=1]
    \end{tikzcd}\]
\end{toappendix}

Following the universal-algebraic definition, in HoTT, we could use a naive higher inductive type to define the free
group, which enforces the group axioms by adding path constructors (see~\cref{def:free-group}). Using the induction
principle, we can easily verify the universal property. However, since this definition of $F(A)$ has lots of path
constructors corresponding to each group axiom, characterising its path space is difficult.

Instead, we will think about elements of the free group as words over an alphabet of letters drawn from the generating
set \emph{and} the set of their formal inverses. If we take the disjoint union of $A$ with itself, that is, $A + A$ as
the group's underlying set, we can use $\inl/\inr$ to mark the elements -- $\inl{a}$ means $a$ and $\inr{a}$ means
$\inv{a}$. Then, we can encode the free group using the free monoid, that is, lists of $A + A$. Additionally, we need to
ensure that the inverse laws hold, so we have to coalesce adjacent occurences of $a$ and $\inv{a}$.

\begin{definition}[Free group]
  \label{def:presentation}
  Let $A$ be a set, and $\List[\blank]$ the free monoid. The free group $F(A)$ on $A$ is the set-quotient of $\List[A +
      A]$ by the congruence closure of the relation $a \cons \inv{a} \cons \nil \sim \nil$ and $\inv{a} \cons a \cons \nil
    \sim \nil$.
\end{definition}

\begin{proposition}
  $F(A) \defeq \quot{\List[A + A]}{\sim^{\ast}}$ has a group structure, with the empty list $\nil$ for the neutral
  element, multiplication given by list append $\append$, and inverse given by flipping $\inl$ and $\inr$, followed by
  reversing the list. Further, $F(A)$ with $\eta_A(a) \defeq \inl(a) \cons \nil$ satisfies the universal property of
  free groups, as stated in~\Cref{prop:free-groups}.
\end{proposition}

\subsection{Group presentations}

A presentation of a group builds it by starting from the free group $F(A)$ and introducing additional equations that are
satisfied in the resulting group. For example, if we take $F(\unit) \defeq \mathbb{Z}$ and add an equation $1 + 1 = 0$,
the resulting group would be $\mathbb{Z}_2 \eqv \Aut[\Bool]$. Note that not all groups have finite (or computable)
presentations, and, a group can have any number of different presentations.

\begin{definition}[Group presentation]
  Let $A$ be a set and $R : \List[A + A] \to \List[A + A] \to \UU$ a binary relation on $\List[A + A]$. The group
  $F(\langle A ; R \rangle)$ presented by $A$ and $R$, is given by the set-quotient of the free group $F(A)$ by the
  congruence closure of $R$.
\end{definition}

The universal property of the above definition is similar to~\Cref{prop:free-groups} except the relation has to be
preserved by the function mapping out of the generating set.

\begin{proposition}[Universal property of $F(\langle A ; R \rangle)$]
  Given a group $G$ and a map $f : A \to G$, such that $f$ extended to $F(A)$ respects $R$, there is a unique group
  homomorphism $\extend{f} : \term{Hom}(F(\langle A ; R \rangle), G)$ such that $\extend{f} \comp \eta_A \htpy f$.
\end{proposition}

Before, the only way to decide the equality of two elements in a group was to evaluate and check them on the nose, but
in a group presentation, this is reduced to deciding whether one word -- a representative of the equivalence class of
the group's elements, can be reduced to another word, using the group's relations. However, these equations are not
directed, so it is not always possible to construct a well-behaved rewriting system. In general, the word problem for
groups is undecidable.

\paragraph{Coxeter Presentation} To present the group $\Sn$, the primitive operations we use will be adjacent swaps.
When dealing with permutations on an $n + 1$-element set, there are $n$ adjacent transpositions -- transposition number
$k$ swaps elements at indices $k$ and $k+1$. Thus, the generating set is $\Fin[n]$. There are three relations that we're
going to specify for this presentation -- we visualise them as braid diagrams in~\Cref{fig:coxeter-braid}.

\begin{enumerate}[leftmargin=*]
  \item[\labelcref{fig:coxeter-a}] Swapping the same two elements twice in a row should be the same as doing nothing.
  \item[\labelcref{fig:coxeter-b}] When swapping two distinct pairs of elements, the order in which swapping happens
        should not matter, that is, we can slide the wires freely.
  \item[\labelcref{fig:coxeter-c}] There are two equivalent ways of swapping the first and last elements in a sequence
        of three elements.
\end{enumerate}

\begin{figure}
  \centering
  \begin{subfigure}[b]{0.25\textwidth}
    \centering
    \begin{tikzpicture}
      \pic[local bounding box=my braid,braid/.cd,
        number of strands = 2,
        width = 0.4cm,
        height = 0.4cm,
        border height = 0.3cm,
        thick] at (0, 0)
      {braid={ s_1, s_1}};
      \node[font=\large] at (1, 0.7) {\(=\)};
      \pic[local bounding box=my braid,braid/.cd,
        number of strands = 2,
        width = 0.4cm,
        height = 0.4cm,
        border height = 0.3cm,
        thick] at (1.6, 0)
      {braid={ 1, 1}};
    \end{tikzpicture}
    \caption{$\cancel$}
    \label{fig:coxeter-a}
  \end{subfigure}
  \begin{subfigure}[b]{0.4\textwidth}
    \centering
    \begin{tikzpicture}
      \pic[local bounding box=my braid,braid/.cd,
        number of strands = 2,
        width = 0.4cm,
        height = 0.4cm,
        border height = 0.3cm,
        thick] at (0, 0)
      {braid={ s_1, 1}};
      \node[] at (0.7, 0.7) {\(\dots\)};
      \pic[local bounding box=my braid,braid/.cd,
        number of strands = 2,
        width = 0.4cm,
        height = 0.4cm,
        border height = 0.3cm,
        thick] at (1, 0)
      {braid={ 1, s_1 }};
      \node[font=\large] at (1.9, 0.7) {\(=\)};
      \pic[local bounding box=my braid,braid/.cd,
        number of strands = 2,
        width = 0.4cm,
        height = 0.4cm,
        border height = 0.3cm,
        thick] at (2.4, 0)
      {braid={ 1, s_1}};
      \node[] at (3.1, 0.7) {\(\dots\)};
      \pic[local bounding box=my braid,braid/.cd,
        number of strands = 2,
        width = 0.4cm,
        height = 0.4cm,
        border height = 0.3cm,
        thick] at (3.4, 0)
      {braid={ s_1, 1}};
    \end{tikzpicture}
    \caption{$\swap$}
    \label{fig:coxeter-b}
  \end{subfigure}
  \begin{subfigure}[b]{0.25\textwidth}
    \centering
    \begin{tikzpicture}
      \pic[local bounding box=my braid,braid/.cd,
        number of strands = 3,
        width = 0.4cm,
        height = 0.4cm,
        border height = 0.1cm,
        thick] at (0, 0)
      {braid={ s_2, s_1, s_2}};
      \node[font=\large] at (1.4, 0.7) {\(=\)};
      \pic[local bounding box=my braid,braid/.cd,
        number of strands = 3,
        width = 0.4cm,
        height = 0.4cm,
        border height = 0.1cm,
        thick] at (2, 0)
      {braid={ s_1, s_2, s_1}};
    \end{tikzpicture}
    \caption{$\braid$}
    \label{fig:coxeter-c}
  \end{subfigure}
  \caption{Braiding diarams for Coxeter relations.}
  \label{fig:coxeter-braid}
\end{figure}

This construction is called a Coxeter presentation of $\Sn$. Writing it formally, we encode the rules discussed above
using a relation $\cox$ on $\List[\Fin[n]]$ (\cref{def:cox}), and then take its congruence closure $\cox*$
(\cref{def:coxstar}).

\begin{definition}[$\cox : {\List[\Fin[n]]} \to {\List[\Fin[n]]} \to {\UU}$]
  \label{def:cox}
  \begin{align*}
    \cancel
     & : \forall n \to (n \cons n \cons \nil) \cox \nil                                                     \\
    \swap
     & : \forall k, n \to (\suc[k] < n) \to (n \cons k \cons \nil) \cox (k \cons n \cons \nil)              \\
    \braid
     & : \forall n \to (\suc[n] \cons n \cons \suc[n] \cons \nil) \cox (n \cons \suc[n] \cons n \cons \nil)
  \end{align*}
\end{definition}

\begin{toappendix}
  \begin{definition}[$\cox* : {\List[\Fin[n]]} \to {\List[\Fin[n]]} \to {\UU}$]
    \label{def:coxstar}
    \begin{align*}
      \reflr{\cox}
       & : \forall w \to w \cox* w                                                                                                           \\
      \symr{\cox}
       & : \forall w_{1}, w_{2} \to w_{1} \cox* w_{2} \to w_{2} \cox* w_{1}                                                                  \\
      \transr{\cox}
       & : \forall w_{1}, w_{2}, w_{3} \to  w_{1} \cox* w_{2} \to w_{2} \cox* w_{3} \to w_{1} \cox* w_{3}                                    \\
      \congrf{\cox}{\append}
       & : \forall w_{1}, w_{2}, w_{3}, w_{4} \to  w_{1} \cox* w_{2} \to w_{3} \cox* w_{4} \to w_{1} \append w_{3} \cox* w_{2} \append w_{4} \\
      \relr{\cox}
       & : \forall w_{1}, w_{2} \to w_{1} \cox w_{2} \to w_{1} \cox* w_{2}
    \end{align*}
  \end{definition}
\end{toappendix}

The idea for solving the word problem for $\Sn$ is to turn these undirected relations into a rewriting
system $(\List[\Fin[n]],\cox*)$, so that, by repeatedly applying the reduction rules as long as possible, any two
$\cox*$-equal terms would eventually converge to the same normal form.

For this to work, we first need the system to have the termination property, meaning that there are no infinite
reductions. We observe that after throwing out reflexivity and symmetry, the right hand sides of the relations $\cox*$
are strictly smaller than the left hand sides, in terms of the lexicographical ordering on words in $\Fin[n]$ (which is
well-founded). Thus, by directing the relation from left to right, we would get the termination property out of the box.
Second, we need the normal forms to be unique, so that we can get a normalisation function. This will be true if the
rewriting system is confluent -- meaning that all critical pairs, that is, terms with overlapping possible reduction
rules, have to converge. For example, in our system, the pair in~\Cref{fig:critical-pairs-converging} converges.
Unfortunately, this is not true for all critical pairs -- an  example is in~\Cref{fig:critical-pairs-non-converging},
where left and right endpoints are normal with respect to the $\cox*$ relation.

\begin{figure}
  \centering
  \begin{subfigure}[b]{0.35\textwidth}
    \adjustbox{scale=\scalef,center}{
      \begin{tikzcd}
        & {\UOLoverline{\tau_2\tau_1}[\tau_2]\UOLunderline{\tau_1\tau_2}} \\
        {\tau_1\tau_2\overline{\tau_1\tau_1}\tau_2} && {\tau_2\underline{\tau_1\tau_1}\tau_2\tau_1} \\
        {\tau_1\overline{\tau_2\tau_2}} && {\underline{\tau_2\tau_2}\tau_1} \\
        & {\tau_1}
        \arrow["\braid"', squiggly, from=1-2, to=2-1]
        \arrow["\braid", squiggly, from=1-2, to=2-3]
        \arrow["\cancel"', squiggly, from=2-1, to=3-1]
        \arrow["\cancel"', squiggly, from=3-1, to=4-2]
        \arrow["\cancel", squiggly, from=2-3, to=3-3]
        \arrow["\cancel", squiggly, from=3-3, to=4-2]
      \end{tikzcd}
    }
    \caption{Converging critical pair}
    \label{fig:critical-pairs-converging}
  \end{subfigure}
  \begin{subfigure}[b]{0.55\textwidth}
    \adjustbox{scale=\scalef,center}{
      \begin{tikzcd}
        & {\UOLoverline{\tau_3\tau_2}[\tau_3]\UOLunderline{\tau_1}} \\
        &&& {\underline{\tau_3\tau_2\tau_1\tau_3}} \\
        {\tau_2\tau_3\tau_2\tau_1}
        \arrow["\braid"', squiggly, from=1-2, to=3-1]
        \arrow["\swap", squiggly, from=1-2, to=2-4]
        \arrow["\longbraid", squiggly, from=2-4, to=3-1]
      \end{tikzcd}
    }
    \caption{Non-converging critical pair that converges with $\longbraid$}
    \label{fig:critical-pairs-non-converging}
  \end{subfigure}

  \caption{Critical Pairs in $\cox*$}
  \label{fig:critical-pairs}
\end{figure}

\subsection{Rewriting via Coxeter}

Because of this counter-example, the relations have to be changed. In this section, we formally define a rewriting
system $(\List[\Fin[n]], \longcox)$, partially based on the Coxeter relations, and prove that it has the desired
properties of confluence and termination. We prove that the new relation defined by this system is equivalent, in a
technical sense, to the standard Coxeter relation $\cox*$. First, we need to define a function $n \downf k$.

\begin{definition}[$\downf : (n : \Nat) \to (k : \Nat) \to {\List[\Fin[k + n]]}$]
  \begin{align*}
    n \downf \zero   & \defeq \nil                       \\
    n \downf \suc[k] & \defeq (k + n) \cons (n \downf k)
  \end{align*}
\end{definition}

\noindent The result of this function is the sequence \([k + n - 1, k + n - 2, k + n - 3, \ldots, n]\), which is a
sequence of transpositions that moves the element at index $k + n$ left by $k$ places, shifting all the elements in
between one place right (see~\cref{fig:downf}). Then, the directed relation $\longcox$ is defined with the following
generators (inlining the congruence closure in $\longcox$, allowing arbitrary reductions inside the list).

\begin{definition}[$\longcox : {\List[\Fin[n]]} \to {\List[\Fin[n]]} \to {\UU}$]
  \begin{align*}
    \longcancel
     & : \forall n, l, r \to (l \append n \cons n \cons r) \longcox (l \append r)                                                                      \\
    \longswap
     & : \forall n, k, l, r \to (\suc[k] < n) \to (l \append n \cons k \cons r) \longcox (l \append k \cons n \append r)                               \\
    \longbraid
     & : \forall n, k, l, r \to (l \append (n \downf 2 + k) \append (1 + k + n) \cons r) \longcox (l \append (k + n) \cons (n \downf 2 + k) \append r)
  \end{align*}
\end{definition}

\noindent Constructors $\longcancel$ and $\longswap$ correspond directly to the appropriate constructors of $\cox$ and
can be visualised in the same way as before. The remaining constructor $\longbraid$ uses the $\downf$ function to
exchange the order of a long sequence of transpositions and a single transposition afterwards. For example, for $n = 0$
and $k = 3$, it allows for the reduction $[4, 3, 2, 1, 0, 4] \longcox [3, 4, 3, 2, 1, 0]$ (see~\cref{fig:longcox} --
note the distinction between wires, where numbers represent the values, and transpositions, where numbers represent
which wires are crossing).

\begin{figure}
  \begin{subfigure}[b]{0.3\textwidth}
    \centering
    \begin{tikzpicture}
      \def\nstrandsdf{7}
      \pic[local bounding box=my braid,braid/.cd,
        number of strands = \nstrandsdf,
        width = 0.4cm,
        height = 0.4cm,
        border height = 0.1cm,
        thick,
        name prefix=braid]
      {braid={s_1, s_2, s_3, s_4, s_5, s_6}};
      \foreach \n in {1,...,\nstrandsdf}{
          \pgfmathtruncatemacro{\nminusone}{\n - 1}
          \node at (braid-\n-s)[yshift = 2.8cm] {\nminusone};
        }
      \node at (braid-1-e)[yshift = -2.8cm] {5};
      \node at (braid-2-e)[yshift = -2.8cm] {6};
      \node at (braid-3-e)[yshift = -2.8cm] {0};
      \node at (braid-4-e)[yshift = -2.8cm] {1};
      \node at (braid-5-e)[yshift = -2.8cm] {2};
      \node at (braid-6-e)[yshift = -2.8cm] {3};
      \node at (braid-7-e)[yshift = -2.8cm] {4};
    \end{tikzpicture}
    \caption{$0 \downf 6$}
    \label{fig:downf}
  \end{subfigure}
  \hfill
  \begin{subfigure}[b]{0.6\textwidth}
    \centering
    \begin{tikzpicture}
      \def\nstrandsbl{6}
      \pic[local bounding box=my braid,braid/.cd,
        number of strands = \nstrandsbl, 
        width = 0.4cm,
        height = 0.4cm,
        border height = 0.1cm,
        thick,
        name prefix=braid] at (0, 0)
      {braid={s_5, s_1, s_2, s_3, s_4, s_5}};
      \node[font=\large] at (3, 1.3) {\(\rightarrow\)};
      \foreach \n in {1,...,\nstrandsbl}{
          \pgfmathtruncatemacro{\nminusone}{\n - 1}
          \node at (braid-\n-s)[yshift = 2.8cm] {\nminusone};
        }
      \node at (braid-1-e)[yshift = -2.8cm] {3};
      \node at (braid-2-e)[yshift = -2.8cm] {5};
      \node at (braid-3-e)[yshift = -2.8cm] {0};
      \node at (braid-4-e)[yshift = -2.8cm] {1};
      \node at (braid-5-e)[yshift = -2.8cm] {4};
      \node at (braid-6-e)[yshift = -2.8cm] {2};

      \def\nstrandsbr{6}
      \pic[local bounding box=my braid,braid/.cd,
        number of strands = \nstrandsbr,
        width = 0.4cm,
        height = 0.4cm,
        border height = 0.1cm,
        thick,
        name prefix=braid] at (4, 0)
      {braid={s_1, s_2, s_3, s_4, s_5, s_4}};
      \foreach \n in {1,...,\nstrandsbl}{
          \pgfmathtruncatemacro{\nminusone}{\n - 1}
          \node at (braid-\n-s)[yshift = 2.8cm] {\nminusone};
        }
      \node at (braid-1-e)[yshift = -2.8cm] {3};
      \node at (braid-2-e)[yshift = -2.8cm] {5};
      \node at (braid-3-e)[yshift = -2.8cm] {0};
      \node at (braid-4-e)[yshift = -2.8cm] {1};
      \node at (braid-5-e)[yshift = -2.8cm] {4};
      \node at (braid-6-e)[yshift = -2.8cm] {2};
    \end{tikzpicture}
    \caption{ $\longbraid$ with $n = 0$ and $k = 3$}
    \label{fig:longcox}
  \end{subfigure}
  \caption{Braiding diagrams for modified Coxeter relations.}
  \label{fig:coxeter-braid-mod}
\end{figure}

\begin{toappendix}
  \begin{definition}[$\longcoxstar : {\List[\Fin[n]]} \to {\List[\Fin[n]]} \to {\UU}$]
    \label{def:longcox-star}
    \begin{align*}
      \reflr{\longcox}
       & : \forall w \to w \longcoxstar w                                                                                  \\
      \transr{\longcox}
       & : \forall w_{1}, w_{2}, w_{3} \to  w_{1} \longcox w_{2} \to w_{2} \longcoxstar w_{3} \to w_{1} \longcoxstar w_{3}
    \end{align*}
  \end{definition}

  \begin{definition}[$\longcoxplus : {\List[\Fin[n]]} \to {\List[\Fin[n]]} \to {\UU}$]
    \label{def:longcox-plus}
    \begin{align*}
      \relr{\longcoxplus}
       & : \forall w_{1}, w_{2} \to w_{1} \longcox w_{2} \to w_{1} \longcoxplus w_{2}                                      \\
      \transr{\longcoxplus}
       & : \forall w_{1}, w_{2}, w_{3} \to  w_{1} \longcox w_{2} \to w_{2} \longcoxplus w_{3} \to w_{1} \longcoxplus w_{3}
    \end{align*}
  \end{definition}
\end{toappendix}

Note that the previous $\braid$ rule is a special case of $\longbraid$, with $k = 0$. As before, the left-hand sides of
the relation are (lexicographically) strictly larger than the right-hand sides. We define the transitive closure of
$\longcox$ to be $\longcoxplus$ (\cref{def:longcox-plus}) and its reflexive-transitive closure to be $\longcoxstar$
(\cref{def:longcox-star}).

Despite the increased complexity of the generators, the rewriting system $(\List[\Fin[n]],\longcox)$ has the properties
we desire. It satisfies local confluence, that is, the Church-Rosser (diamond) property -- for example, using
$\longbraid$, we can now show the problematic critical pair \cref{fig:critical-pairs-non-converging} converges, and, it
is terminating, so by Newman's lemma, it produces a unique normal form. We follow the terminology
of~\citet*{huetConfluentReductionsAbstract1980a,krausCoherenceWellFoundednessTaming2020} to state our results formally.

\begin{theorem}
  \label{prop:nf}
  \leavevmode
  \begin{enumerate}
    \item $\longcox$ is (locally) confluent. For every span $\coxspan{w_{1}}{w_{2}}{w_{3}}$, there is a matching
          extended cospan $\coxcospan*{w_{2}}{w_{3}}{w}$.
    \item $\longcoxplus$ is terminating. For every $w \longcoxplus v$, it holds that $v < w$, where $<$ is the (well-founded)
          lexicographic ordering on $\List[\Fin[n]]$.
    \item $\longcoxstar$ is confluent. For every extended span $\coxspan*{w_{1}}{w_{2}}{w_{3}}$, there is a matching
          extended cospan $\coxcospan*{w_{2}}{w_{3}}{w}$.
    \item For every $w$, there exists a unique normal form $v$ such that $w \longcoxstar v$.~\label{prop:nf-uniqueness}
  \end{enumerate}
\end{theorem}

The modified form of the Coxeter relations are unwieldy and difficult to prove properties about by induction. However,
we can make the following observation relating it to $\cox*$.

\begin{proposition}
  \label{prop:coxlongcox}
  The relations $\cox*$ and $\longcoxstar$ are equivalent in the following sense: for every $w$ and $v$, $w \cox* v$ iff
  there is a $u$ such that $\coxcospan*{w}{v}{u}$.
\end{proposition}

By~\Cref{prop:nf}~\ref{prop:nf-uniqueness}, we get a unique choice function $\normf : {\List[\Fin[n]]} \to
  {\List[\Fin[n]]}$ that produces a normal form for terms of $\List[\Fin[n]]$. We state two important properties enjoyed
by $\normf$.

\begin{proposition}
  \leavevmode
  \begin{enumerate}
    \item For all $l : \List[\Fin[n]]$, we have that $l \cox* \normf(l)$.
    \item $\normf$ is idempotent, that is, $\normf \comp \normf \htpy \normf$.
  \end{enumerate}
\end{proposition}

\noindent Finally, we define the type $\Sn$ as the set-quotient of $\List[\Fin[n]]$ by $\cox*$,

\begin{definition}[$\Sn$]
  \(\Sn \defeq \quot{\List(\Fin[n])}{\cox*}\)
\end{definition}

\noindent Note that reductions need not be unique, hence $w \cox* v$ is not necessarily a proposition. So, the quotient
$\Sn$ is not effective, that is, $\quotrel : w \cox* v \to q(w) \id q(v)$ is not an equivalence. Using the $\normf$
function, we could instead define a new relation $\approx$ to equate those terms that have the same normal form, $(w
  \approx v) \defeq (\normf(w) \id \normf(v))$. This relation is prop-valued, and we could quotient $\List[\Fin[n]]$ by
$\approx$, obtaining an equivalent definition for $\Sn$.

\begin{proposition}
  \leavevmode
  \begin{enumerate}
    \item $\normf$ splits into a section-retraction pair, that is, we have ${\List[\Fin[n]]} \xrightarrow{s} \Sn[n]
            \xrightarrow{r} {\List[\Fin[n]]}$ such that $s \comp r \htpy \normf$ and $r \comp s \htpy \idfunc_{\Sn[n]}$.
    \item \(\im{\quotinc} \eqv \Sn \eqv \im{\normf} \), where $\quotinc$ is the mapping into the quotient, and
          $\im{f}$ denotes the image of $f$ (see~\cref{def:im}).
  \end{enumerate}
\end{proposition}

Notice however, that a group presentation, as defined in~\Cref{def:presentation}, requires the relation to be on the set
of words $A + A$, where the right copy corresponds to the set of formal inverses of the generators. The constructor
$\cancel$ specifies that the inverse of each element is again the same element, using which we can show that our
definition of $\Sn$ is equivalent to the definition of a presented group, by lifting the $\cox*$ relation.

\begin{theorem}
  \leavevmode
  \begin{enumerate}
    \item There is a group structure on $\Sn$, where the identity element is $\nil$, multiplication is given by list
          append, and inverse is given by list reversal.
    \item $\Sn$ is equivalent to the group presented by generators $\Fin[n]$ with the relations given by $\cox*$
          extended to $\List(\Fin[n] + \Fin[n])$ along the codiagonal map $\nabla_{A} : A + A \to A$.
  \end{enumerate}
\end{theorem}

To decide if two words in $\List[\Fin[n]]$ are $\cox*$-equal, we simply have to compute their normal forms using
$\normf$. They correspond to the same permutation if and only if these normal forms are equal, which is decidable for
$\List[\Fin[n]]$.

\subsection{Lehmer Codes}
\label{subsec:lehmer}

To prove the equivalence between $\Aut[\Fin[n]]$ and $\Sn[n]$, we will need to define functions back and forth between
the two types. The terms in $\Sn$ can be identified with equivalence classes of terms in $\List[\Fin[n]]$ with respect
to the Coxeter relation $\cox*$. The easiest way to define a function out of this presentation is to define it on the
representatives. We know that these are the unique normal forms in the set-quotient given by $\quotinc \comp \normf$,
but now we will explicitly describe what these representatives look like, using an encoding called Lehmer
codes~\cite{lehmerTeachingCombinatorialTricks1960}.

Lehmer codes are known in Combinatorial Analysis~\cite{bellmanCombinatorialAnalysis1960} where they are sometimes called
"subexcedant sequences", or "factoriadics". They can be written as a decimal number in the factorial number system, or
as a tuple encoding the digits~\cite*{knuthArtComputerProgramming1997,laisantNumerationFactorielleApplication1888}. This
gives a convenient way of representing permutations on a computer, partly because they are
bitwise-optimal~\cite{bergerTeachingOrdinalPatterns2019a}.

Formally, we define $\Lehmer[n]$ to be an $n+1$-element tuple, where the position $k \leq n$ stores an element of
$\Fin[k]$. Since the 0-th position is trivial, in practice it is
ignored~\cite{duboisTestsProofsCustom2018,vajnovszkiNewEulerMahonian2011}.

\begin{definition}[$\Lehmer : \Nat \to \UU$]
  \begin{align*}
    \Lehmer[\zero]   & \defeq \Fin[\suc[\zero]]                     \\
    \Lehmer[\suc[n]] & \defeq \Lehmer[n] \times \Fin[\suc[\suc[n]]]
  \end{align*}
\end{definition}

Given a permutation $\sigma : \Aut[\Fin[n]]$, for any element $i: \Fin[n]$, we can define the inversion count of $i$ as
the number of smaller elements appearing after it in the permutation.

\begin{definition}[Inversion count]
  Given a permutation $\sigma : \Aut[\Fin[n]]$, the inversion count of $i: \Fin[n]$ is given by
  \[ \invcount{i} = \#\Set{j < i | \sigma(j) > \sigma(i)}. \]
\end{definition}

Just knowing the inversion counts for all the elements, one can reconstruct the starting permutation. Also, observe that
$\invcount{i} < i$, thus fitting in the $i$-th place of a Lehmer code tuple. As an example, consider the following
tabulated presentation of a permutation of $\Fin[5]$.
\[
  \sigma =
  \begin{pmatrix}
    0      & 1      & 2      & 3      & 4      \\
    \el{2} & \el{1} & \el{4} & \el{0} & \el{3} \\
  \end{pmatrix}
\]


\noindent The Lehmer code for the permutation $\sigma$ is then the 5-tuple
\[
  l = (\invcount{0}, \invcount{1}, \invcount{2}, \invcount{3}, \invcount{4}) = (0, 1, 2, 0, 2)
\]

\noindent To decode the permutation back from this Lehmer code, we perform an algorithm similar to \emph{insertion
  sort}. The element of the Lehmer code being currently processed is highlighted in the left column of the table below.
Starting from a sorted list, the element at index $k$ has to be given $l[k]$ inversions. Because of the invariant
that all the elements before newly processed one are smaller than it, the proper number of inversions is created by
simply shifting the element $l[k]$ places left.

\begin{center}
  \begin{tabular}{l|r}
    (\highlight{{0}}, 1, 2, 0, 2) & $[\highlightAlt{\el{0}}, \el{1}, \el{2}, \el{3}, \el{4}]$ \\
    (0, \highlight{{1}}, 2, 0, 2) & $[\highlightAlt{\el{1}}, \el{0}, \el{2}, \el{3}, \el{4}]$ \\
    (0, 1, \highlight{{2}}, 0, 2) & $[\highlightAlt{\el{2}}, \el{1}, \el{0}, \el{3}, \el{4}]$ \\
    (0, 1, 2, \highlight{{0}}, 2) & $[\el{2}, \el{1}, \el{0}, \highlightAlt{\el{3}}, \el{4}]$ \\
    (0, 1, 2, 0, \highlight{{2}}) & $[\el{2}, \el{1}, \highlightAlt{\el{4}}, \el{0}, \el{3}]$ \\
  \end{tabular}
\end{center}

Writing formally, to turn a Lehmer code into a word in $\Sn$, we define a function $\immersion$. As described above, the
number $r$ at position $k$ in the tuple describes how many inversions the element $\el{k}$ has. Thus, we need to perform
$r$ many adjacent transpositions to get to the desired position, which is given by $(\suc[n] - r) \downf r$.

\begin{definition}[$\immersion : (n : \Nat) \to {\Lehmer[n]} \to {\List[\Fin[\suc[n]]]}$]
  \begin{align*}
    \immersion_{\zero}(\zero)    & \defeq \nil                                               \\
    \immersion_{\suc[n]}((r, l)) & \defeq \immersion_{n}(l) \append ((\suc[n] - r) \downf r)
  \end{align*}
\end{definition}

\noindent We can show that the function $\immersion_{n}$ gives an equivalence betweeen $\Lehmer[n]$ and $\im{\normf}$.

\begin{theoremrep}
  \leavevmode
  \begin{enumerate}
    \item For any Lehmer code $c$, $\immersion_{n}(c)$ is a normal form with respect to $\longcoxstar$, that is,
          $\immersion_{n}(c)$ is in $\im{\normf}$.
    \item Any element of $\im{\normf}$ can be constructed from a unique Lehmer code by $\immersion$, that is, the fibers
          of $\immersion_{n} : \Lehmer[n] \to {\im{\normf}}$ are contractible.
  \end{enumerate}
  Therefore, there is an equivalence between $\Lehmer[n]$ and $\im{\normf}$.
\end{theoremrep}

\begin{proof}
  For any code $c : \Lehmer[n]$, we have that
  \[
    \immersion_{n}(c) = (r_0 \downf k_0) \append (r_1 \downf k_1) \append \dots \append (r_{m-1} \downf k_{m-1})
  \]
  is a concatenation of $n \geq m \geq 0$ non-empty strictly decreasing lists. Reductions in $\longcox$ cannot happen
  inside any of the strictly decreasing $(r \downf k)$: $\longcancel$ requires repeating elements, $\longswap$ acts when
  consecutive numbers differ by at least 2, and $\longbraid$ acts on a non-monotone sequence. This leaves the case of
  reduction happening on a fragment that borders two (or more) subsequences. Again, $\longcancel$ requires two equal
  consecutive numbers, which would then have to be the last one in some $(r_i \downf k_i)$ sequence, and the first one
  in the next $(r_{i+1} \downf k_{i+1})$. But the first number in a sequence $(r_{i+1} \downf k_{i+1})$ is larger than
  every number in $(r_{i} \downf k_i)$ -- which also shows why $\longswap$ cannot happen. The remaining case of
  $\longbraid$ follows similarly, since its argument is a decreasing sequence followed immediately by a number equal to
  the first element of this sequence.
\end{proof}


\begin{corollary}
  \label{prop:sn-im-lehmer-equiv}
  For all $n : \Nat$, \( \Sn \eqv \im{\normf} \eqv \Lehmer[n] \).
\end{corollary}

\subsection{Running Lehmer codes}

Finally, it is time to complete our goal of characterising the permutation groups. Having produced a Lehmer code by
normalising words in $\Sn$, we need to run it to produce a concrete bijection of finite sets, and, given a bijection
between finite sets, we need to encode it as a Lehmer code. We will prove that these maps construct an equivalence
between the types $\Lehmer[n]$ and $\Aut[\Fin[\suc[n]]]$. The idea for this proof is borrowed
from~\cite{Molzer-cubical}. ~\footnote{Note that the indices for the type of permutations are off-by-one, because we
  chose $\Fin[n]$ to represent generators for permutations on $\Fin[\suc[n]]$.}

\begin{definition}[${\FinExcept{n}} : {\Fin[n]} \to \UU$]
  For $n : \Nat$, the type family $\FinExcept{n}$ picks out all elements in $\Fin[n]$ except the one in the argument.
  \[
    \FinExcept{n}[i] \defeq \dsum*{j : \Fin[n]}{i \neq j}
  \]
\end{definition}

\noindent Note that $\FinExcept{n}[i]$ for $i : \Fin[n]$ is a subtype of $\Fin[n]$ and is hence a set. We state and
prove a few auxiliary lemmas about how $\FinExcept{n}$ interacts with $\Fin$ -- these are obtained by counting arguments
using the fact that $\Fin[n]$ and $\FinExcept{n}$ both have decidable equality.

\begin{lemmarep}
  \label{prop:fin-finexcept}
  \leavevmode
  \begin{enumerate}
    \item For any $k : \Fin[\suc[n]]$, we have $\FinExcept{\suc[n]}[k] \eqv \Fin[n]$. \label{prop:fin-finexcept-2}
    \item For any $n : \Nat$, we have \( \Aut[\Fin[\suc[n]]] \eqv \dsum*{k : \Fin[\suc[n]]}{(\FinExcept{\suc[n]}[n] \eqv
            \FinExcept{\suc[n]}[n - k])} \). \label{prop:fin-finexcept-3}
  \end{enumerate}
\end{lemmarep}

\begin{proof}
  The first and second propositions follow from simply constructing the bijections using the decidable equality of
  $\Fin[n]$, and making sure to punch-in and punch-out the element $k$ at the right place.

  The third proposition performs some combinatorial tricks. On the left, we have the type of automorphisms of
  $\Fin[\suc[n]]$. Assume a particular $\sigma : \Fin[\suc[n]] \xrightarrow{\sim} \Fin[\suc[n]]$. Pick $k$ to be the
  inversion count of $n$, the largest element in $\Fin[\suc[n]]$. Then, the image of $n$ under $\phi$ has to be $n - k$,
  since all other elements in the set are smaller. Removing those two from the domain and codomain of $\phi$, the rest
  of the elements are fixed by $\sigma$, so we compute the bijection between the rest of the elements.

  For the other direction, if we are given a $k$ and a bijection $\pi$ between $\FinExcept{\suc[n]}[n]$ and
  $\FinExcept{\suc[n]}[n - k]$, we can extend $\pi$ to $\sigma : \Fin[\suc[n]] \eqv \Fin[\suc[n]]$ by inserting the
  element $n$ at the position $n - k$, resulting in the element $n$ having inversion count $k$.
\end{proof}

\noindent Using these facts, we can now prove the main result of this section.

\begin{theorem}
  \label{prop:lehmer-aut-equiv}
  For all $n:\Nat$, \( \Lehmer[n] \eqv \Aut[\Fin[\suc[n]]] \).
\end{theorem}

\begin{proof}
  For $n = \zero$, note that $\Lehmer[\zero]$ is contractible, and so is $\Aut[\Fin[\suc[\zero]]]$. For $n = \suc[m]$,
  we compute a chain of equivalences.
  \begin{gather*}
    \arraycolsep=0.5em\def\arraystretch{1.5}
    \begin{array}{rl}
           & \Lehmer[\zero]          \\
      \eqv & \unit                   \\
      \eqv & \Aut[\Fin[\suc[\zero]]]
    \end{array}
    \qquad\qquad
    \begin{array}{rlr}
           & \Lehmer[\suc[m]]                                                                                    &                                                                 \\
      \eqv & \Fin[\suc[\suc[m]]] \times \Lehmer[m]                                                               & \text{by definition}                                            \\
      \eqv & \Fin[\suc[\suc[m]]] \times \Aut[\Fin[\suc[m]]]                                                      & \text{induction hypothesis}                                     \\
      \eqv & \dsum*{k : \Fin[\suc[\suc[m]]]}{\Fin[\suc[m]] \eqv \Fin[\suc[m]]}                                   & \text{$\Sigma$ over a constant family}                          \\
      \eqv & \dsum*{k : \Fin[\suc[\suc[m]]]}{\FinExcept{\suc[\suc[m]]}[m] \eqv \Fin[\suc[m]]}                    & \text{by~\cref{prop:fin-finexcept}~\cref{prop:fin-finexcept-2}} \\
      \eqv & \dsum*{k : \Fin[\suc[\suc[m]]]}{\FinExcept{\suc[\suc[m]]}[m] \eqv \FinExcept{\suc[\suc[m]]}[m - k]} & \text{by~\cref{prop:fin-finexcept}~\cref{prop:fin-finexcept-2}} \\
      \eqv & \Aut[\Fin[\suc[\suc[m]]]]                                                                           & \text{by~\cref{prop:fin-finexcept}~\cref{prop:fin-finexcept-3}}
    \end{array}
  \end{gather*}
\end{proof}

\noindent By composing~\Cref{prop:lehmer-aut-equiv} and~\Cref{prop:sn-im-lehmer-equiv}, we obtain the final equivalence.

\begin{corollary}
  \label{prop:sn-lehmer-fin-equiv}
  For all $n : \Nat$,
  \(
  \Sn \eqv \Lehmer[n] \eqv \Aut[\Fin[\suc[n]]]
  \).
\end{corollary}

\section{Correspondence between \texorpdfstring{$\PiLang$}{Pi} and \texorpdfstring{$\UFin$}{UFin}}
\label{sec:equivalence}

In this section, we first translate $\PiLang$ to its additive fragment $\PiPlusLang$. This is the language that we
interpret to $\UFin$ and back, using the tools developed in the previous sections. Further, we go through an
intermediate step of the language $\PiHatLang$, which is a simplified variant of $\PiPlusLang$ that uses adjacent
transpositions for combinators, while preserving all the required structure.

\[\begin{tikzcd}
    \PiLang && \PiPlusLang && \PiHatLang && \UFin
    \arrow["\evalt", from=1-1, to=1-3]
    \arrow["\evalp", curve={height=-24pt}, from=1-3, to=1-5]
    \arrow["\evalh", curve={height=-24pt}, from=1-5, to=1-7]
    \arrow["\quotep", curve={height=-24pt}, from=1-5, to=1-3]
    \arrow["\quoteh", curve={height=-24pt}, from=1-7, to=1-5]
  \end{tikzcd}\]

We present the types and 1-combinators of $\PiPlusLang$ and $\PiHatLang$ in~\Cref*{fig:piplus,fig:pihat} respectively,
eliding the 2-combinators for brevity. We enforce that there is a unique 2-combinator between compatible 1-combinators,
by relating them with a truncation. These can be found in~\cref{app:leveltwo} and in the accompanying Agda code.

The translations between the languages are defined separately on types, 1-combinators, and 2-combinators. Following the
terminology of Normalisation by Evaluation, the translations from the left to the right, going from the syntax towards
the semantics, are called $\evalt$ and the translations the other way are called $\quotet$.

To state our results formally, we organise the syntax for each language using a technical device, called a syntactic
category. We define them formally in the appendix, and only state our results here. For each of the $\PiLang$,
$\PiPlusLang$ and $\PiHatLang$ languages, their syntactic categories, respectively $\PiCat$, $\PiPlusCat$ and
$\PiHatCat$, have 0-cells for types, 1-cells for 1-combinators, and 2-cells for 2-combinators. We can show that these
syntactic categories here are actually $(2,0)$-categories, since all the 1-cells and 2-cells are invertible. They are
also locally-strict, or locally-posetal, because there is at most one 2-cell between compatible 1-cells.

\begin{toappendix}
  \begin{proposition}
    We can form a weak 2-category $\PiCat$ with
    \begin{itemize}
      \item $\PiLang$ types ($U$) for 0-cells,
      \item for $X, Y : U$, a collection of 1-cells $X \iso Y$,
      \item for $p, q : X \iso Y$, a collection of 2-cells $p \Iso q$.
    \end{itemize}
  \end{proposition}

  \begin{proposition}
    We can form a weak 2-category $\PiPlusCat$ with
    \begin{itemize}
      \item $\PiPlusLang$ types ($\UPlus$) for 0-cells,
      \item for $X, Y : \UPlus$, a collection of 1-cells $X \isop Y$,
      \item for $p, q : X \isop Y$, a collection of 2-cells $p \Isop q$.
    \end{itemize}
  \end{proposition}

  \begin{proposition}
    We can form a weak 2-category $\PiHatCat$ with
    \begin{itemize}
      \item $\PiHatCat$ types ($\UHat$) for 0-cells,
      \item for $X, Y : \UHat$, a collection of 1-cells $X \isoh Y$,
      \item for $p, q : X \isoh Y$, a collection of 2-cells $p \Isoh q$.
    \end{itemize}
  \end{proposition}
\end{toappendix}

We use the $\evalt/\quotet$ translation maps to construct functors between these categories. We only name the maps on
the 0, 1, and 2-cells -- the coherences hold by definition or by calculation, which is shown in our accompanying Agda
code. We use these functors to state our results establishing the equivalences between the languages.

\subsection{$\PiLang$ to $\PiPlusLang$}

First, we show how to translate $\PiLang$ programs to $\PiPlusLang$, which is the additive fragment of $\PiLang$. The
syntax for 1-combinators is given in~\Cref{fig:piplus}.

\begin{figure}[t]
  {\scalebox{\scalef}{$
        \begin{array}{rrcll}
          \idc :     & A           & \iso & A           & : \idc     \\
          \identlp : & \zerot + A  & \iso & A           & : \identrp \\
          \swapp :   & A + B       & \iso & B + A       & : \swapp   \\
          \assoclp : & A + (B + C) & \iso & (A + B) + C & : \assocrp \\ [1.5ex]
        \end{array}$}}

  {\scalebox{\scalef}{
      \Rule{}
      {\jdg{}{}{c_1 : A \iso B} \quad \vdash c_2 : B \iso C}
      {\jdg{}{}{c_1 \fatsemi c_2 : A \iso C}}
      {}

      \Rule{}
      {\jdg{}{}{c_1 : A \iso B} \quad \vdash c_2 : C \iso D}
      {\jdg{}{}{c_1 \oplus c_2 : A + C \iso B + D}}
      {}
    }}
  \caption{$\PiPlusLang$ syntax}
  \label{fig:piplus}
\end{figure}

$\PiLang$ has two 0-ary type constructors, and two binary type constructors -- the additive tensor product and the
multiplicative one. $\PiPlusLang$ has all the type constructors of $\PiLang$ except multiplication. However, we will
show how to recover the multiplicative structure, by defining multiplication as repeated addition. We encode $\times$ in
terms of $+$ as follows.

\begin{definition}[$\times : \UPlus \to \UPlus \to \UPlus$]
  \begin{align*}
    \zerot \times Y      & \defeq \zerot                      \\
    \onet \times Y       & \defeq Y                           \\
    (X_1 + X_2) \times Y & \defeq X_1 \times Y + X_2 \times Y
  \end{align*}
\end{definition}

\begin{lemma}
  There are two symmetric monoidal structures on $\PiPlusCat$, given by $(\zerot, +)$ and $(\onet, \times)$, with
  $\times$ distributing over $+$, giving it a rig structure.
\end{lemma}

\noindent Using this rig structure, we translate $\PiLang$ to $\PiPlusLang$, constructing a rig equivalence from
$\PiCat$ to $\PiPlusCat$.

\begin{definition}[$\evalt$]
  \begin{align*}
    \evalt_{0} & : U \to \UPlus                                             \\
    \evalt_{1} & : (c : X \iso Y) \to \evalt_{0}(X) \iso \evalt_{0}(Y)      \\
    \evalt_{2} & : (\alpha : p \Iso q) \to \evalt_{1}(p) \Iso \evalt_{1}(q)
  \end{align*}
\end{definition}

\begin{theorem}
  $\evalt$ gives a rig equivalence between $\PiCat$ and $\PiPlusCat$.
\end{theorem}

\subsection{$\PiPlusLang$ to $\PiHatLang$}

Next, we show how to translate $\PiPlusLang$ programs to $\PiHatLang$ and back. $\PiHatLang$ is a simplified variant of
$\PiPlusLang$, with (unary) natural numbers for 0-cells, 1-combinators generated by adjacent transpositions, and an
appropriate set of 2-combinators. We give the syntax, again omitting 2-combinators, in~\Cref{fig:pihat}.

\begin{figure}[t]
  {\scalebox{\scalef}{$
        \begin{array}{rrcll}
          \idc :   & n             & \isoh & n             & : \idc   \\
          \swapc : & \suc[\suc[n]] & \isoh & \suc[\suc[n]] & : \swapc \\
        \end{array}
      $}}

  {\scalebox{\scalef}{
      \Rule{}
      {\jdg{}{}{c_1 : n \isoh m} \quad \vdash c_2 : m \isoh o}
      {\jdg{}{}{c_1 \fatsemi c_2 : n \isoh o}}
      {}

      \Rule{}
      {\jdg{}{}{c : n \isoh m}}
      {\jdg{}{}{\oplus(c) : \suc[n] \isoh \suc[m]}}
      {}
    }}
  \caption{$\PiHatLang$ syntax}
  \label{fig:pihat}
\end{figure}

As described, $\PiHatLang$ doesn't have a tensor product, but we can build it simply by adding up natural numbers, and,
we need to verify that this indeed equips $\PiHatCat$ with a symmetric monoidal structure.~\footnote{Since each object
is a natural number, this makes $\PiHatCat$ a \emph{PROP}, that is, a products and permutations category.}

To produce a braiding $n + m \isoh m + n$ from adjacent tranpositions, we recursively traverse the left subexpression,
swapping each element using adjacent transpositions along the elements on the right, placing it in the right position.
The computational content of this translation can be visualised using tree transformations, for the recursive case
in~\cref{fig:plusplusswap}. The challenging part is showing that these moves are coherent with respect to 2-combinators.

\begin{figure}
   \[
      \Tree [ [ {\tiny 1} {\tiny n} ] {\tiny m} ] ~\xrightarrow{}~
      \Tree [ {\tiny 1} [ {\tiny n} {\tiny m} ] ] ~\xrightarrow{}~
      \Tree [ [ {\tiny n} {\tiny m} ] {\tiny 1} ] ~\xrightarrow{}~
      \Tree [ [ {\tiny m} {\tiny n} ] {\tiny 1} ] ~\xrightarrow{}~
      \Tree [ {\tiny m} [ {\tiny n} {\tiny 1} ] ] ~\xrightarrow{}~
      \Tree [ {\tiny m} [ {\tiny 1} {\tiny n} ] ] ~
    \]
    \caption{Braiding from transpositions, recursive case}
    \label{fig:plusplusswap}
\end{figure}

\begin{lemma}
  $\PiHatCat$ has a symmetric monoidal structure, with the unit given by 0 and the tensor product given by natural
  number addition.
\end{lemma}

\noindent Using this symmetric monoidal structure, we translate from $\PiPlusLang$ to $\PiHatLang$.

\begin{definition}[$\evalp$]
  \begin{align*}
    \evalp_{0} & : \UPlus \to \UHat                                                           \\
    \evalp_{1} & : (c : t_{1} \isop t_{2}) \to \evalp_{0}(t_{1}) \isoh \evalp_{0}(t_{2})      \\
    \evalp_{2} & : (\alpha : c_{1} \Isop c_{2}) \to \evalp_{1}(c_{1}) \Isoh \evalp_{1}(c_{2})
  \end{align*}
\end{definition}

To go back from $\PiHatLang$ to $\PiPlusLang$, we turn a natural number into a $\PiPlusLang$ type, using right-biased
addition, that is, the natural number $n$ gets mapped to the type $\onet + (\onet + (\onet + \ldots + \zerot))$. Since
the types are already right-biased, an adjacent transposition in $\PiHatLang$ is easily encoded by using the braiding in
$\PiPlusLang$, as shown in~\cref{fig:transpfrombraid}. Again, these are shown to be coherent.

\begin{figure}
  \[
    \Tree [ {\tiny $\onet$} [ {\tiny $\onet$} {\tiny X} ] ] ~\xrightarrow{\assoclp}~
    \Tree [ [ {\tiny $\onet$} {\tiny $\onet$} ] {\tiny X} ] ~\xrightarrow[\swapp\phantom{xx}]{\phantom{xx}\idc}~
    \Tree [ [ {\tiny $\onet$} {\tiny $\onet$} ] {\tiny X} ] ~\xrightarrow{\assocrp}~
    \Tree [ {\tiny $\onet$} [ {\tiny $\onet$} {\tiny X} ] ] ~
  \]
  \caption{Transpositions from braiding}
  \label{fig:transpfrombraid}
\end{figure}

\begin{definition}[$\quotep$]
  \begin{align*}
    \quotep_{0} & : \UHat \to \UPlus                                                            \\
    \quotep_{1} & : (p : X_{1} \isoh X_{2}) \to \quotep_{0}(X_{1}) \iso \quotep_{0}(X_{2})      \\
    \quotep_{2} & : (\alpha : p_{1} \Isoh p_{2}) \to \quotep_{1}(p_{1}) \Iso \quotep_{1}(p_{2})
  \end{align*}
\end{definition}

\begin{theorem}
  $\evalp/\quotep$ give a symmetric monoidal equivalence between $\PiPlusCat$ and $\PiHatCat$.
\end{theorem}

\subsection{$\PiHatLang$ to $\UFin$}

Finally, we show how to interpret $\PiHatLang$ to $\UFin$, and back from $\UFin$ to $\PiHatLang$. Types in $\PiHatLang$
are interpreted as 0-cells in $\UFin$, that is, a natural number $n$ is mapped to $\Fin[n]$. The 1-combinators in
$\PiHatLang$ are mapped to 1-paths in $\UFin$, that is, 1-loops in each connected component, equivalent to
$\Aut[\Fin[n]]$. In $\PiHatLang$, the 1-combinators are generated by adjacent transpositions, so these can be mapped to
words in $\Sn$ and then to automorphisms using~\Cref{prop:sn-lehmer-fin-equiv}. Finally, 2-combinators are mapped to
2-paths between loops in $\UFin$.

\begin{definition}[$\evalh$]
  \begin{align*}
    \evalh_{0} & : \UHat \to \UFin                                                          \\
    \evalh_{1} & : (c : t_{1} \isoh t_{2}) \to \evalh_{0}(t_{1}) \id \evalh_{0}(t_{2})      \\
    \evalh_{2} & : (\alpha : c_{1} \Isoh c_{2}) \to \evalh_{1}(c_{1}) \id \evalh_{1}(c_{2})
  \end{align*}
\end{definition}

\noindent 0-cells in $\UFin$ are mapped to their cardinalities in $\PiHatLang$, 1-loops are decoded to words in $\Sn$ to
generate a sequence of adjacent transpositions, producing a 1-combinator in $\PiHatLang$. Finally, 2-paths are quoted
back to 2-combinators in $\PiHatLang$.

\begin{definition}[$\quoteh$]
  \begin{align*}
    \quoteh_{0} & : \UFin \to \UHat                                                            \\
    \quoteh_{1} & : (p : X_{1} \id X_{2}) \to \quoteh_{0}(X_{1}) \isoh \quoteh_{0}(X_{2})      \\
    \quoteh_{2} & : (\alpha : p_{1} \id p_{2}) \to \quoteh_{1}(p_{1}) \Isoh \quotet_{1}(p_{2})
  \end{align*}
\end{definition}


\begin{theorem}
  $\evalh/\quoteh$ give a symmetric monoidal equivalence between $\PiHatCat$ and $\UFin$.
\end{theorem}

The semantics that we presented here takes a different route to constructing the permutation from a $\PiLang$
combinator, compared to the direct interpretation given using the big-step interpreter in~\Cref{subsec:denotational}. We
verify that the two semantics agree, establishing that the semantics is adequate and fully abstract.

\begin{definition}[${\gdenot{\blank}}$]
  \begin{gather*}
    \begin{aligned}
      \gdenot{\blank}_{0} & : U \to \UFin                                       \\
      \gdenot{\blank}_{0} & \defeq \evalh_{0} \comp \evalp_{0} \comp \evalt_{0}
    \end{aligned}
    \qquad
    \begin{aligned}
      \gdenot{\blank}_{1} & : (c : X \iso Y) \to \gdenot{X}_{0} \id_{\UFin} \gdenot{Y}_{0} \\
      \gdenot{\blank}_{1} & \defeq \evalh_{1} \comp \evalp_{1} \comp \evalt_{1}
    \end{aligned}
  \end{gather*}
\end{definition}

\begin{theorem}[Full Abstraction and Adequacy]
  For any $c_1, c_2 : X \iso Y$, we have that
  \[
    \denot{c_1} = \denot{c_2} \text{ if and only if } \gdenot{c_1}_{1} = \gdenot{c_2}_{1}
  \]
\end{theorem}

\section{Normalisation of Reversible Circuits}
\label{sec:applications}

Using our semantics, we can normalise, synthesise, prove equivalence, and generally reason about~$\PiLang$ programs. The
two key definitions are presented below.

\begin{definition}[Normalisation of $\PiLang$ programs]
  \begin{gather*}
    \begin{aligned}
      \normt_{0} & : U \to \UPlus                                            \\
      \normt_{0} & = \quotep_{0} \comp \quoteh_{0} \comp \gdenot{\blank}_{0}
    \end{aligned}
    \qquad
    \begin{aligned}
      \normt_{1} & : (c : X \iso Y) \to \normt_{0}(X) \iso \normt_{0}(Y)     \\
      \normt_{1} & = \quotep_{1} \comp \quoteh_{1} \comp \gdenot{\blank}_{1}
    \end{aligned}
  \end{gather*}
\end{definition}

\noindent Normalisation involves translating $\PiLang$ programs to $\PiHatLang$, computing a permutation, and quoting
back to $\PiPlusLang$. Note that the normalisation happens in the step from $\PiHatLang$ to $\UFin$ and back to
$\PiHatLang$ and that normalisation also provides a decision procedure for program equivalence. Synthesis happens by
quoting permutations. More general, user-guided, reasoning can be done using the sound and complete level-2
combinators to rewrite $\PiLang$ programs.

Recall the specification of reversible disjunction from \Cref{sec:examples}, the two Qiskit circuits for implementing
it, and the corresponding $\PiLang$ definitions \Afun{reversibleOr1} and \Afun{reversibleOr2}. The normal forms for both
circuits compute to the following, establishing their equivalence:

\medskip
\resetnormtwo{}

Instead of manually producing $\Pi$-programs to implement the reversible disjunction specification, it is also possible
to simply quote the desired permutation:

\medskip
\resetperm{}

\noindent The permutation uses the canonical encoding of sequences of bits as natural numbers (e.g., (\textsf{false},
\textsf{true},\textsf{true}) is encoded as 011 or 3).  The second entry maps index 1 (= 001) to the value 5 (= 101)
which states that since one of the right bits is set in 001 then the leftmost bit in the output is set. Quoting this
permutation generates the same normalised program which can then be composed with a map from $\PiLang$ to
$\PiPlusLang$ to produce a program matching the desired structured types.

\section{Discussion \& Related Work}
\label{sec:discussion}


The main theme of our work is the semantic foundation of reversible languages. We prove that a programming language
presentation of reversible programming based on algebraic types matches---exactly---the categorified group-theoretic
semantics, thereby closing the circle on a complete Curry-Howard-Lambek correspondence for reversible languages.
Historically, the first such correspondence was between the $\lambda$-calculus, intuitionistic logic, and
cartesian-closed categories~\cite{curryCurryEssaysCombinatory1980}. For reversible languages, the Curry-Howard
correspondence was established by~\citet{sparksSuperstructuralReversibleLogic2014} and the Lambek correspondence
suggested by~\citet{caretteComputingSemiringsWeak2016} and~\citet{threemodels} and established in this work. In the
remainder of this section, we discuss some broader related work.

\paragraph{Coherence and Rewriting}

Higher-order term-rewriting systems and word problems have a long history of being formalised in proof assistants like
{homotopy.io}, Agda, Coq and Lean~\cite{krausCoherenceWellFoundednessTaming2020}. As part of the proof of our main
result, we developed a rewriting system for the Coxeter relations for $\Sn$ to solve its word problem. \citet{Hiver-coq}
describes an explicit algorithm for producing normal forms. It could provide an alternative to our rewriting system. Other
encodings of permutations as listed vectors, matrices, inductively generated trees (Motzkin trees), Young diagrams, or
string diagrams, proved either difficult to formalise in type theory or difficult to relate directly to the primitive type
isomorphisms of $\PiLang$. The automatic \citet{knuthSimpleWordProblems1970} completion produced too many equations
making proving correctness and termination intractable.

Coherence theorems are famous problems in category theory, and Mac Lane's coherence
theorem~\cite{maclaneNaturalAssociativityCommutativity1963,joyalBraidedTensorCategories1993,gurskiInfiniteLoopSpaces2013}
for monoidal categories is a particular one. The use of rewriting and proof assistants to prove coherence theorems
for higher categorical structures has a long history, see~\cite*{forestCoherenceGrayCategories2018,beylinExtractingProofCoherence1996}.

\paragraph{Computing with Univalence} In HoTT, univalence characterises the path type in the universe as equivalences of
types. The map $\term{idtoeqv} : A \id_{\UU} B \to A \eqv B$ can be easily constructed using path induction. The term
$\term{ua} : A \eqv B \to A \id_{\UU} B$, its computation rule
$\term{ua_\beta} : (e : A \eqv B) \to \term{idtoeqv}(\term{ua}(e)) \id e$, and its extensionality rule
$\term{ua_\eta} : (p : A \id_{\UU} B) \to p \id \term{ua}(\term{idtoeqv(p)})$ are generally added as postulates when
formalising in Agda. Together, $\term{ua}$ and $\term{ua}_\beta$ give the full univalence axiom
$(A \eqv B) \eqv (A \id_{\UU} B)$.
By giving a computable presentation for a univalent subuniverse, we are able to describe its path space syntactically
via a complete equational axiomatisation of the equivalences between types in the subuniverse.
In the subuniverse of finite types, $\term{idtoeqv}$ corresponds to giving a denotation for a program (1-combinator),
which is easily done by induction. The $\term{ua}$ map corresponds to synthesising a program from an equivalence (which,
in general, is of course undecidable~\cite{krogmeierDecidableSynthesisPrograms2020}). In case of reversible boolean
circuits, it is decidable, as we have shown, but still far from trivial, which matches the need to assert the existence
of $\term{ua}$ without giving a constructive argument. Then, the computation rule $\term{ua_\beta}$ expresses the fact
that program synthesis is sound, while $\term{ua_\eta}$ corresponds to the soundness of the equational theory ($\PiLang$
2-combinators). Thus, our results suggest a new computational interpretation of the univalence principle, which
provides an intuition on why certain constructions are hard (or impossible in the general case). There are other,
different approaches to computing with univalence,
in~\cite*{angiuliInternalizingRepresentationIndependence2021,tabareauMarriageUnivalenceParametricity2021}, and in
Cubical Type
Theory\cite*{angiuliComputationalSemanticsCartesianCubical2019,vezzosiCubicalAgdaDependently2019}.












\paragraph{Algebraic Theories} In universal algebra, algebraic theories are used to describe structures such as groups
or rings. A specific group or ring is a model of the appropriate algebraic theory. Algebraic theories are usually
\emph{presented} in terms of logical syntax, that is, as first-order theories whose signatures allow only functional
symbols, and whose axioms are universally quantified equations. In his seminal thesis,
\citet{lawvereFUNCTORIALSEMANTICSALGEBRAIC1963} defined a presentation-free categorical notion of universal algebraic
structure, called a Lawvere theory. Programming Languages, such as the $\lambda$-calculus, can be viewed as algebraic
structures with variable-binding operators, which can be formalised using second-order algebraic
theories~\cite{fioreSecondOrderAlgebraicTheories2010}, or algebraic theories with closed
structure~\cite{hylandClassicalLambdaCalculus2017}, called $\lambda$-theories, making the $\lambda$-calculus the
presentation of the initial $\lambda$-theory $\Lambda$. Our family of reversible languages $\PiLang$ have been presented
as first-order algebraic
2-theories~\cite{cohenCoherenceRewriting2theories2009,bekeCategorificationTermRewriting2011,yanofskySyntaxCoherence2000},
which are a categorification of algebraic theories. The types $\zerot$ and $\onet$ are nullary function symbols, the
type formers $+$ and $\times$ are binary function symbols, the 1-combinators are invertible reduction rules, and the
2-combinators are equations or coherence diagrams of compositions of reduction rules. Just like models of Lawvere
theories are given by algebras of (finitary) monads on $\SetCat$, models of 2-theories are given by algebras of 2-monads
on $\CatCat$. Our development is related to the free symmetric monoidal completion 2-monad.

\paragraph{Free Symmetric Monoidal Category} The free symmetric monoidal category has been used to study
concurrency~\cite{hylandSymmetricMonoidalSketches2004}, Petri nets~\cite{baezCategoriesNets2021}, combinatorial
structures~\cite{fioreCartesianClosedBicategory2008}, quantum mechanics~\cite{abramskyAbstractScalarsLoops2005}, and
bicategorical models of (differential) linear logic~\cite{melliesTemplateGamesDifferential2019}. The forgetful functor
from $\SymMonCat$, the 2-category of (small) symmetric monoidal categories, strong symmetric monoidal functors, and
monoidal natural transformations, to the 2-category $\CatCat$, has a left adjoint giving the free symmetric monoidal
category $\FSM[\CCat{C}]$ on a category $\CCat{C}$. This is a 2-monad on
$\CatCat$~\cite{blackwellTwodimensionalMonadTheory1989}, whose algebras are (strict) symmetric monoidal categories. Its
construction is known in the literature~\cite{abramskyAbstractScalarsLoops2005}.  Concretely, the objects of
$\FSM[\CCat{C}]$ are given by lists of objects of $\CCat{C}$, that is, a pair $(n:\Nat, A:[n] \to \CCat{C}_{0})$;
morphisms between $(n,A)$ and $(n,B)$ are pairs $(\pi,\lambda)$ where $\pi$ is a permutation of $[n]$, and
$\lambda_{i} : A_{i} \to B_{\pi(i)}$ for $1 \leq i \leq n$. Abstractly, this is given by the Grothendieck construction
$\int F$ of the functor $F : \BFin \to \CatCat$ from the groupoid of finite sets and bijections to $\CatCat$, assigning
each natural number $n$ to the $n$-power $C^{n}$ of $C$, and each permutation on $[n]$ inducing an endofunctor on
$C^{n}$ by action. The groupoid $\BFin$ is the free symmetric monoidal category (groupoid) on one generator, $\FSM[\unit]$.

Coherence and normalisation problems for monoids in constructive type theory using coherence for monoidal categories
were studied by \citet{beylinExtractingProofCoherence1996}. In HoTT, coherence for the free monoidal groupoid over a
groupoid and the proof of its universal property has been considered
by~\citet{piceghelloCoherenceMonoidalGroupoids2020}. Free commutative monoids in type theory have been studied by
\citet{gylterudMultisetsTypeTheory2020}, and using HoTT by \citet{choudhuryFinitemultisetConstructionHoTT2019}. The free
symmetric monoidal groupoid $\FSM[A]$ over a groupoid $A$ can be given by $\dsum*{X:\UFin}{A^{X}}$, or it can be
presented as an algebraic 2-theory using 1-HITs. These HITs and the proof of their universal property have been
considered by \citet{piceghelloCoherenceSymmetricMonoidal2019,choudhuryFinitemultisetConstructionHoTT2019}. The proof of
the universal property of $\FSM$ is asserted by appealing to Mac Lane's coherence theorem for symmetric monoidal
categories, and using the fact that the finite symmetric group $\Sn$ encodes the permutation group on a finite set. The
existence of the proof is folklore. We have produced a proof and formalised it in constructive type theory. A stronger
universal property that $\BFin$ is biinitial in $\term{RigCat}$, is in~\cite{elguetaGroupoidFiniteSets2021}.

\paragraph{Reversible Programming Languages} The practice of programming languages is replete with \emph{ad hoc}
instances of reversible computations: database transactions, mechanisms for data provenance, checkpoints, stack and
exception traces, logs, backups, rollback recoveries, version control systems, reverse engineering, software
transactional memories, continuations, backtracking search, and multiple-level undo features in commercial
applications. In the early nineties, \citet{Baker:1992:LLL,Baker:1992:NFT} argued for a systematic, first-class,
treatment of reversibility. But intensive research in full-fledged reversible models of computations and reversible
programming languages was only sparked by the discovery of deep connections between physics and
computation~\cite{Landauer:1961,PhysRevA.32.3266,Toffoli:1980,bennett1985fundamental,Frank:1999:REC:930275,Hey:1999:FCE:304763,fredkin1982conservative},
and by the potential for efficient quantum computation~\cite{springerlink:10.1007/BF02650179}. The early developments of
reversible programming languages started with a conventional programming language, e.g., an extended $\lambda$-calculus,
and either (i) extended the language with a history
mechanism~\cite{vanTonder:2004,Kluge:1999:SEMCD,lorenz,danos2004reversible}, or (ii) imposed constraints on the control
flow constructs to make them reversible~\cite{Yokoyama:2007:RPL:1244381.1244404}.  More modern approaches recognize that
reversible programming languages require a fresh approach and should be designed from first principles without the
detour via conventional irreversible
languages~\cite{Yokoyama:2008:PRP,Mu:2004:ILRC,abramsky2005structural,DiPierro:2006:RCL:1166042.1166047}. The version of
$\PiLang$ studied in this paper is restricted to finite types and terminating total computations. It would be
interesting to understand which versions of free monoidal structures correspond to extensions of $\Pi$ with recursive
types~\cite{jamesInformationEffects2012,rc2011} and negative/fractional
types~\cite{chenComputationalInterpretationCompact2021}.

\paragraph{Permutations} Finding formal systems for expressing various flavors of computable functions has been a major
focus of logic and computer science since its inception. Permutations, being at the core of reversible computing, are an
interesting class of functions, for which there are few formal systems. We develop such a system bringing in all the
associated benefits of syntactic calculi, notably, their calculational flavor. Instead of comparing two reversible
programs by extensional equality of the underlying bijective functions, a calculus offers more nuanced techniques that
can enforce additional intensional constraints on the desired equality relation.

\bibliography{2dtypesZot,survey} 

\end{document}